\documentclass[aps,twocolumn, prX,nofootinbib, superscriptaddress]{revtex4}
\usepackage[usenames, dvipsnames]{xcolor}
\usepackage{graphicx, array}
\usepackage{graphics,epsfig}
\usepackage{amssymb}
\usepackage{amsmath}
\usepackage{ulem}
\usepackage{multirow}
\usepackage{ulem}
\usepackage{gensymb}
\usepackage{natbib}
\usepackage{appendix}
\usepackage{siunitx}
\usepackage{float}
\usepackage{orcidlink}

\makeatletter
\def\p@subsection{}
\makeatother

\usepackage{listings}
\usepackage{aas_macros}
\usepackage{comment}
\usepackage{soul}

\newcommand{\rmn}{\mathrm}
\usepackage{hyperref}

\DeclareSIUnit{\jansky}{Jy}
\usepackage{subcaption}
\usepackage{ragged2e}

\DeclareCaptionFormat{justified}{%
  \justifying
  #1#2#3
}

\begin{document}

\input{epsf}

\title{Optical Depths from the Thermal Sunyaev-Zel’dovich Effect with ACT DR6 and DESI DR1 Spectroscopic Galaxies and Optically-Selected Clusters}


\author{J.~E.~Moore\orcidlink{0000-0002-7340-9291}}
\affiliation{Department of Physics, Duke University, Durham, NC, 27708, USA}

\author{C.~Popik\orcidlink{0000-0002-6612-2524}}
\affiliation{Department of Astronomy, Cornell University, Ithaca, NY 14853, USA}

\author{Y.~Gong\orcidlink{0000-0003-4624-795X}}
\affiliation{Department of Astronomy, Cornell University, Ithaca, NY 14853, USA}

\author{Y-H.~Hsu\orcidlink{0000-0003-0381-562X}}
\affiliation{University Observatory, Faculty of Physics, Ludwig-Maximilians-Universit\"{a}t, Scheinerstr. 1, 81679 Munich, Germany}
\affiliation{Max-Planck-Institut f\"{u}r extraterrestrische Physik\,(MPE), Giessenbachstrasse 1, 85748 Garching bei M\"{u}nchen, Germany}
\affiliation{Institute of Astronomy and Astrophysics, Academia Sinica\,(ASIAA),  Taipei 10617, Taiwan}

\author{E.~M.~Vavagiakis\orcidlink{0000-0002-2105-7589}}
\affiliation{Department of Physics, Duke University, Durham, NC, 27708, USA}

\author{N.~Battaglia\orcidlink{0000-0001-5846-0411}}
\affiliation{Department of Astronomy, Cornell University, Ithaca, NY 14853, USA}

\author{R.~Bean\orcidlink{0009-0004-3640-061X}}
\affiliation{Department of Astronomy, Cornell University, Ithaca, NY 14853, USA}

\author{D.~Gruen\orcidlink{0000-0003-3270-7644}}
\affiliation{University Observatory, Faculty of Physics, Ludwig-Maximilians-Universit\"{a}t, Scheinerstr. 1, 81679 Munich, Germany}
\affiliation{Excellence Cluster ORIGINS, Boltzmannstr. 2, 85748 Garching, Germany}

\author{P.~Gallardo\orcidlink{0000-0003-3270-7644}}
\affiliation{Department of Physics and Astronomy, University of Pennsylvania, 209 South 33rd Street, Philadelphia, PA, USA 19104}

\author{B.~Hadzhiyska\orcidlink{0000-0002-2312-3121}}
\affiliation{Institute of Astronomy, University of Cambridge, Madingley Road, Cambridge CB3 0HA, UK}
\affiliation{University of California, Berkeley, 110 Sproul Hall \#5800 Berkeley, CA 94720, USA}

\author{G.~Ockert}
\affiliation{Department of Physics, Duke University, Durham, NC, 27708, USA}

\author{J.~Aguilar}
\affiliation{Lawrence Berkeley National Laboratory, 1 Cyclotron Road, Berkeley, CA 94720, USA}

\author{S.~Ahlen\orcidlink{0000-0001-6098-7247}}
\affiliation{Department of Physics, Boston University, 590 Commonwealth Avenue, Boston, MA 02215 USA}

\author{A.~Aviles\orcidlink{0000-0001-5998-3986}}
\affiliation{Instituto Avanzado de Cosmolog\'{\i}a A.~C., San Marcos 11 - Atenas 202. Magdalena Contreras. Ciudad de M\'{e}xico C.~P.~10720, M\'{e}xico}
\affiliation{Instituto de Ciencias F\'{\i}sicas, Universidad Nacional Aut\'onoma de M\'exico, Av. Universidad s/n, Cuernavaca, Morelos, C.~P.~62210, M\'exico}

\author{F.~Beutler\orcidlink{0000-0003-0467-5438}}
\affiliation{Institute for Astronomy, University of Edinburgh, Royal Observatory, Blackford Hill, Edinburgh EH9 3HJ, UK}

\author{D.~Bianchi\orcidlink{0000-0001-9712-0006}}
\affiliation{Dipartimento di Fisica ``Aldo Pontremoli'', Universit\`a degli Studi di Milano, Via Celoria 16, I-20133 Milano, Italy}
\affiliation{INAF-Osservatorio Astronomico di Brera, Via Brera 28, 20122 Milano, Italy}

\author{J.~R.~Bond\orcidlink{0000-0003-2358-9949}}
\affiliation{Canadian Institute for Theoretical Astrophysics, University of Toronto, Toronto, ON, Canada M5S 3H8}

\author{D.~Brooks}
\affiliation{Department of Physics \& Astronomy, University College London, Gower Street, London, WC1E 6BT, UK}

\author{E.~Bulbul}
\affiliation{Max-Planck-Institut f\"{u}r extraterrestrische Physik\,(MPE), Giessenbachstrasse 1, 85748 Garching bei M\"{u}nchen, Germany}
\affiliation{Faculty of Physics, Fakult\"at f\"ur Physik, Ludwig-Maximilians-Universit\"at, Scheinerstr. 1, 81679 M\"unchen, Germany}

\author{A.~Carnero Rosell\orcidlink{0000-0003-3044-5150}}
\affiliation{Departamento de Astrof\'{\i}sica, Universidad de La Laguna (ULL), E-38206, La Laguna, Tenerife, Spain}
\affiliation{Instituto de Astrof\'{\i}sica de Canarias, C/ V\'{\i}a L\'{a}ctea, s/n, E-38205 La Laguna, Tenerife, Spain}

\author{E.~Chaussidon\orcidlink{0000-0001-8996-4874}}
\affiliation{Lawrence Berkeley National Laboratory, 1 Cyclotron Road, Berkeley, CA 94720, USA}

\author{T.~Claybaugh}
\affiliation{Lawrence Berkeley National Laboratory, 1 Cyclotron Road, Berkeley, CA 94720, USA}

\author{J.~Comparat}
\affiliation{Univ. Grenoble Alpes, CNRS, Grenoble INP, LPSC-IN2P3, 53, Avenue des Martyrs, 38000, Grenoble, France}

\author{A.~de la Macorra\orcidlink{0000-0002-1769-1640}}
\affiliation{Instituto de F\'{\i}sica, Universidad Nacional Aut\'{o}noma de M\'{e}xico,  Circuito de la Investigaci\'{o}n Cient\'{\i}fica, Ciudad Universitaria, Cd. de M\'{e}xico  C.~P.~04510,  M\'{e}xico}

\author{Biprateep~Dey\orcidlink{0000-0002-5665-7912}}
\affiliation{Department of Astronomy \& Astrophysics, University of Toronto, Toronto, ON M5S 3H4, Canada}
\affiliation{Department of Physics \& Astronomy and Pittsburgh Particle Physics, Astrophysics, and Cosmology Center (PITT PACC), University of Pittsburgh, 3941 O'Hara Street, Pittsburgh, PA 15260, USA}

\author{P.~Doel}
\affiliation{Department of Physics \& Astronomy, University College London, Gower Street, London, WC1E 6BT, UK}

\author{S.~Ferraro\orcidlink{0000-0003-4992-7854}}
\affiliation{Lawrence Berkeley National Laboratory, 1 Cyclotron Road, Berkeley, CA 94720, USA}
\affiliation{University of California, Berkeley, 110 Sproul Hall \#5800 Berkeley, CA 94720, USA}

\author{A.~Font-Ribera\orcidlink{0000-0002-3033-7312}}
\affiliation{Instituci\'{o} Catalana de Recerca i Estudis Avan\c{c}ats, Passeig de Llu\'{\i}s Companys, 23, 08010 Barcelona, Spain}
\affiliation{Institut de F\'{i}sica d’Altes Energies (IFAE), The Barcelona Institute of Science and Technology, Edifici Cn, Campus UAB, 08193, Bellaterra (Barcelona), Spain}

\author{J.~E.~Forero-Romero\orcidlink{0000-0002-2890-3725}}
\affiliation{Departamento de F\'isica, Universidad de los Andes, Cra. 1 No. 18A-10, Edificio Ip, CP 111711, Bogot\'a, Colombia}
\affiliation{Observatorio Astron\'omico, Universidad de los Andes, Cra. 1 No. 18A-10, Edificio H, CP 111711 Bogot\'a, Colombia}

\author{E.~Gaztañaga\orcidlink{0000-0001-9632-0815}}
\affiliation{Institut d'Estudis Espacials de Catalunya (IEEC), c/ Esteve Terradas 1, Edifici RDIT, Campus PMT-UPC, 08860 Castelldefels, Spain}
\affiliation{Institute of Cosmology and Gravitation, University of Portsmouth, Dennis Sciama Building, Portsmouth, PO1 3FX, UK}
\affiliation{Institute of Space Sciences, ICE-CSIC, Campus UAB, Carrer de Can Magrans s/n, 08913 Bellaterra, Barcelona, Spain}

\author{Satya~{Gontcho A Gontcho}\orcidlink{0000-0003-3142-233X}}
\affiliation{University of Virginia, Department of Astronomy, Charlottesville, VA 22904, USA}

\author{G.~Gutierrez}
\affiliation{Fermi National Accelerator Laboratory, PO Box 500, Batavia, IL 60510, USA}

\author{K.~Honscheid\orcidlink{0000-0002-6550-2023}}
\affiliation{Center for Cosmology and AstroParticle Physics, The Ohio State University, 191 West Woodruff Avenue, Columbus, OH 43210, USA}
\affiliation{Department of Physics, The Ohio State University, 191 West Woodruff Avenue, Columbus, OH 43210, USA}
\affiliation{The Ohio State University, Columbus, 43210 OH, USA}

\author{M.~Ishak\orcidlink{0000-0002-6024-466X}}
\affiliation{Department of Physics, The University of Texas at Dallas, 800 W. Campbell Rd., Richardson, TX 75080, USA}

\author{S.~Juneau\orcidlink{0000-0002-0000-2394}}
\affiliation{NSF NOIRLab, 950 N. Cherry Ave., Tucson, AZ 85719, USA}

\author{T.~Karim\orcidlink{0000-0002-5652-8870}}
\affiliation{Department of Astronomy \& Astrophysics, University of Toronto, Toronto, ON M5S 3H4, Canada}

\author{R.~Kehoe}
\affiliation{Department of Physics, Southern Methodist University, 3215 Daniel Avenue, Dallas, TX 75275, USA}

\author{M.~Kluge}
\affiliation{Max-Planck-Institut f\"{u}r extraterrestrische Physik\,(MPE), Giessenbachstrasse 1, 85748 Garching bei M\"{u}nchen, Germany}

\author{A.~Kremin\orcidlink{0000-0001-6356-7424}}
\affiliation{Lawrence Berkeley National Laboratory, 1 Cyclotron Road, Berkeley, CA 94720, USA}

\author{A.~Kusiak\orcidlink{0000-0002-1048-7970}}
\affiliation{Institute of Astronomy, University of Cambridge, Cambridge, CB3 0HA, UK}
\affiliation{Kavli Institute for Cosmology, University of Cambridge, Cambridge CB3 0HA, UK}

\author{O.~Lahav\orcidlink{0000-0002-1134-9035}}
\affiliation{Department of Physics \& Astronomy, University College London, Gower Street, London, WC1E 6BT, UK}

\author{M.~Landriau\orcidlink{0000-0003-1838-8528}}
\affiliation{Lawrence Berkeley National Laboratory, 1 Cyclotron Road, Berkeley, CA 94720, USA}

\author{L.~Le~Guillou\orcidlink{0000-0001-7178-8868}}
\affiliation{Sorbonne Universit\'{e}, CNRS/IN2P3, Laboratoire de Physique Nucl\'{e}aire et de Hautes Energies (LPNHE), FR-75005 Paris, France}

\author{M.~Manera\orcidlink{0000-0003-4962-8934}}
\affiliation{Departament de F\'{i}sica, Serra H\'{u}nter, Universitat Aut\`{o}noma de Barcelona, 08193 Bellaterra (Barcelona), Spain}
\affiliation{Institut de F\'{i}sica d’Altes Energies (IFAE), The Barcelona Institute of Science and Technology, Edifici Cn, Campus UAB, 08193, Bellaterra (Barcelona), Spain}

\author{A.~Meisner\orcidlink{0000-0002-1125-7384}}
\affiliation{NSF NOIRLab, 950 N. Cherry Ave., Tucson, AZ 85719, USA}

\author{R.~Miquel}
\affiliation{Instituci\'{o} Catalana de Recerca i Estudis Avan\c{c}ats, Passeig de Llu\'{\i}s Companys, 23, 08010 Barcelona, Spain}
\affiliation{Institut de F\'{i}sica d’Altes Energies (IFAE), The Barcelona Institute of Science and Technology, Edifici Cn, Campus UAB, 08193, Bellaterra (Barcelona), Spain}

\author{T.~Mroczkowski\orcidlink{0000-0003-3816-5372}}
\affiliation{Institute of Space Sciences, ICE-CSIC, Campus UAB, Carrer de Can Magrans s/n, 08913 Bellaterra, Barcelona, Spain}
\affiliation{Institut d'Estudis Espacials de Catalunya (IEEC), c/ Esteve Terradas 1, Edifici RDIT, Campus PMT-UPC, 08860 Castelldefels, Spain}

\author{S.~Nadathur\orcidlink{0000-0001-9070-3102}}
\affiliation{Institute of Cosmology and Gravitation, University of Portsmouth, Dennis Sciama Building, Portsmouth, PO1 3FX, UK}

\author{J.~ A.~Newman\orcidlink{0000-0001-8684-2222}}
\affiliation{Department of Physics \& Astronomy and Pittsburgh Particle Physics, Astrophysics, and Cosmology Center (PITT PACC), University of Pittsburgh, 3941 O'Hara Street, Pittsburgh, PA 15260, USA}

\author{M.~D.~Niemack\orcidlink{0000-0001-7125-3580}}
\affiliation{Department of Physics, Cornell University, Ithaca, NY, 14853, USA}
\affiliation{Department of Astronomy, Cornell University, Ithaca, NY 14853, USA}

\author{H.~E.~Noriega\orcidlink{0000-0002-3397-3998}}
\affiliation{Instituto de Ciencias F\'{\i}sicas, Universidad Nacional Aut\'onoma de M\'exico, Av. Universidad s/n, Cuernavaca, Morelos, C.~P.~62210, M\'exico}
\affiliation{Instituto de F\'{\i}sica, Universidad Nacional Aut\'{o}noma de M\'{e}xico,  Circuito de la Investigaci\'{o}n Cient\'{\i}fica, Ciudad Universitaria, Cd. de M\'{e}xico  C.~P.~04510,  M\'{e}xico}

\author{E.~Paillas\orcidlink{0000-0002-4637-2868}}
\affiliation{Instituto de Estudios Astrof\'isicos, Facultad de Ingenier\'ia y Ciencias, Universidad Diego Portales, Av. Ej\'ercito Libertador 441, Santiago, Chile}
\affiliation{Steward Observatory, University of Arizona, 933 N. Cherry Avenue, Tucson, AZ 85721, USA}

\author{W.~J.~Percival\orcidlink{0000-0002-0644-5727}}
\affiliation{Department of Physics and Astronomy, University of Waterloo, 200 University Ave W, Waterloo, ON N2L 3G1, Canada}
\affiliation{Perimeter Institute for Theoretical Physics, 31 Caroline St. North, Waterloo, ON N2L 2Y5, Canada}
\affiliation{Waterloo Centre for Astrophysics, University of Waterloo, 200 University Ave W, Waterloo, ON N2L 3G1, Canada}

\author{F.~Prada\orcidlink{0000-0001-7145-8674}}
\affiliation{Instituto de Astrof\'{i}sica de Andaluc\'{i}a (CSIC), Glorieta de la Astronom\'{i}a, s/n, E-18008 Granada, Spain}

\author{I.~P\'erez-R\`afols\orcidlink{0000-0001-6979-0125}}
\affiliation{Departament de F\'isica, EEBE, Universitat Polit\`ecnica de Catalunya, c/Eduard Maristany 10, 08930 Barcelona, Spain}

\author{C.~Ravoux\orcidlink{0000-0002-3500-6635}}
\affiliation{Universit\'{e} Clermont-Auvergne, CNRS, LPCA, 63000 Clermont-Ferrand, France}

\author{B.~Ried~Guachalla\orcidlink{0000-0002-0418-6258}}
\affiliation{Kavli Institute for Particle Astrophysics and Cosmology, Stanford University, 452 Lomita Mall, Stanford, CA, 94305, USA}
\affiliation{Department of Physics, Stanford University, 382 Via Pueblo Mall, Stanford, CA, 94305, USA}
\affiliation{SLAC National Accelerator Laboratory, 2575 Sand Hill Road, Menlo Park, California 94025, USA}

\author{G.~Rossi}
\affiliation{Department of Physics and Astronomy, Sejong University, 209 Neungdong-ro, Gwangjin-gu, Seoul 05006, Republic of Korea}

\author{R.~Ruggeri\orcidlink{0000-0002-0394-0896}}
\affiliation{Queensland University of Technology,  School of Chemistry \& Physics, George St, Brisbane 4001, Australia}

\author{L.~Samushia\orcidlink{0000-0002-1609-5687}}
\affiliation{Abastumani Astrophysical Observatory, Tbilisi, GE-0179, Georgia}
\affiliation{Department of Physics, Kansas State University, 116 Cardwell Hall, Manhattan, KS 66506, USA}

\author{E.~Sanchez\orcidlink{0000-0002-9646-8198}}
\affiliation{CIEMAT, Avenida Complutense 40, E-28040 Madrid, Spain}

\author{C.~Saulder\orcidlink{0000-0002-0408-5633}}
\affiliation{Max Planck Institute for Extraterrestrial Physics, Gie\ss enbachstra\ss e 1, 85748 Garching, Germany}

\author{D.~Schlegel}
\affiliation{Lawrence Berkeley National Laboratory, 1 Cyclotron Road, Berkeley, CA 94720, USA}

\author{J.~Silber\orcidlink{0000-0002-3461-0320}}
\affiliation{Lawrence Berkeley National Laboratory, 1 Cyclotron Road, Berkeley, CA 94720, USA}

\author{M.~Siudek\orcidlink{0000-0002-2949-2155}}
\affiliation{Institute of Space Sciences, ICE-CSIC, Campus UAB, Carrer de Can Magrans s/n, 08913 Bellaterra, Barcelona, Spain}
\affiliation{Instituto de Astrof\'{\i}sica de Canarias, C/ V\'{\i}a L\'{a}ctea, s/n, E-38205 La Laguna, Tenerife, Spain}

\author{G.~Tarl\'{e}\orcidlink{0000-0003-1704-0781}}
\affiliation{University of Michigan, 500 S. State Street, Ann Arbor, MI 48109, USA}

\author{B.~A.~Weaver}
\affiliation{NSF NOIRLab, 950 N. Cherry Ave., Tucson, AZ 85719, USA}

\author{E.~J.~Wollack\orcidlink{0000-0002-7567-4451}}
\affiliation{NASA Goddard Space Flight Center, 8800 Greenbelt Road, Greenbelt, MD 20771, USA}

\author{H.~Zou\orcidlink{0000-0002-6684-3997}}
\affiliation{National Astronomical Observatories, Chinese Academy of Sciences, A20 Datun Road, Chaoyang District, Beijing, 100101, P.~R.~China}

\begin{abstract}

We present stacked thermal Sunyaev-Zel'dovich (tSZ) effect measurements for three samples of galaxy groups and clusters: those traced by the Dark Energy Spectroscopic Instrument Data Release 1 (DESI DR1) luminous red galaxies (LRG) and the DESI DR1 Bright Galaxy Sample (BGS), and an \texttt{eROMaPPer} optically-selected sample from the DESI Legacy Imaging Survey. We use the latest Atacama Cosmology Telescope DR6 (ACT)+\textit{Planck} component-separated internal linear combination (ILC) Compton-$y$ maps and ACT+\textit{Planck} coadded 90, 150, and 220 GHz temperature maps to extract the tSZ signal within a $\sim2'$ disk aperture for sources binned by luminosity, richness, or mass. We measure the average tSZ signal with high statistical significance, with signal-to-noise ratios surpassing 38 for LRG, 27 for BGS, and 39 for the \texttt{eROMaPPer} sample using the 90 GHz ACT DR6+\textit{Planck} map. We conduct a detailed study of systematics and foregrounds such as dust and cosmic infrared background (CIB) contamination, which remain a core challenge for tSZ analysis. For the LRG and BGS samples, we find that dust and radio source emission dominate the tSZ signal at scales near and below the disk aperture radius. Large-scale ($R>4'$) contamination from the CIB is less significant. We mitigate these contaminants to isolate the tSZ signal and use a combination of simulated and real measurements to develop Compton-$y-$optical depth ($\bar y-\bar \tau$) scaling relations to infer optical depths, which are found to be in agreement with values measured using the pairwise kinematic SZ effect for the same tracer samples. The $\bar y-\bar \tau$ scaling relation for the \texttt{eROMaPPer} sample is the first such relationship to be derived directly from SZ measurements.
\end{abstract}

\maketitle

\section{Introduction}\label{sec:intro}
Much is still unknown about the formation and evolution of galaxy groups and clusters. In order to better constrain theoretical models, we must improve our understanding of the distribution of baryons within these structures and the physical processes that govern them.  One method of probing the baryon content of these structures observationally is via the Sunyaev-Zel'dovich (SZ) effects  \cite{zeldovich69, sunyaev72}, which are secondary cosmic microwave background (CMB) anisotropies that arise as a result of interactions between CMB photons and high-energy free electrons within the circumgalactic medium (CGM) and intracluster medium (ICM). As CMB photons travel through these reservoirs of hot gas surrounding galaxies and clusters, these interactions induce spectral shifts and distortions that can be quantified via measurements of the CMB temperature.

The thermal SZ (tSZ) effect arises from the inverse Compton scattering of CMB photons off highly energetic electrons as they travel through hot gas. This scattering induces a characteristic frequency-dependent and redshift-independent signature in the CMB as collisions with hot electrons boost the energy of the CMB photons, resulting in CMB temperature fluctuation due to the tSZ effect is given by
\begin{equation}\label{eq:tsz}
    \frac{\delta T_{\mathrm{tSZ}}}{T_{\mathrm{CMB}}} = f_{\mathrm{SZ}} y,
\end{equation}
where $f_{\mathrm{SZ}}$ is the frequency dependence, which in the non-relativistic limit is given by
\begin{equation}
    f_{\mathrm{SZ}} = x\frac{e^x+1}{e^x-1}-4,
\end{equation}
with $x=h\nu/k_BT_{\mathrm{CMB}}$. This leads to an excess of photons above the null frequency of 217 GHz, and a deficit of photons below. In CMB temperature maps (at frequencies below the null), this signature manifests as a temperature decrement. The amplitude of the signal is described by the dimensionless Compton-$y$ parameter, which is defined as
\begin{equation} \label{eq:comptony}
    y = \frac{\sigma_T}{m_ec^2}\int_{los}n_e k_B T_e dl,
\end{equation}
where $\sigma_T$ is the Thomson cross-section, $m_e$ is the electron mass, $c$ is the speed of light, $k_B$ is the Boltzmann constant, and $n_e$ and $T_e$ are the electron number density and temperature, integrated along the line-of-sight ($los$). Because the Compton-$y$ parameter is sensitive to the electron temperature as well as the optical depth of the gas $\tau$, defined as
\begin{equation}
    \tau = \sigma_T \int_{los}n_e dl,
\end{equation}
 measurements of the tSZ effect around galaxy clusters provide an an important probe of the underlying gas distribution and thermodynamics, which can in turn inform models of galaxy formation and evolution \cite{battaglia17, mroczkowski19}. 

The kinematic Sunyaev-Zel'dovich (kSZ) effect arises from the peculiar motion of galaxy clusters relative to the rest frame of the CMB. When CMB photons travel through clusters, this motion imparts a Doppler shift on the CMB. Unlike the tSZ effect, the kSZ effect does not have a characteristic frequency dependence and is an order of magnitude smaller, making it significantly more difficult to detect. A pairwise technique that uses differential temperature measurements to extract the relative momentum of pairs of galaxy clusters has emerged as a method of detecting the kSZ effect that is unaffected by contamination from foregrounds and the tSZ effect \cite{hand12}. When combined with estimates of the average optical depth $\bar{\tau}$, one may extract the pairwise velocity, which can be used to constrain cosmological parameters and address fundamental questions related to dark energy, neutrinos, growth of structure, and more \cite{dedeo05, bhattacharya08, mueller15.1, mueller15.2}.

High-resolution maps from CMB observatories such as the Atacama Cosmology Telescope (ACT) \cite{swetz11} and the South Pole Telescope (SPT) \cite{carlstrom11} have been used to detect individual galaxy clusters using the SZ effect \cite{kornoelje25, act26}. CMB maps have also been used in tandem with galaxy survey data to measure the average SZ effects from stacks of known sources. In \cite{amodeo21, schaan21, vavagiakis21, calafut21}, ACT+\textit{Planck} data is used to measure the stacked tSZ and kSZ effects of galaxy groups selected from the Baryon Oscillation Spectroscopic Survey (BOSS). In \cite{meinke21,meinke23}, the stacked tSZ effect around quiescent galaxies selected from the Dark Energy Survey (DES) and Wide-Field Infrared Survey Explorer (WISE) is measured using data from SPT and ACT+\textit{Planck}. Studies like these have been used to improve our understanding of the baryonic content of galaxies and clusters and to constrain models of galaxy evolution and test models of active galactic nuclei (AGN) feedback \cite{grayson23, hadzhiyska25.2}. This work builds on the work of \cite{vavagiakis21, calafut21}, which used earlier versions of the ACT maps to extract measurements of the tSZ and kSZ effects around BOSS galaxies and estimate their optical depth. 

In this work, we analyze the stacked tSZ signal from three samples: galaxies from the DESI DR1 Luminous Red Galaxy (LRG) sample and the DESI DR1 Bright Galaxy Sample (BGS), which we use as proxies for the centers of massive galaxy groups and clusters, as well as a sample of \texttt{eROMaPPer}-selected optical clusters. Pairwise kSZ measurements are presented for each of these samples in three companion papers: Gong et al. (LRG) \cite{gong26}, Hadzhiyska et al. (BGS) \cite{hadzhiyska25}, and Hsu et al. (\texttt{eROMaPPer}) \cite{hsu26}. The primary aim of the joint tSZ+kSZ analysis of these samples is to develop empirical scaling relationships between the average Compton-y ($\bar y$; derived from the tSZ measurements) and the average optical depth ($\bar \tau$, derived from the kSZ measurements). Such scaling relationships will allow the optical depth to be inferred from the tSZ measurements, which in turn will enable the extraction of the pairwise velocity from the kSZ measurements.

We use the latest high-resolution ACT DR6+\textit{Planck} component-separated internal linear combination (ILC) maps \cite{coulton24} and the ACT DR6+\textit{Planck} co-added single-frequency maps \cite{naess25} to extract the tSZ signal around each of our sources, which are binned according to luminosity (LRG), stellar mass (BGS), and richness (\texttt{eROMaPPer} sample). For each catalog, we stack the binned measurements to generate highly significant measurements of the average Compton-$y$ signal ($\bar y$). We closely follow the methods presented in \cite{liu25}, a recent study of the tSZ effect with ACT and DESI LRGs, for analysis and removal of foreground contamination from dust and the Cosmic Infrared Background (CIB). We use a forward-modeling approach to address the two-halo effect and beam systematics. We compare our tSZ measurements to the kSZ-derived optical depth estimates ($\bar \tau$) and use model $\bar y-\bar \tau$ scaling relations to convert Compton-$y$ to optical depth.

We further describe the samples and maps analyzed in this work in Section \ref{sec:data}. In Section \ref{sec:methods} we describe the map processing and aperture photometry and error estimation methodology. In Section \ref{sec:results} we present our results and explore tSZ measurement systematics such as contamination from dust, CIB, and radio sources, as well as other systematics such as the two-halo effect. In Section \ref{sec:taucomparisons} we develop $\bar y-\bar \tau$ scaling relationships for each sample and compare tSZ and kSZ-derived optical depths. In Section \ref{sec:conclusion} we present our conclusions and look toward future improvements in tSZ analysis with next-generation millimeter and submillimeter observatories.

\begin{figure*}
\begin{center}
\includegraphics[width=17.2cm]{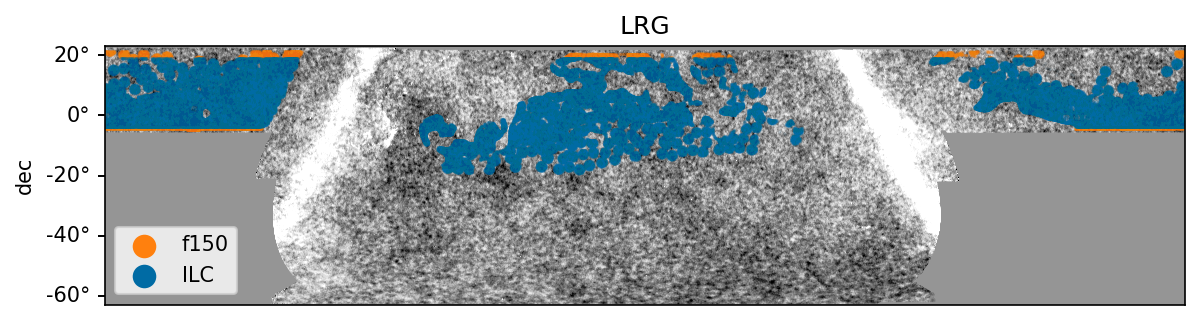}
\includegraphics[width=17.2cm]{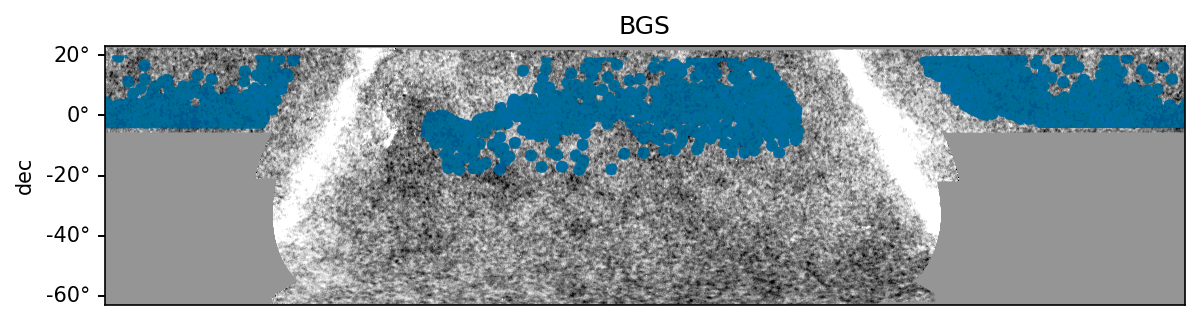}
\hspace*{0.3cm}\includegraphics[width=17.5cm]{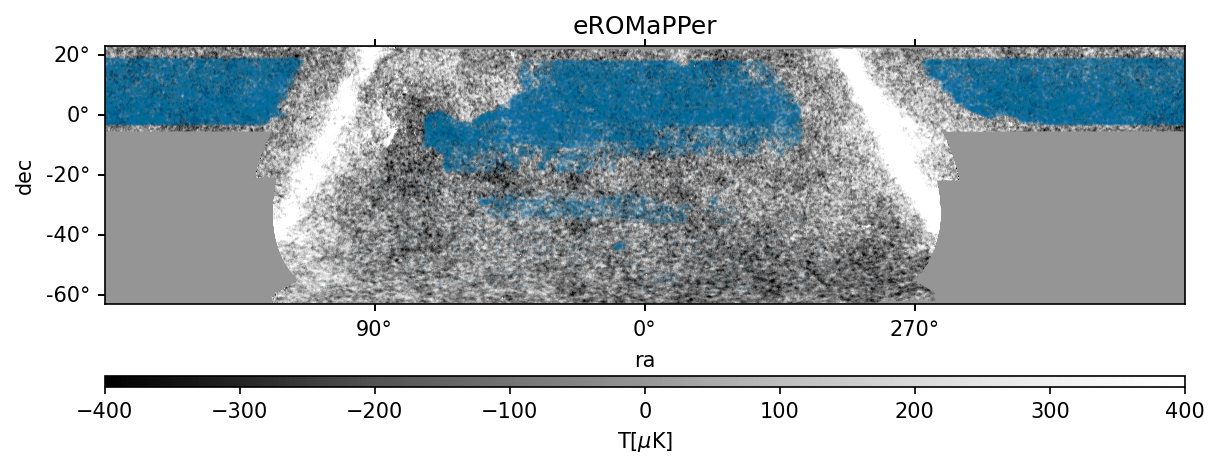}
\caption{Sources from each of the three catalogs overplotted on the ACT + \textit{Planck} DR6 f150 map. We note that due to apodization, the ACT ILC maps have a slightly smaller footprint than the f150 map. As a result, approximately $5\%$ of the f150 LRG sample [orange] is cut from the ILC analysis [blue]. This cut does not affect the average redshift or luminosity of the sample. The BGS and \texttt{eROMaPPer} samples were selected based on the ILC map footprint.}
\label{fig:catalogs}
\end{center}
\end{figure*}

\begin{figure}
\begin{center}
\hspace*{-0.25cm}
\includegraphics[width=7.6cm]{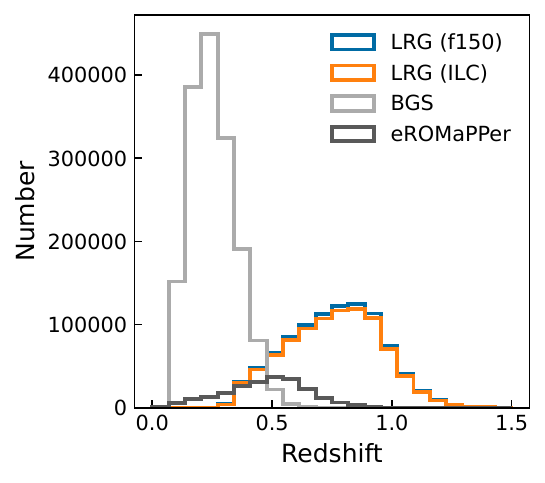}
\caption{The spectroscopic redshift distribution for each sample analyzed in this work. The LRG sample [blue and orange] is comprised of 957,095 and 913,286 sources in the f150 and ILC footprints, respectively, with an average redshift of 0.76. The BGS sample [light gray] consists of 1.6 million sources with an average redshift of 0.25. The \texttt{eROMaPPer} sample [dark gray] contains 220,698 sources with an average redshift of 0.48.}
\label{fig:redshift}
\end{center}
\end{figure}

\section{Data}\label{sec:data}
\subsection{DESI data}
\label{sec:desidata}
The Dark Energy Spectroscopic Instrument (DESI), mounted on the 4 meter Mayall telescope at Kitt Peak National Observatory in Arizona, is a spectroscopic survey instrument designed to simultaneously measure the spectra of thousands of objects over a $3^{\circ}$ field of view \cite{desi16.2,desi22,miller23,schlafly23,poppett24}. Over the course of the 8 year survey, DESI will identify over 60 million galaxies and quasars over approximately 17,000 square degrees of the sky. The primary aim of the DESI survey is to elucidate the nature of dark energy through high-precision measurements of the expansion rate of the universe \cite{desi16.1,guy23,desi25,desi25.2,desi25.3}.

This work analyzes three separate samples, described in detail below. The footprint of each sample is shown in Figure \ref{fig:catalogs}. For each of these samples, pairwise kSZ measurements are presented in companion papers: Gong et al., \cite{gong26} Hadzhiyska et al., \cite{hadzhiyska25}, and Hsu et al. (in prep.) \cite{hsu26}. The redshift distribution of each sample is shown in Figure \ref{fig:redshift}. 

As in our previous analysis \cite{vavagiakis21}, we subject each sample to a series of cuts and masks based on the ACT data as described in Gong et al. Sec II.B. \cite{gong26}. For the LRG and BGS samples, we apply a 45 $\mu$K per pixel white noise variance cut, followed by a 50\% galactic plane mask to reduce contamination from the Milky Way. We use the ACT point source and cluster catalog to mask all sources brighter than 15 mJy and within 10 arcminutes in the lower noise Deep 56 region, and all sources brighter than 100 mJy and within 5 arcminutes in all other map areas. Finally, following \cite{liu25}, we use the ACT DR6 SZ cluster catalog \cite{act26} to identify and remove all targets within $10'$ of known tSZ clusters with SNR $>6$, as the exceptionally bright signal of these very massive clusters can bias the stacked measurements. The \texttt{eROMaPPer} sample is subjected to the same noise and point source cuts, but as the sample is less impacted by the Milky Way, we employ a less aggressive 70\% galactic plane mask in order to preserve a larger number of sources.

\begin{table*}
\begin{center}
\setlength{\tabcolsep}{6pt}
\renewcommand{\arraystretch}{1.25}
\begin{tabular}{|c|c|c|c|c|c|c|c|}

\cline{7-8}
\multicolumn{6}{l|}{} & ILC & f150 \\

\cline{5-8}
\hline
Bin & Luminosity cut/$10^{10} L_{\odot}$ & $M_{\rmn{vir}}$ cut/$10^{13} M_{\odot}$ & $\langle \log_{10}{(M_*/M_{\odot})} \rangle$ & $\langle L\rangle/10^{10} L_{\odot}$ & $\langle z\rangle$ & N & N \\
\hline
L36 & $L > 3.58$ & $M > 0.38$ & 11.32 & 6.89 & 0.76 & 913,286 & 957,095 \\
L48 & $L > 4.78$ & $M > 0.63$ & 11.37 & 7.78 & 0.78 & 685,766 & 718,542 \\
L60 & $L > 5.98$ & $M > 0.96$ & 11.43 & 9.00 & 0.80 & 456,803 & 478,585 \\
L79 & $L > 7.86$ & $M > 1.64$ & 11.53 & 11.18 & 0.84 & 228,650 & 239,567 \\
L98 & $L > 9.8$ & $M > 2.59$ & 11.61 & 13.67 & 0.88 & 114,024 & 119,353 \\
\cline{1-8}
L36D & $3.58 < L < 4.78$ & $0.38 < M < 0.63$ & 11.10 & 4.21 & 0.70 & 227,520 & 238,553 \\
L48D & $4.78 < L < 5.98$ & $0.63 < M < 0.96$ & 11.21 & 5.36 & 0.72 & 228,963 & 239,957 \\
L60D & $5.98 < L < 7.86$ & $0.96 < M < 1.64$ & 11.31 & 6.82 & 0.76 & 228,153 & 239,018 \\
L79D & $7.86 < L < 9.8$ & $1.64 < M < 2.59$ & 11.42 & 8.71 & 0.81 & 114,626 & 120,214 \\
\hline


\end{tabular}
\caption{Overview of the DESI Luminous Red Galaxy catalog from \cite{gong26}. In this work we analyze this catalog using five cumulative (L36, L48, L60, L79, L98) and four disjoint (L36D, L47D, L60D, L78D) luminosity-binned samples using both the ACT DR6 ILC and f150 maps. For each bin we provide the luminosity cut and equivalent virial mass cut, along with the average stellar mass, luminosity, redshift, and sample size for both the f150 and ILC analysis.}\label{lrgtable}
\end{center}
\end{table*}


\begin{table*}
    \begin{center}
    \setlength{\tabcolsep}{5pt}
\renewcommand{\arraystretch}{1.25}
    \begin{tabular}{|c|c|c|c|c|} 
    \hline
    Bin & $\langle \log_{10}{(M_*/M_{\odot})} \rangle$ & $\langle \log_{10}{M_h} \rangle$ & $\langle z \rangle$ & N \\ \hline
    $\log_{10} {(M_*/M_{\odot})}>10$  & 10.72  & 13.135  & 0.25  & 1,610,381 \\
    $\log_{10} {(M_*/M_{\odot})}>10.25$  & 10.82  & 13.184  & 0.27  & 1,379,230 \\
    $\log_{10} {(M_*/M_{\odot})}>10.5$  & 10.94  & 13.266  & 0.28  & 1,084,339 \\
    $\log_{10} {(M_*/M_{\odot})}>10.75$  & 11.08  & 13.31  & 0.30  & 744,562 \\
    $\log_{10} {(M_*/M_{\odot})}>11$  & 11.25  & 13.37  & 0.33  & 421,150 \\
    $\log_{10} {(M_*/M_{\odot})}>11.25$  & 11.42  & 13.456  & 0.36  & 178,206 \\\hline
    \end{tabular}
    \caption{Overview of the DESI Bright Galaxy Survey catalog from \cite{hadzhiyska25}. The sample is split into 6 cumulative stellar mass bins with a minimum $\log_{10} (M_*/M_{\odot})$ cut of 10. For each mass bin we provide the average stellar mass, halo mass, redshift, and number of galaxies used in the analysis.}  
    \label{bgstable}
    \end{center}
\end{table*}

\begin{table}[]
    \begin{center}
    \setlength{\tabcolsep}{4pt}
\renewcommand{\arraystretch}{1.25}
    \begin{tabular}{|c|c|c|c|c|} 
    \hline
     Bin & $\langle \lambda_{\mathrm{DES}} \rangle $ & $\langle \log M_{200m} \rangle (h^{-1}M_{\odot})$&  $\langle z \rangle$ & N \\ \hline
$\lambda > 5$  & 11.27  & 13.77  & 0.48  & 220,698 \\
$\lambda > 10$  & 18.92  & 14.01  & 0.48  & 80,447 \\
$\lambda > 20$  & 32.71  & 14.26  & 0.47  & 22,320 \\ \hline

    \end{tabular}
    \caption{Overview of the \texttt{eROMaPPer} optically-selected catalog from \cite{hsu26}. In addition to the full sample, we also examine two cumulative richness bins. For each bin we provide the average DES-Y1 richness $\lambda$, halo mass, redshift, and number of clusters. The halo masses are calculated from the richness values using the DES Y1 richness-mass relation \cite{mcclintock19}.}
    \label{opttable}
    \end{center}
\end{table}

\subsubsection{Luminous Red Galaxy (LRG) sample}\label{sec:lrg}
The LRG catalog (Table \ref{lrgtable}; described in detail in \cite{gong26}) is derived from the DESI DR1 LRG sample \cite{zhou20, zhou23.1, zhou23.2}. Of the 1,689,748 DESI LRGs in the ACT DR6 150 GHz (``f150'') map footprint, 1,197,347 remain after performing the noise cut and applying masks for the galactic plane, point sources, and known tSZ clusters. The final number of LRGs with luminosity greater than our lower bound of $3.6 \times 10^{10} L_{\odot}$ is 957,095. While the base LRG catalog is defined using the f150 map footprint, the multi-frequency component-separated ILC maps (described in detail in Section \ref{sec:act}) have a slightly smaller footprint due to apodization. When stacking this sample on the ILC maps, we remove an additional, 240,252 LRGs that lie near the boundary (see Figure \ref{fig:catalogs}). 

The LRG sample has an average $M_*/10^{11}M_{\odot} = 2.07$ and covers a redshift range of $0 < z < 1.5$ with an average redshift of 0.76. The sample is divided into 5 cumulative and 4 disjoint luminosity bins according to the 20th, 40th, 60th, and 80th percentile luminosities (3.6, 4.8, 6.0, and 7.9 $\times 10^{10} L_{\odot})$. To facilitate comparison with previous analyses, we also include a $L > 9.8\times10^{10} L_{\odot}$ bin.

\subsubsection{Bright Galaxy Sample (BGS) catalog}\label{sec:bgs}
The BGS catalog (Table \ref{bgstable}; described in detail in \cite{hadzhiyska25}) is derived from the DESI DR1 BGS\_BRIGHT\_full and DESI DR1 value-added catalogs \cite{desi24, desi25, abdul25, siudek24, siudek25}. After applying all aforementioned cuts, the final sample includes 1.61 million BGS-BRIGHT galaxies with $\log_{10}(M_*/M_{\odot}) > 10$ in the ACT ILC footprint, with an average $\log_{10}(M_*/M_{\odot}) \approx 10.75$ and an average $z \approx 0.25$. The sample is divided into 6 cumulative stellar mass bins from $\log_{10} (M_*/M_{\odot})_* > 10$ to $\log_{10} (M_*/M_{\odot})_* > 11.25$, in increments of 0.25. The average redshift increases with increasing stellar mass threshold, up to $\langle z \rangle \approx 0.36$ for the $\log_{10} (M_*/M_{\odot})_* > 11.25$ bin. The average halo (virial) masses of the binned samples span a range of of $13.135 < \log_{10} M_h < 13.456$.

\subsubsection{\texttt{eROMaPPer} sample}\label{sec:opt}
We use the \texttt{eROMaPPer} catalog (Table \ref{opttable}; described in detail in \cite{hsu26}), which consists of clusters selected with the \texttt{eROMaPPer} cluster finder \cite{kluge24}. The \texttt{eROMaPPer} cluster catalogs were generated in two different configurations, blind-mode and scan-mode. In blind-mode the finder only considers optical information, and in scan-mode it considers external positional prior such as X-ray positions. The full \texttt{eROMaPPer} survey covers over 13,000 square degrees of the sky. Using the cluster finder in blind-mode, the clusters are selected from the DR10 South area of the DESI Legacy Imaging Surveys (LS10; \cite{dey19}) using photometry from the \textit{grz} filters. 257,456 sources are cross-matched with DESI Y1 spectroscopic redshifts \cite{desi24, desi25} and an additional 186,411 redshifts are sourced from the LS10 catalog. Cluster richness (mass proxy) is converted from \texttt{eROMaPPer} richness to DES-Y1 richness following \cite{kluge24}. We select clusters with richness $\lambda > 5$ in DES-Y1 scale and redshift between 0.05 and 1.01, reducing the number of sources to 333,585. After applying additional cuts for DESI redshift quality as well as the previously described ACT cuts, the final \texttt{eROMaPPer} sample contains 220,698 clusters with an average redshift of 0.48 and an average richness of 11.3 (DES-Y1 richness), which corresponds to a richness of 8.91 in the \texttt{eROMaPPer} scale. We note this sample was generated using a less aggressive 70\% galactic plane mask. A smaller sample (174,174 clusters), generated using the 50\% galactic plane mask, was analyzed, and no significant difference in measured signal was found. {Lastly, we note that we forego the SZ cluster mask for this sample. In \cite{hsu26} we have tested the impact of masking the most massive clusters on the pairwise kSZ signal by imposing an upper richness limit on the sample. As these are optical clusters, masking the clusters by richness is more self-consistent than using the SZ cluster mask. We have found the pairwise kSZ-derived optical depths do not significantly change when this richness limit is  imposed and therefore choose to include the high-richness clusters in our sample.

\subsection{ACT data}\label{sec:act}

The Atacama Cosmology Telescope (ACT) was a ground-based millimeter telescope located on Cerro Toco in the Atacama Desert of Chile. Over the course of fifteen years, ACT made high-resolution, multi-frequency (90/150/220 GHz) observations of the CMB, enabling the precise analysis of CMB temperature and polarization as well as the detection of massive galaxy clusters via the SZ effect \cite{swetz11}. In the final five years of operations, ACT observed with the Advanced ACTPol camera, which extended the frequency coverage down to 20 GHz \cite{henderson16}. 

\begin{figure*}[ht!]
\begin{center}
\includegraphics[width=17.9cm]{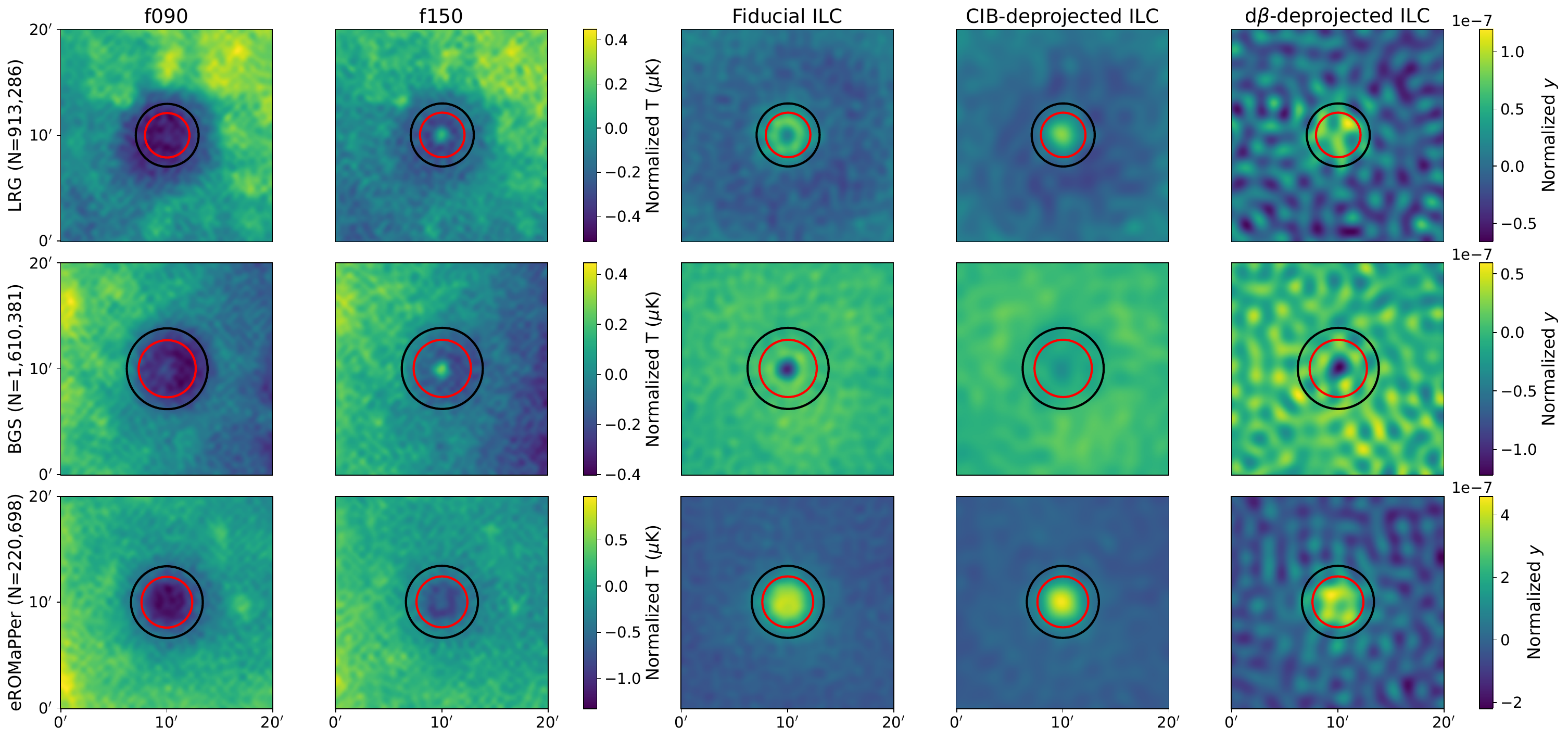}
\caption{Stacked $20'\times20'$  submaps of the complete LRG [top row], BGS [middle row], and \texttt{eROMaPPer} [bottom row] samples. Each image has been normalized to its mean value and the aperture photometry disk and ring radii are indicated by red and black circles, respectively. The first two columns show the 90 and 150 GHz ACT+\textit{Planck} co-added temperature maps \cite{naess25}. Columns 3-5 show the ACT DR6 ILC Compton-$y$ maps \cite{coulton24}. In the 150 GHz LRG and BGS stacks, the tSZ signal is overwhelmed by centralized contamination from foreground signals. These submaps qualitatively motivate the quantitative radial profile comparisons presented in Sections \ref{sec:profiles}-\ref{sec:radio}.}
\label{fig:submaps}
\end{center}
\end{figure*}

In this work we use the ACT DR6 + \textit{Planck} co-added single-frequency temperature maps: f090 (77-112 GHz), f150 (124-172 GHz), and f220 (182-277 GHz) with beam sizes of 2.07, 1.42, and 1.01 arcminutes, respectively \cite{naess25}. We also use the ACT DR6 component-separated needlet ILC (NILC) Compton-$y$ maps, which are described in detail in \cite{coulton24}. The NILC method combines multi-frequency data from both ACT (93, 148, 225 GHz) and \textit{Planck} (30-545 GHz) to generate arcminute-resolution component-separated maps of the Compton-$y$ signal. Briefly, the maps are generated by convolving the pre-processed input maps with needlet kernels, applying the NILC component separation method at each needlet scale, and then re-convolving with the needlet kernels to transform the maps back to real space. We use the fiducial ILC Compton-$y$ tSZ map, as well two derivative maps that remove contamination from the cosmic infrared background. These maps mitigate contamination from dust and the CIB through the deprojection of a theoretical CIB spectral energy distribution (SED) from the fiducial ILC map. The first of these deprojected maps (hereafter ``CIB-deprojected ILC'') removes the CIB signature, assuming the SED follows a modified blackbody of the form

\begin{equation}\label{eq:mbb}
    f_{\text{CIB}} (\nu) = \frac{A(\frac{\nu}{\nu_0})^{3+\beta}}{\exp \frac{h\nu}{k_B T_{\text{CIB}}} -1}\bigg(\frac{dB(\nu, T)}{dT}\bigg|_{T=T_{\text{CMB}}}\bigg)^{-1},
\end{equation}
where $T_{\text{CIB}}$ and $\beta$ are chosen parameters. The value of $\beta$ has a substantial impact on the deprojection, so in samples with significant contamination, measured signals may vary greatly from map to map. To address this, \cite{coulton24} offers a second set of deprojected ILC maps (hereafter ``d$\beta$-deprojected ILC'') which deproject an additional template that represents the first-order Taylor expansion about the spectral index:

\begin{equation}\label{eq:db_deproj}
    f_{\text{CIB}-\delta\beta} (\nu) = \frac{A\ln(\nu/\nu_0)(\frac{\nu}{\nu_0})^{3+\beta}}{\exp \frac{h\nu}{k_B T_{\text{CIB}}} -1}\bigg(\frac{dB(\nu, T)}{dT}\bigg|_{T=T_{\text{CMB}}}\bigg)^{-1},
\end{equation}

The deprojection of this second template helps to mitigate the sensitivity to the chosen spectral index. One limitation of this procedure is that it assumes all of the contamination may be described by one SED. In reality, the central contamination, originating from dust emission by our target galaxies, may differ from the large scale CIB contamination, which stems from many different sources over a wide redshift range. It is also important to note that each deprojection steps increases the overall noise of the map, which can clearly be seen in the stacked ILC submaps shown in Fig.~\ref{fig:submaps}. We examine the different ILC maps and their sensitivity to assumed SED parameters in Section \ref{sec:dustilc}.

\section{Methods}\label{sec:methods}
In this section we detail the methods used to compute the average SZ signal for each of our stacked samples. In Section \ref{sec:filt}, we describe the CMB map filtering. In Section \ref{sec:tSZextraction}, we discuss the tSZ signal extraction and error estimation. In Section \ref{sec:beamcorr} we model the systematic effects of the different beam sizes for the different ACT maps, and in Section \ref{sec:kSZsignals} we summarize the pairwise kSZ methodology used in the companion papers.

\subsection{Filtering CMB maps}\label{sec:filt}

For each source, we generate $20 \times 20$ arcminute cutouts centered at the target coordinates. These map cutouts are reprojected to a pixel size of 0.1 arcminutes per pixel using \texttt{pixell} \cite{naess21} and averaged together to form a stacked image for each bin. The stacked images for the full samples using the various maps are shown in Fig.~ \ref{fig:submaps}. 

\subsection{tSZ signal extraction}\label{sec:tSZextraction}
We compute disk-minus-ring aperture photometry (AP; alternatively ``compensated aperture photometry'' or CAP) for each source using source-centered submaps excised from the ILC and single-frequency maps. We use \texttt{iskay2} \cite{gallardo25} to extract the tSZ signal for each source as follows: the average submap pixel value is computed within a circular disk aperture of radius $R$, centered on the source RA and DEC. From the disk average we subtract the average value computed within a surrounding circular ring of equal area, with inner radius $R$ and outer radius $\sqrt{2}R$. For a single source, this may be written as

\begin{equation}
    y_{AP}= \frac{\sum y_d}{N_{\text{d}}} - \frac{\sum y_r}{N_{\text{r}}},
\end{equation}

where $N_{\text{d}}$ and $N_{\text{r}}$ are the number of pixels in the disk and ring. To match the companion analyses, disk apertures of $2.1'$, $2.7'$, and $2.4'$ are used for the LRG, BGS, and \texttt{eROMaPPer} samples, respectively. This approach is identical to that of \cite{vavagiakis21}, though slightly different from the methodology of \cite{liu25}, which computes the sum instead of the average of the pixels within the disk or ring. For the single-frequency temperature maps, we convert the average tSZ measurement from native map units of differential CMB temperature ($\mu$K) to Compton-$y$ following Equation \ref{eq:tsz}.

For each binned sample, we compute the average tSZ signal $\bar{y}_{AP}$ by averaging over the individual aperture photometry measurements $y_{AP}$ in the bin:

\begin{equation}
    \bar{y}_{AP} = \frac{\sum{y_{AP}}}{N_{\text{sources}}}.
\end{equation}

We perform a jackknife uncertainty estimation using 2,000 iterations to estimate the $1\sigma$ uncertainty of each measurement. We randomly split each sample into N smaller subsamples and iteratively remove one sample at a time, compute the average signal over the remaining N-1 subsamples, and compute the uncertainty as

\begin{equation}
    \sigma_{jk} = \sqrt{\frac{N-1}{N}\sum_{i=1}^{N}(\bar{y}_i-\bar{y})^2}.
\end{equation}

\noindent We have varied the value of $N$ to check for convergence of $\sigma_{jk}$ for all samples and found $N=2000$ to be an appropriate choice. Details about the null test performed to check for pipeline-induced systematics are presented in Appendix \ref{sec:nullgrad}.

In addition to our main apertures of interest, we compute radial profiles of the stacked samples by extracting aperture photometry measurements at $0.5'$ increments out to $R\sim7'$. The radial profiles allow us to identify systematics that may not be immediately apparent in the single aperture measurements. We discuss and account for these systematics in Section \ref{sec:results}.

\subsection{Beam correction}\label{sec:beamcorr}

The single frequency maps and various ILC deprojected maps have their own respective effective beam profiles, which should contribute some difference in measurements \cite{vavagiakis21}. We employ a profile-to-measurement forward model \texttt{CAPPIBARAS}\footnote{\href{https://github.com/chadpopik/CAPPIBARAS}{https://github.com/chadpopik/CAPPIBARAS}} [Popik et al 2026 in prep] (based on the work of \cite{amodeo21}) to capture the extent of this effect, incorporating different beams into an otherwise identical process to isolate their influence. As with \citet{amodeo21}, we use a modified generalized Navarro-Frenk-White (gNFW) model from \citep{battaglia12} to describe the 3D radial electron pressure profile $P_e(r, z, M_h)$, and include a two-halo component (further described in Section \ref{sec:2halo}) constructed using linear theory \cite{moser22, vikram17} with a halo mass function and halo bias from \cite{tinker08, tinker10}. We calculate a survey averaged election pressure profile $\bar{P}_e(r)$ using the halo mass function (bounded between the mass range of the sample) and the redshift distribution for each respective sample (described in \ref{sec:desidata}). To account for contamination from satellites, which occupy higher mass halo and  offset from the halo center, we incorporate the halo occupancy distributions of \cite{hadzhiyska25, hadzhiyska25.2} following the methodology outlined in \citet{popik_impacts_2025}. We project the average pressure to a 2D Compton-$y$ profile $\bar{y}(R)$ at the mean redshift of the sample using Equation \ref{eq:comptony}, and then convolve it with the effective beam $B(\ell)$ of the measurement's map:
\begin{equation} \begin{aligned}
    \tilde{\bar{y}}(R) = \mathcal{F}^{-1}\bigg( \mathcal{F} \Big(\bar{y}(R)\Big) \cdot B(\ell) \bigg)\\
\end{aligned} \ , \end{equation}
where $\mathcal{F}$ and $\mathcal{F}^-1$ are the Fourier transform and its inverse, respectively, performed using a radial Hankel Transform as described in \cite{moser23}. We then apply the aperture photometry described in \ref{sec:methods}. 

For each subsample (shown in Tables \ref{lrgtable}, \ref{bgstable}, and \ref{opttable}), we fit the gNFW pressure profile model to their respective ILC CIB deprojected measurements (shown in orange in Fig \ref{fig:ilc_deproj_profiles}), finding a unique $P_e(r|z,M)$ which we forward model to a final predicted measurement $\tilde{\bar{y}}_\text{AP}(R)$ using the beams of each of the different maps. We use the provided ILC beam from \cite{coulton24} for the ILC maps, and a Gaussian beam for the f150, f220, and f090 single frequency maps with a FWHM of 1.42, 1.01, and 2.07 arcmin, respectively. The ratio of these predicted measurements $\tilde{\bar{y}}_{\text{AP}, \nu_i}/\tilde{\bar{y}}_{\text{AP}, \nu_j}(R)$ quantifies the beam impact at each aperture, and can be used as a beam correction factor that accounts for differences arising solely from the beam of the map used to construct the measurements. We find that for each sample, this factor remains consistent across all its subsamples, for both the full and radio-cleaned samples. We show results of the beam correction factor normalized to f150 in \ref{beamcorrtab}.

\subsection {kSZ signal extraction}
\label{sec:kSZsignals}

At separations on the order of 25-50 Mpc, groups and clusters of galaxies tend to move, on average, toward one another due to gravity. This motion leads to a correlation in the signature Doppler shift of the CMB induced by the kSZ effect. \cite{gong26, hadzhiyska25, hsu26} use a pairwise kSZ statistic that measures the average relative momentum of pairs of sources to detect the underlying kSZ signal. These measurements are then compared to theoretically predicted pairwise velocities from linear theory to infer the optical depth $\bar{\tau}$, which is related to the pairwise kSZ momentum $\hat{\textbf{p}}_{th}$ and the pairwise velocity $\hat{\textbf{V}}$ by

\begin{equation}
    \hat{\textbf{p}}_{th}(r, z) = -\frac{T_{\text{CMB}}}{c} \bar{\tau}\hat{\textbf{V}}(r, z).
\end{equation}

\noindent The methodology for extracting the kSZ signal using this pairwise estimator is described in depth in Gong et al. \cite{gong26} Section III. One of the goals of the present work is to derive a tSZ-based estimate of the optical depth for the samples, which may then be used along with the pairwise kSZ momentum measurements to recover the pairwise velocities.

\section{Results: tSZ systematics}\label{sec:results}
In the following section we explore numerous foregrounds and systematic effects that bias our tSZ measurements. The pairwise differencing of the kSZ measurements is expected to effectively cancel out many foregrounds, thus care must be taken to correct for foreground biases in the tSZ measurements before attempting any interpretation of $\bar{y}$ and $\bar{\tau}$ measurements. In Section~\ref{sec:profiles} we present the radial profiles for each sample and map (Figs.~\ref{fig:raw_profiles} and \ref{fig:ilc_deproj_profiles}) and apply beam corrections to account for the differing beam sizes. In Section~\ref{sec:dustilc} we discuss the efficacy of the ILC map deprojections at removing contamination originating from dust and large-scale CIB and examine their sensitivity to assumed SED parameters. In Section \ref{sec:dustsf}, we adopt the methodology of \cite{liu25} to estimate and remove contamination from the single-frequency data using the 220 GHz data as a tracer of dust and CIB emission (Figure \ref{fig:deproj_sf_profiles}). In Section \ref{sec:radio}, we examine the impact of emission from nearby radio sources by cross-matching our samples with the NRAO VLA Sky Survey catalog (Figures \ref{fig:radio_cleaned_stacks} and \ref{fig:radio_clean_profiles}). Dust, CIB, and radio source contamination all bias our Compton-$y$ measurements low and, in some cases, completely overwhelm the tSZ signal. Finally, in Section \ref{sec:2halo}, we address the two-halo effect, which biases the tSZ measurement high on scales larger than the virial radius.

\begin{figure}
\begin{center}
\includegraphics[width=8.4cm]{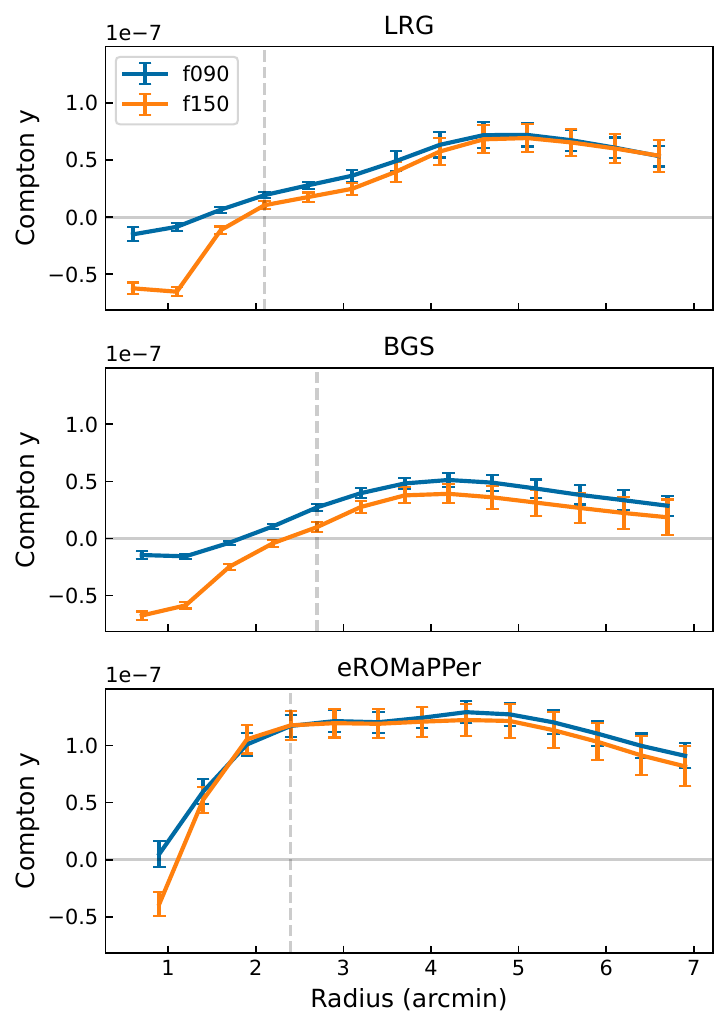}
\caption{Stacked radial profiles of the three samples computed using the ACT DR6 f090 map [blue] and f150 map [orange]. The radial profiles are composed of disk-ring aperture photometry measurements as described in Section \ref{sec:methods}. Beam corrections (described in Section \ref{sec:beamcorr}) have been applied to the 90 GHz profiles. Lines are used to guide the eye and do not indicate a fit to the data. The nominal aperture for each sample ($2.1'$, $2.7'$, and $2.4'$ for the LRG, BGS, and \texttt{eROMaPPer} samples, respectively) is indicated by a vertical dashed line. Profiles for all binned samples are provided in Appendix \ref{sec:binnedprofs} Figure \ref{fig:raw_profiles_binned}.}
\label{fig:raw_profiles}
\end{center}
\end{figure}

\begin{figure}[h!]
\begin{center}
\includegraphics[width=8.6cm]{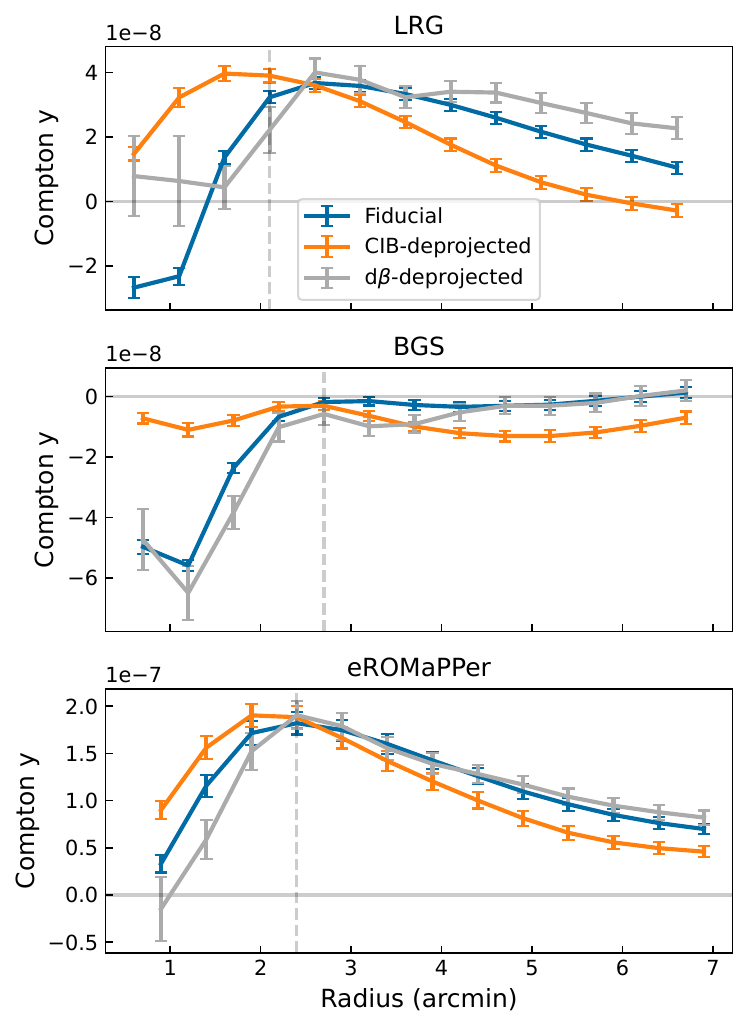}
\caption{Stacked radial profiles of the three samples computed using the fiducial [blue], CIB-deprojected [orange], and d$\beta$-deprojected [gray] ACT DR6 ILC Compton-$y$ maps. For the LRG sample, the CIB-deprojection gives an improved measurement near the aperture radius, while the d$\beta$-deprojection introduces additional noise and instability. For the BGS sample, all ILC variants remain severely contaminated. Differences are comparatively minor for the \texttt{eROMaPPer} sample. Profiles for all binned samples are provided in Appendix \ref{sec:binnedprofs} Figure \ref{fig:ilc_deproj_profiles_binned}.}
\label{fig:ilc_deproj_profiles}
\end{center}
\end{figure}

\subsection{Radial profiles}\label{sec:profiles}
We present radial profiles for the complete stacked samples using the single-frequency maps in Figure \ref{fig:raw_profiles} and the ILC maps in Figure \ref{fig:ilc_deproj_profiles}. We provide radial profiles for each binned sample in Appendix \ref{sec:binnedprofs}  Figures \ref{fig:raw_profiles_binned} and \ref{fig:ilc_deproj_profiles_binned}. These radial profiles show the average disk-ring aperture photometry measurements computed as described in Section \ref{sec:tSZextraction} at a range of aperture sizes. The 90 GHz and ILC profiles are beam corrected as described in Section \ref{sec:beamcorr}. We find the corrections are most significant at the lowest aperture radii around and below $1'$, but become less important with increasing aperture radii beyond $2'$. The beam correction factors are consistent among all bins of a given sample. Approximate values of the factor are shown in Table \ref{beamcorrtab} for the aperture size used in this paper.

\begin{table}
\centering
    \setlength{\tabcolsep}{5pt}
\renewcommand{\arraystretch}{1.25}
\begin{tabular}{|l|c|l|l|}
\hline
        & LRG  & BGS  & \texttt{eROMaPPer} \\ \hline
ILC/150 & +5\%  & +2\%  & +3\%     \\ \hline
090/150 & +21\% & +10\% & +16\%     \\ \hline
220/150 & -6\%  & -4\%  & -5\%     \\ \hline
\end{tabular}
\caption{Approximate beam correction factors at the nominal aperture sizes, obtained by taking the ratio of modeled signals with the only difference being the beam used. The sign indicates the direction of the correction: the 90 GHz and ILC beams are larger than the 150 GHz beam, so the measurements are scaled up by the percentages shown, while the 220 GHz measurements are scaled down. These values apply to all subsamples within these samples, as deviations between subsamples are below 1\%, and also hold for the radio cleaned samples discussed in Section \ref{sec:radio}.}\label{beamcorrtab}
\end{table}

A negative 150 GHz signal relative to the 90 GHz signal, particularly near the center, is a strong indicator of dust/CIB leakage. The 150 GHz radial profiles for the LRG and BGS samples are substantially negative below the nominal aperture size, indicating significant contamination is present in these samples on small scales. Although the signals become positive near the aperture size, the 150 GHz measurements are still slightly suppressed relative to 90 GHz, further indicating the presence of foreground contaminants. Beyond $R\sim4'$, the 90 and 150 GHz profiles are more closely aligned with one another, which suggests the dominant source of contamination is the centralized dust emission. The \texttt{eROMaPPer} sample, comprised of higher-mass sources, shows much stronger agreement between 90 and 150 GHz signals on all but the smallest scales, indicating that while contamination may be present, it is subdominant to the tSZ signal. We test for the presence of large-scale CIB contamination in these profiles in Section \ref{sec:dustsf}.  

\subsection{Dust and Cosmic Infrared Background: ILC maps}\label{sec:dustilc}
The deprojected ILC maps discussed in Section \ref{sec:act} aim to remove contamination originating from large-scale CIB as well as dust emission from within galaxies by deprojecting a template SED (Equations \ref{eq:mbb} and \ref{eq:db_deproj}) from the fiducial map that depends on chosen values of $\beta$, the spectral index, and $T_{\mathrm{CIB}}$ \cite{coulton24}. In this section, we compare the two deprojection methods and explore the impact of the assumed parameter values on our measurements. We find that for our samples, the CIB deprojection is generally more stable than the d$\beta$ deprojection and the chosen SED parameters have little influence on our aperture photometry measurements.

In Figure \ref{fig:ilc_deproj_profiles} (see Appendix \ref{sec:binnedprofs} Figure \ref{fig:ilc_deproj_profiles_binned} for all binned samples), we see that the CIB-deprojected ILC profiles consistently show an increase in signal relative to the fiducial profiles at small radii, which is to be expected with the removal of dust contamination (which biases Compton-$y$ low). On larger scales, the CIB-deprojected signal is suppressed relative to the fiducial profile. The d$\beta$-deprojected profiles are significantly noisier and do not exhibit a consistent trend across the samples. For the LRG sample, the d$\beta$-deprojected signal is higher than the fiducial signal on the smallest and largest scales. Conversely, the d$\beta$-deprojected measurements are lower than the fiducial measurements on nearly all scales for the BGS sample. Notably, for this sample, all versions of the ILC map give a measured signal that is below or consistent with zero, indicating severe contamination that is not well-handled by either of the deprojections. The d$\beta$-deprojected measurements for the \texttt{eROMaPPer} sample are largely consistent with the fiducial profile on all but the smallest scales, where they fall below the fiducial signal level. The reduction in signal in the d$\beta$-deprojected maps at small scales, where there is visible contamination in the submaps, indicates either this template is not well-matched to the dust and CIB in the sample, or that there is an additional source of contamination present.

In Appendix \ref{sec:compare_ilc}, we examine the stability of the deprojected radial profiles to the assumed CIB SED parameters. Briefly, we find that for all samples and deprojected ILC maps, the measured signal at the default aperture size is insensitive to the assumed SED parameters, consistent with the findings in \cite{coulton26}. On all scales, the measurements are not particularly sensitive to the choice of spectral index $\beta$ (Figure \ref{fig:yprofiles_compare_beta}). We find the choice of CIB temperature does not significantly impact the signal measured from the d$\beta$-deprojected map, however at scales beyond the aperture size, assuming a larger value of $T_{\mathrm{CIB}}$ leads to a significantly higher Compton-$y$ using the CIB-deprojected map (Figure \ref{fig:ilc_deproj_profile_temp_comparison}).

\subsection{Dust and Cosmic Infrared Background: Single-frequency maps}\label{sec:dustsf}
To gain a better understanding of contamination present in our samples, we follow the methodology of \cite{liu25} Appendix A and stack the samples on the 90 and 150 GHz maps. At 90 and 150 GHz, the underlying tSZ signal should be approximately the same, but any dust or CIB contamination will be predominantly evident in the 150 GHz map. This contamination is evident in Figure \ref{fig:raw_profiles}, in which the 150 GHz profiles exhibit a sharp decrease relative to the 90 GHz profiles at small radii, even after accounting for the difference in beam sizes. We do not observe a similar decrease in the 150 GHz signal in the \texttt{eROMaPPer} profiles except on the smallest scales; beyond $1'$, the 150 GHz profile is consistent with or slightly increased relative to 90 GHz. 

To separate any large-scale CIB from the central dust contamination, we use a ring-ring filter, which computes the differential aperture photometry signal in adjacent, equal-area rings. The inner ring is defined as the region $R_0 < R < R_1$. To obtain equal area rings, the outer ring is defined as $R_1 < R < \sqrt{2R_1^2-R_0^2}$. We mask the central region by selecting a constant inner ring inner radius $R_0 = 1'$ and varying $R_1$ in $0.5'$ increments out to approximately $7'$. The ring-ring profiles for the full samples are shown in Figure \ref{fig:rr_profiles}. Aside from the BGS sample, which shows a slight (though still largely consistent) reduction in 150 GHz signal relative to 90 GHz, we find the 90 and 150 GHz measurements to be consistent in all samples on scales beyond the aperture size, indicating there is little large-scale CIB contamination in our samples. 

\begin{figure}
    \begin{center}
\includegraphics[width=8.6cm]{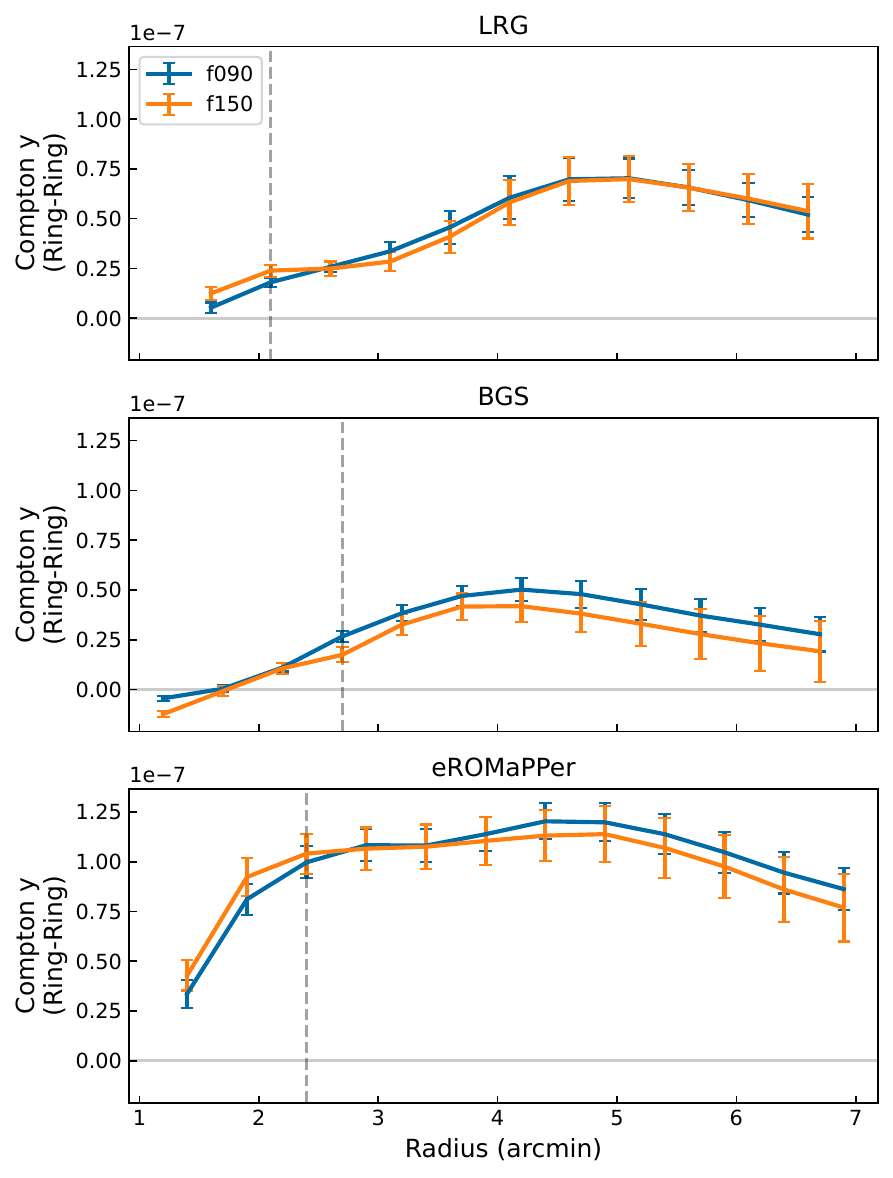}
\caption{90 and 150 GHz radial profiles for the LRG [top], BGS [middle] and \texttt{eROMaPPer} [bottom] galaxy catalogs computed using the ring-ring filter, which masks the central region and is used to test for large-scale contamination. The profiles agree within the error for all samples beyond the fiducial aperture size, indicating large-scale contamination does not significantly impact our measurements.} 
\label{fig:rr_profiles}
\end{center}
\end{figure}

The central dust contamination is best observed in the 220 GHz maps, which we assume to be free of tSZ. In the following analysis we assume the 220 GHz signal is a clean tracer of the dust and CIB signals, however due to bandpass effects we recognize the tSZ contribution may not be exactly zero. In Figure \ref{fig:f220_ims_and_profiles}, we show the stacked 220 GHz submaps and radial profiles for the full samples. In the LRG and BGS samples, we observe a bright central signal, confirming the presence of significant dust. From the radial profiles we can see this signal extends beyond the aperture size, though at large scales the signal is consistent with zero. We also observe a signal in the \texttt{eROMaPPer} sample which is more diffuse, extending beyond $5'$ in the radial profile, which indicates dust and CIB contamination may indeed be present in this sample, though unlike the LRG and BGS samples it does not overwhelm the tSZ signal at 150 GHz. 

\begin{figure}
\begin{center}
\includegraphics[width=8.6cm]{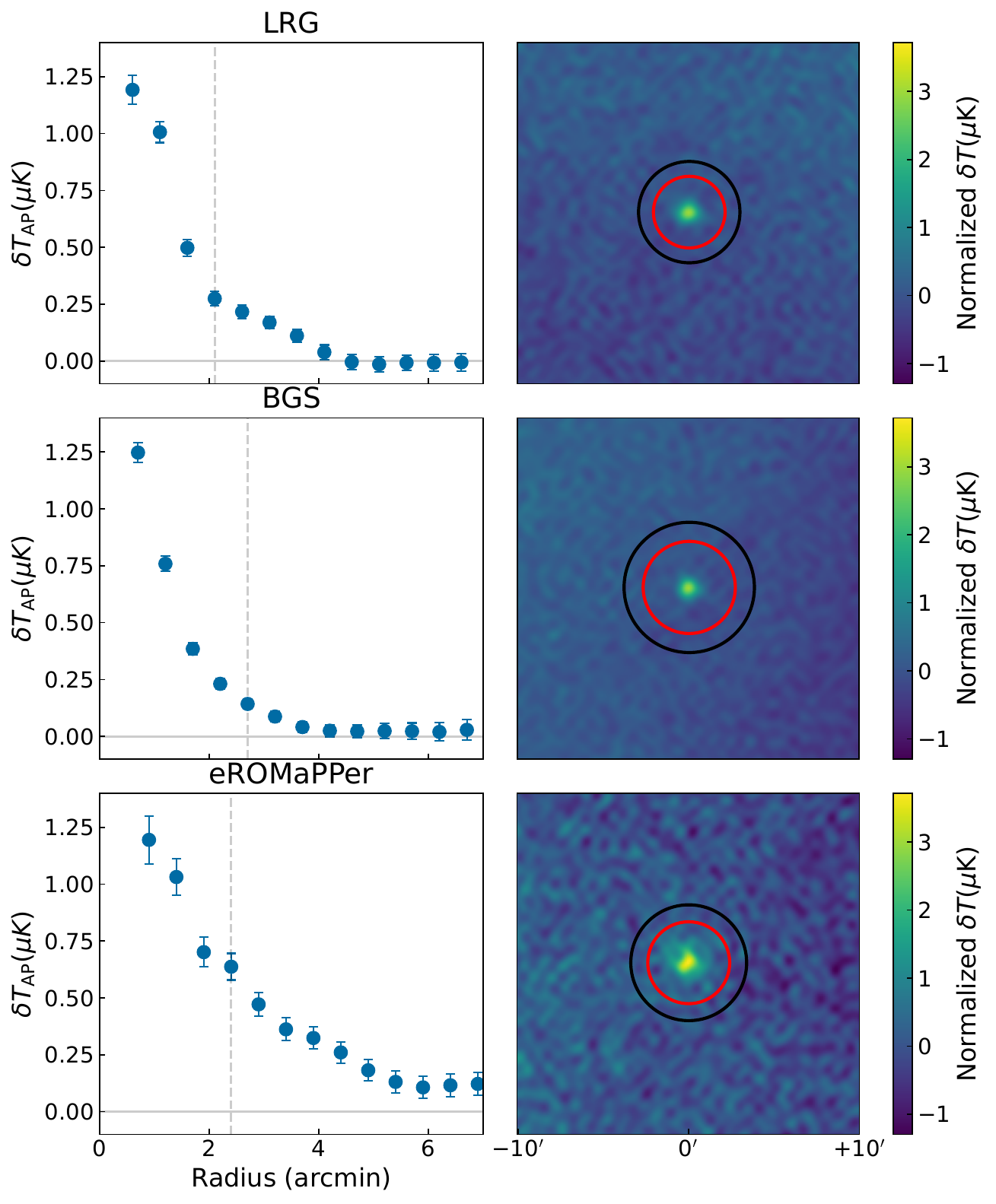}
\caption{Radial profiles and stacked submaps for the full LRG [top], BGS [middle], and \texttt{eROMaPPer} samples [bottom] derived from the ACT DR6 220 GHz map. The images have been mean-subtracted to facilitate comparison across samples. The default aperture photometry disk is indicated in the profiles by the vertical dashed lines and in the stacked images by the red circles. The black circles in the stacked images indicate the aperture photometry ring. All samples show evidence of centralized dust emission.} 
\label{fig:f220_ims_and_profiles}
\end{center}
\end{figure}

To isolate and remove the dust contamination that dominates the LRG and BGS samples, we follow the method presented in Liu et al. 2025 Appendix C \cite{liu25} and use the 220 GHz profiles as a tracer for the contamination. We rescale the stacked 220 GHz signal to 90 and 150 GHz, choosing the value of $\beta$ that minimizes the $\chi^2$ between the corrected 90 and 150 GHz profiles, which are calculated as

\begin{equation}
    y_{\mathrm{deproj}}(\nu) = y_\mathrm{raw}(\nu) - \frac{f(\nu)}{f(220)} y_{\mathrm{raw}}(220)
\end{equation}

where

\begin{equation}
    f(\nu) = \frac{\nu^{3+\beta}}{e^{h\nu/k_BT_{\mathrm{CIB}}}-1} \cdot \frac{dB(\nu, T)}{dT}^{-1}\bigg|_{T=T_{\mathrm{CMB}}}
\end{equation}
is just the frequency-dependent part of Equation \ref{eq:mbb}. We follow the method in \cite{liu25} to compute covariance matrices for the single frequency measurements via bootstrap resampling of the samples and propagate the errors through the deprojection process. Unlike \cite{liu25}, we apply this deprojection technique to our disk-ring profiles instead of the ring-ring profiles to maintain consistency with the kSZ measurements, which use the disk-ring filter. We choose to fit the inner $R<2'$ core, which is dominated by dust emission, separately from the rest of the profile. We find the contamination in the inner core is described by a lower value of $\beta$ compared to the rest of the profile, which results in a more aggressive deprojection in that regime. We show the combined deprojected profiles for the LRG and BGS samples in Figure \ref{fig:deproj_sf_profiles}.  In both samples, we see a significant increase in 150 GHz signal in the central core of the profile, bringing the 150 GHz measurements more in line with the 90 GHz measurements which are expected to be relatively free of dust. Aligning with the observed lack of 220 GHz signal at large scales in Figure \ref{fig:f220_ims_and_profiles}, we see virtually no difference between raw and corrected f150 profiles beyond $R\sim5'$. Although there is evidence of contamination in the \texttt{eROMaPPer} sample from the 220 GHz data shown in Figure \ref{fig:f220_ims_and_profiles}, we see in Figure \ref{fig:raw_profiles} that beyond $R=1'$, the raw 150 GHz Compton-$y$ signal is consistent with than the 90 GHz signal, indicating the contamination is subdominant. Furthermore, in the case of the binned $\lambda>10$ and $\lambda >20$ profiles shown in Appendix Figure \ref{fig:raw_profiles_binned}, the 150 GHz signal surpasses the 90 GHz signal near the fiducial aperture size. As a result, this deprojection procedure fails to induce any meaningful shift in the profiles for this sample.

\begin{figure}
\begin{center}
\includegraphics[width=8.6cm]{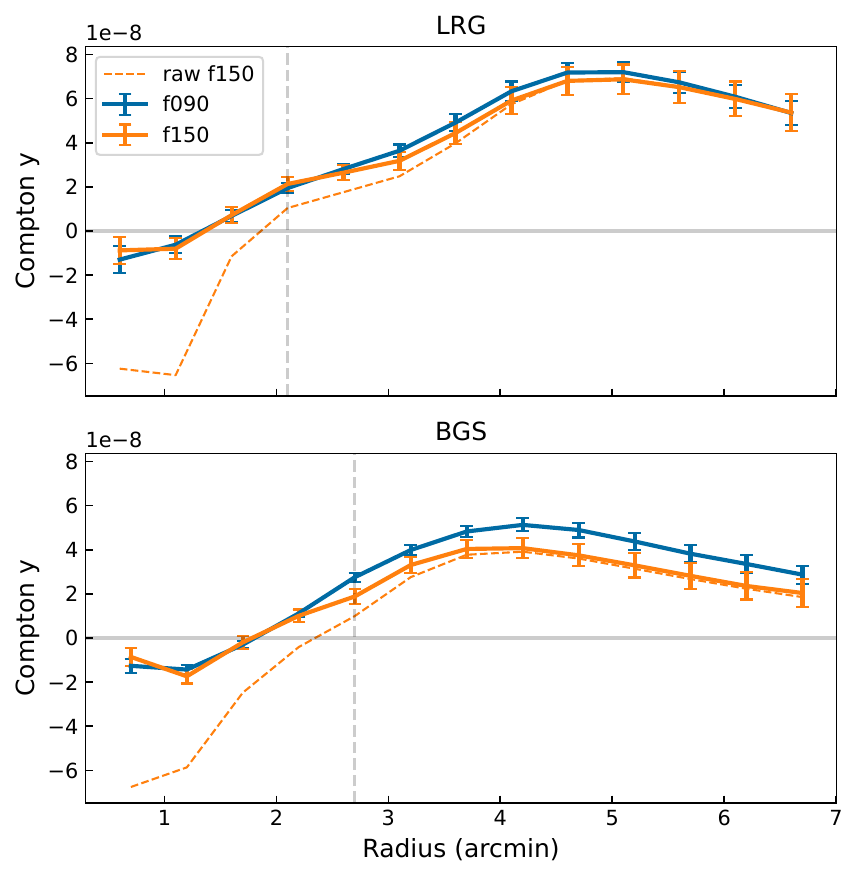}
\caption{Corrected f090 and f150 disk-ring radial profiles for the LRG and BGS samples, in units of Compton-$y$. The uncorrected 150 GHz profiles are shown as orange dashed lines. The deprojected 90 and 150 GHz profiles are shown as solid blue and orange lines, respectively. For both samples, the deprojection successfully recovers a clean 150 GHz signal at and below the default aperture size.}
\label{fig:deproj_sf_profiles}
\end{center}
\end{figure}

\subsection{Radio contamination}\label{sec:radio}
The residual contamination observed in the deprojected ILC submaps and profiles for both the LRG and BGS samples (Figures \ref{fig:submaps} and \ref{fig:ilc_deproj_profiles}) implies the presence of additional contamination that is not well described by the deprojection templates. Another potential source of contamination in tSZ measurements is emission from co-located radio sources \cite{cooray98}. To assess the extent of radio contamination in our samples, we follow the methodology of \cite{battaglia26} to cross-match our catalogs with the NRAO VLA Sky Survey (NVSS) catalog of radio sources \cite{condon98} and search for targets that are located within 1 arcminute of an NVSS source. We find approximately 7\% of the LRGs, 6\% of the BGS, and 13\% of the \texttt{eROMaPPer} sources have nearby radio sources, which is comparable to the contamination level observed in \cite{battaglia26}. If we expand the search area to the aperture size ($2.1'$ for LRG, $2.7'$ for BGS, and $2.4'$ for the \texttt{eROMaPPer} sample), the number of affected sources rises to between $20-30\%$ of the sample size. 

\begin{figure*}
\begin{center}
\includegraphics[width=14.7cm]{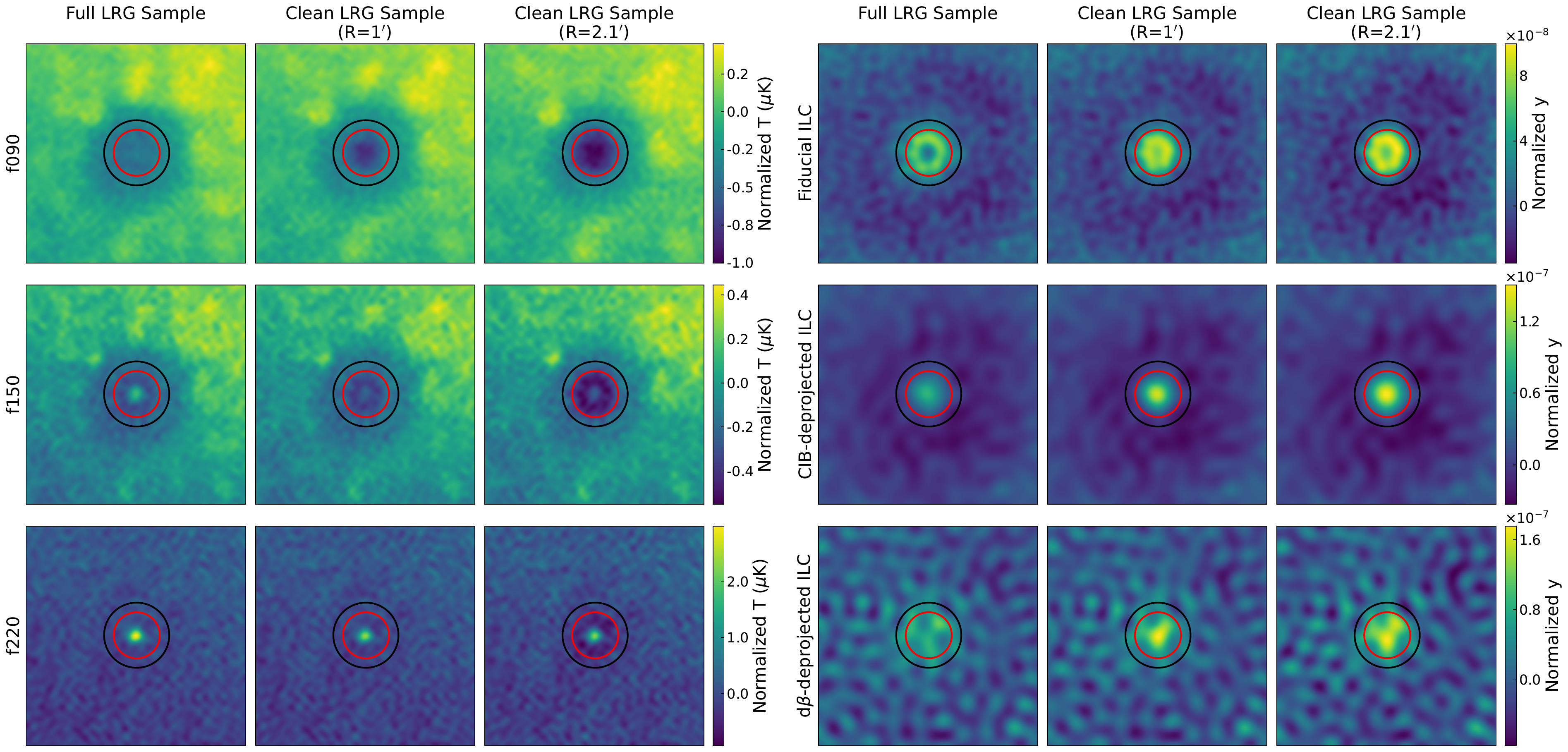}
\includegraphics[width=14.7cm]{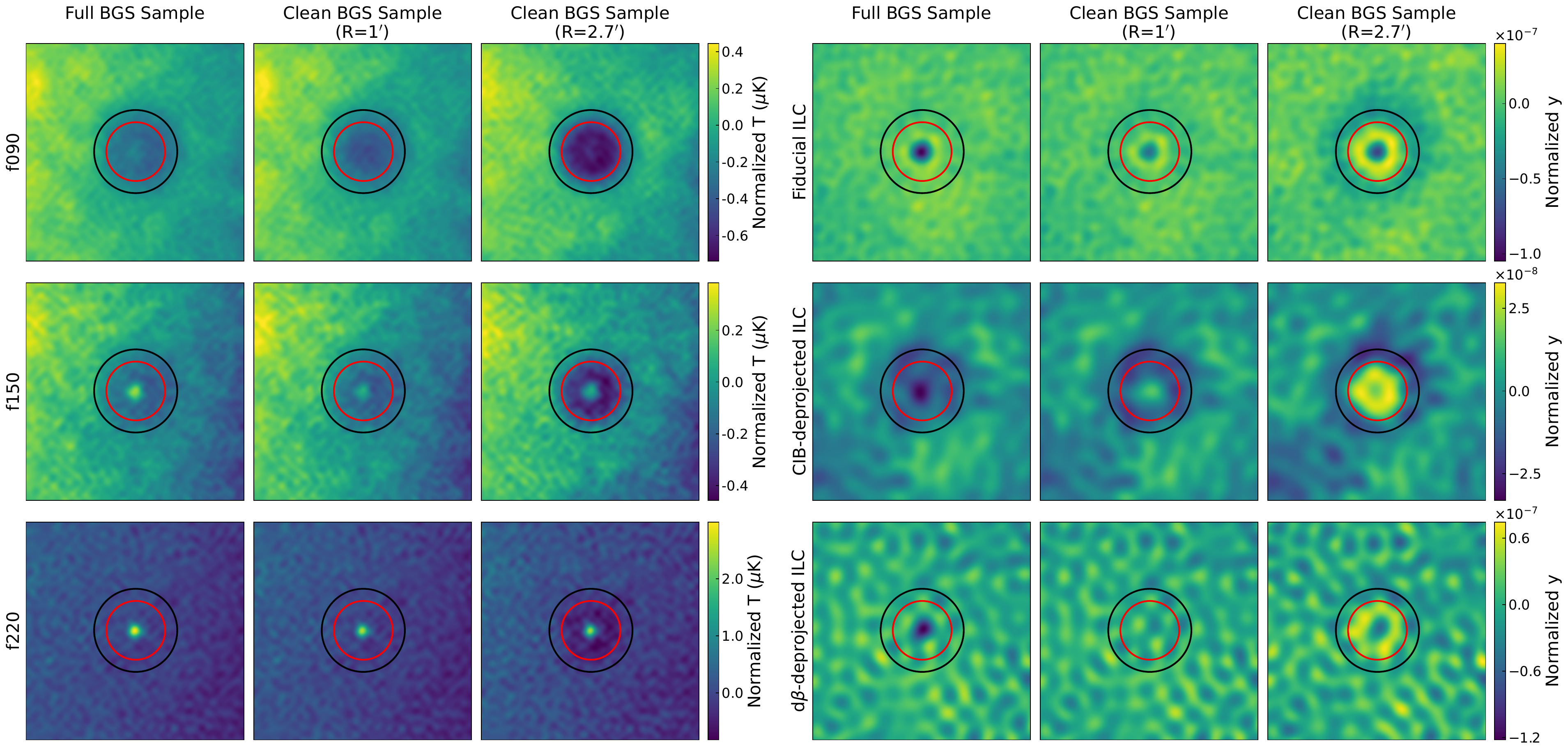}
\includegraphics[width=14.7cm]{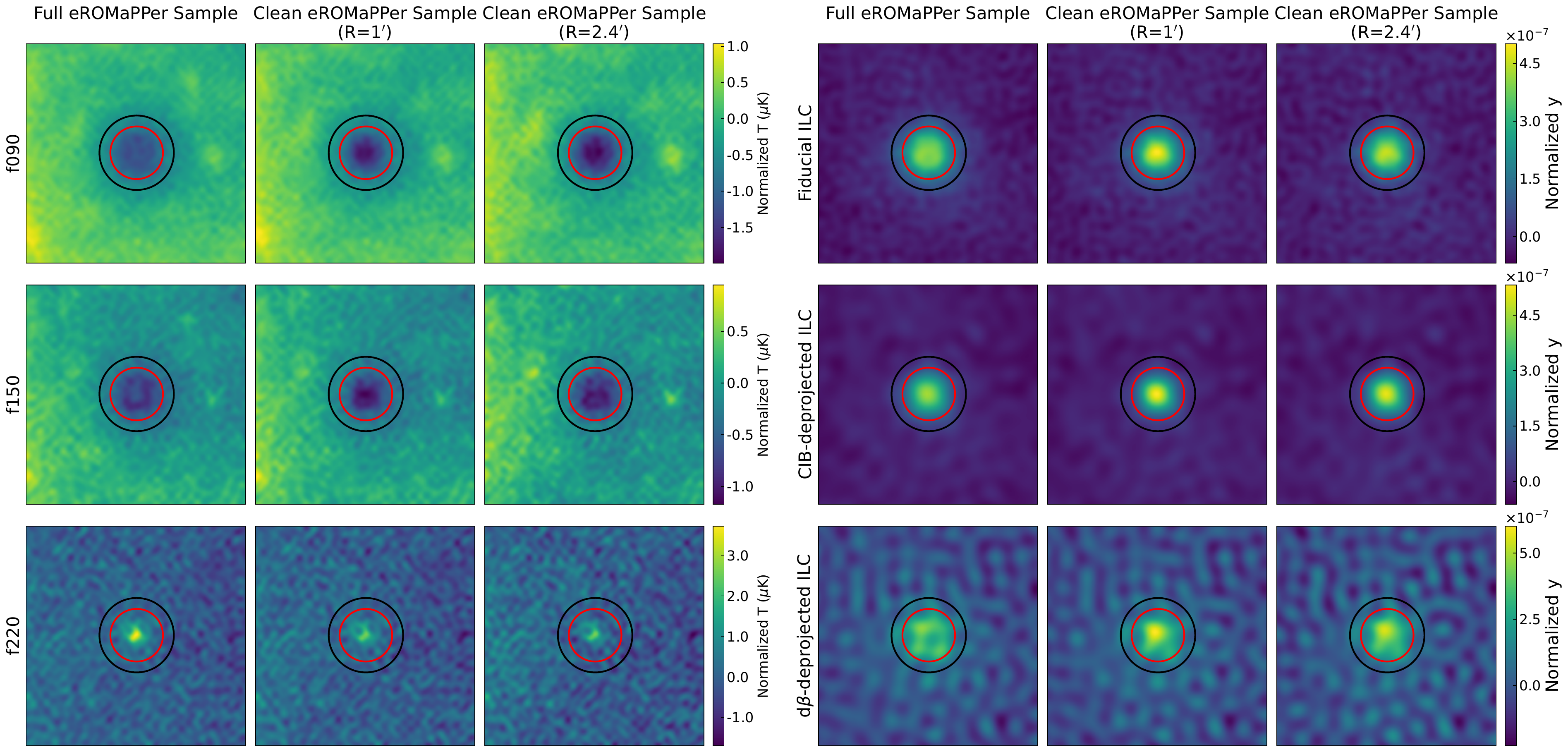}
\caption{Illustrating the impact of radio source contamination on the stacked submaps.  3x3 image grids of the single-frequency [left] and ILC [right] submaps for the LRG [top], BGS [middle], and \texttt{eROMaPPer} [bottom] samples. The tSZ signal becomes more evident as more contaminated sources are removed from the stacks. Centralized dust contamination persists in all samples. }\label{fig:radio_cleaned_stacks}
\end{center}
\end{figure*}

\begin{figure*}
\begin{center}
\includegraphics[width=17.9cm]{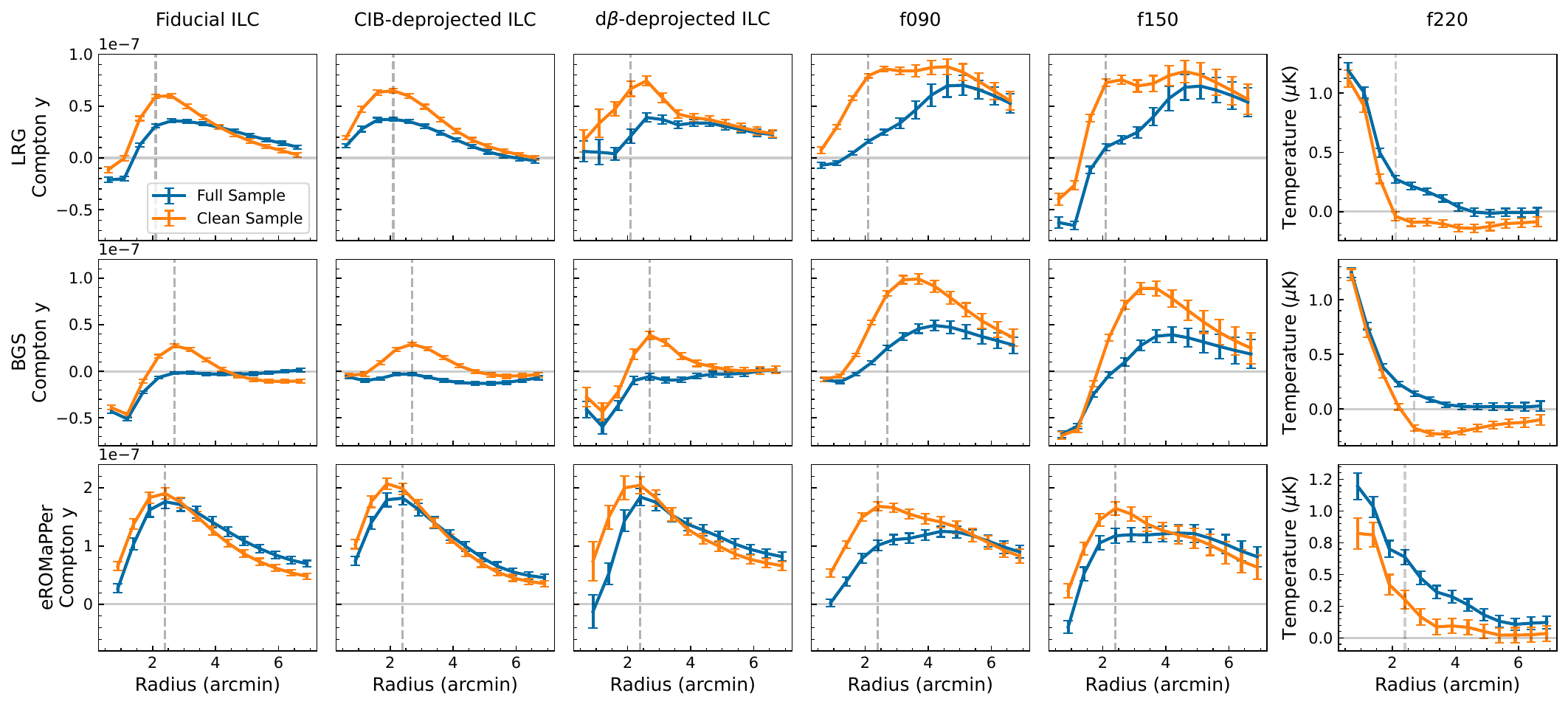}
\caption{The impact of radio contamination on radial profiles for the LRG [top], BGS [middle], and \texttt{eROMaPPer} [bottom] samples, comparing the full samples [blue] and radio-cleaned samples ($R=R_{\mathrm{AP}}$) [orange]. ILC profiles [columns 1-3] and 090/150 GHz profiles [columns 4-5] are presented in units of Compton-$y$. 220 GHz profiles [column 6] are presented in units of $\mu$K. All samples see a significant boost in Compton-$y$ with the removal of radio-contaminated sources, especially in the single-frequency maps. Evidence of centralized dust emission is still present in the 220 GHz profiles, and for the LRG and BGS samples, a negative 220 GHz signal near the default aperture size may be indicative of residual tSZ.}
\label{fig:radio_clean_profiles}
\end{center}
\end{figure*}

To quantify the impact of this contamination on our measurements, we repeat the stacking procedure outlined in \ref{sec:methods} on the radio-free sources. We compare the resulting stacked submaps to the original stacks in Figure \ref{fig:radio_cleaned_stacks}. For the LRG and BGS samples, we see the tSZ signal become much more evident in the 90 and 150 GHz maps as you remove more contaminated sources. Notably, the 220 GHz signal for these samples appears relatively unchanged, and central contamination is still evident in the 150 GHz maps, indicating dust remains prevalent in the cleaned samples. The stacked ILC submaps for the cleaned LRG and BGS samples are also improved, with both samples exhibiting a distinct positive signal in the d$\beta$-deprojected ILC maps. The differences are much more subtle in the \texttt{eROMaPPer} sample. We see enhanced tSZ signal in the 90 and 150 GHz maps, coupled with a decline in signal at 220 GHz, though some diffuse contamination remains visible. There is very little visible change in the \texttt{eROMaPPer} ILC submaps. This is in agreement with our previous conclusion that contamination in this sample is subdominant to the tSZ signal.


We define ``radio-clean'' cuts for each of the three samples by excluding all sources containing NVSS sources within the AP disk radius. Comparing the full-sample and radio-clean radial profiles for all maps and samples in Figure \ref{fig:radio_clean_profiles}, we see a distinct boost in Compton-$y$ at the aperture size. For all samples, the shift is most significant in the single-frequency maps. As noted earlier, we still observe a divergence in 90 and 150 GHz signal at small scales, and the 220 GHz signal for the radio-clean samples is nearly identical to that of the full samples at small scales, indicating dust is still a major factor for these samples. For the BGS sample, we see the d$\beta$-deprojected profile for the cleaned sample is still negative below $R\sim2'$, which may indicate this deprojection template is not well-suited to this sample.

While the 220 GHz temperature profiles for the cleaned LRG and BGS samples are consistent with the signal from the full sample within the innermost region, they become negative beyond $R\sim2'$, indicating that the radio-free sample has had an effect on this presumed foreground-only tSZ null channel, warranting further study. While the 220 GHz profile had been assumed to be a clean tracer of dust and CIB, such as in the deprojection technique described in Section \ref{sec:dustsf}, this radio-clean sample comparison illustrates that there may be subtleties to using 220 GHz data for dust and CIB constraints. Further work is needed to develop a more robust single-frequency deprojection pipeline, however, we note that given the degree of apparent contamination from radio sources in our samples, the uncorrected single-frequency measurements of the radio-clean samples are more representative of the true tSZ signal than the deprojected full sample measurements. Conversely, the 220 GHz profiles are positive at all radii for the \texttt{eROMaPPer} samples, and for the $\lambda > 5$ and $\lambda > 10$ bins, there is a distinct deficit in 150 GHz signal relative to 90 GHz at small radii that was not observed in the full sample profiles. As a result, we are able to deproject those measurements. These deprojected profiles are shown in Figure \ref{fig:opt_deproj_profiles}.

\begin{figure}
\begin{center}
\includegraphics[width=8.6cm]{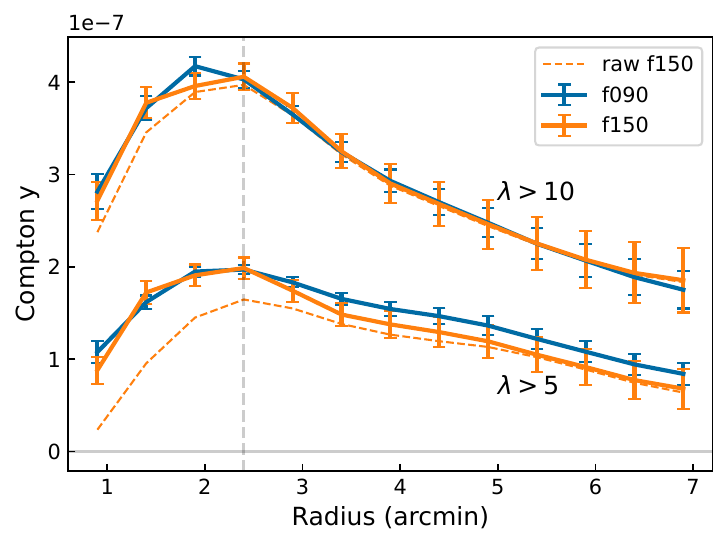}
\caption{Corrected f090 [orange full] and f150 [blue full] and uncorrected (raw) f150 profile [orange dashed] disk-ring radial profiles for the radio-clean \texttt{eROMaPPer} $\lambda>5$ and $\lambda>10$ samples, in units of Compton-$y$.}
\label{fig:opt_deproj_profiles}
\end{center}
\end{figure}

While we have established that contamination from radio sources is non-negligible for all of our samples, the aim of this paper is to compare thermal and kinematic SZ measurements for identical samples. We do not find removing the contaminated sources significantly alters the average redshift or mass of the samples, however the kSZ-derived optical depth values presented in \cite{gong26, hadzhiyska25, hsu26} are estimated from measurements of the full samples as described in Section \ref{sec:data}, so in Section \ref{sec:results} we present results for the full samples as well as the radio-clean samples. Future work will explore the impact of this contamination on the kSZ results. 

\subsection{Two-halo term}\label{sec:2halo}

3D radial electron pressure profiles of clusters $P_e(r)$ can be described as the combination of two different components: the one-halo term $P_e^{1h}(r)$ which describes the pressure contribution coming from the centered object, and the two-halo term $P_e^{2h}(r)$ which describes the contribution of surrounding halos (which becomes relevant around and beyond the virial radius of the centered halo) \cite{hill18, vikram17, moser21}. As the profile-to-measurement forward model described in \ref{sec:beamcorr} (and fully outlined in \citet{popik_impacts_2025}) is a sequence of linear processes, it can therefore be shown that $\tilde{\bar{y}}_{AP}(R)$ can also be broken down into a one-halo component $\tilde{\bar{y}}_{AP}^{1h}(R)$ and two-halo component $\tilde{\bar{y}}_{AP}^{2h}(R)$. Previous studies have shown that the two-halo component can account for between 2-10\% of the observed signal \citep{vavagiakis21}. Using the fitting results off the radio-cleaned measurements from \ref{sec:beamcorr}, we find values of around 5-10\% for each of the samples at nominal apertures. 

While our rough overall values are consistent with \cite{vavagiakis21}, full profiles for all samples were not reliable enough to be displayed in this paper due to the limits of the measurements combined with the nature of the modeling. There are significant degeneracies between the one-halo and two-halo components \citep{moser21} which are mostly only broken by measurement at large aperture values, but we are constrained to fitting apertures below $4'$ as the significant suppression in signal from dust contamination (discussed in more detail in \ref{sec:dustilc}) is infeasible to fit with our correct model. Although the resulting best-fit profiles matched the measurement well for both the full and radio-cleaned subsamples, the individual one-halo and two-halo profiles varied wildly across subsamples, which is notably unexpected as profiles should remain similar within subsamples of the same galaxy selection. We attribute this large variance to the degeneracies caused by the lack of information at large aperture.

\subsection{Summary}

In this section we have detailed numerous systematic effects and sources of contamination in the tSZ measurements. Our key findings are:
\begin{enumerate}
    \item Across the LRG and BGS samples, centralized contamination from dust and radio sources dominates the tSZ signal at small ($R<2'$) radii. Contamination is subdominant to the tSZ signal for the \texttt{eROMaPPer} sample.
    \item Both sources of contamination bias the tSZ measurements low, but can be accounted for through deprojection (dust) or by removing affected sources from the samples (radio).
    \item Large-scale contamination from the CIB is subdominant in all samples.
    \item For the ILC analysis, CIB-deprojection is generally more stable than the d$\beta$-deprojection (especially for the BGS sample).
\end{enumerate}

\begin{figure*}
    \begin{center}
    \includegraphics[width=17.6cm]{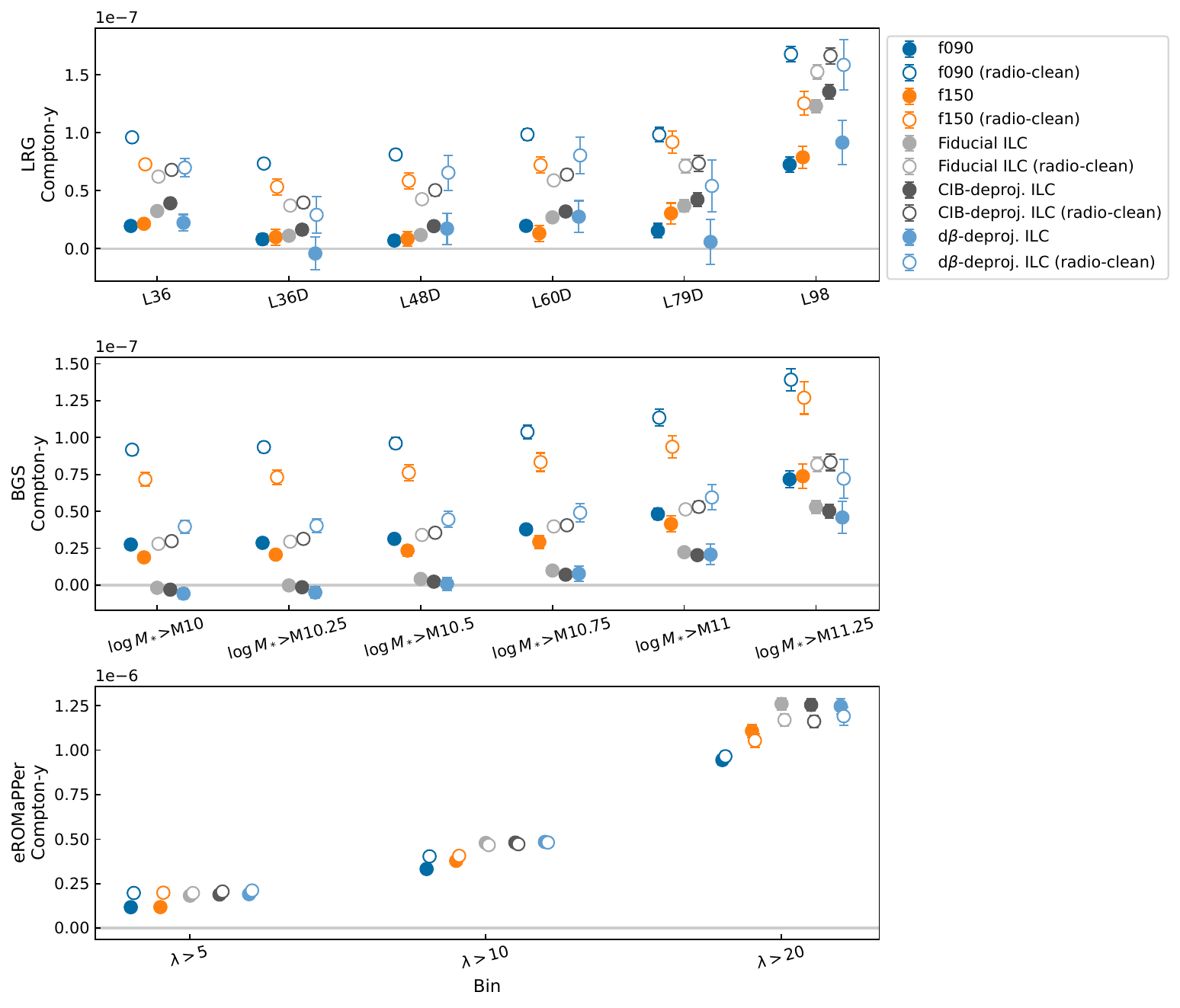}
        \caption{Final corrected Compton-$y$ aperture photometry values for the LRG [top], BGS [middle], and \texttt{eROMaPPer} [bottom] samples. We present both the original full-sample measurements [filled circles] as well as the radio-clean measurements [empty circles]. In general, we see an increase in $\bar  y_{\mathrm{AP}}$ with increased mass, luminosity, and richness. For the full samples, the 90 GHz  and CIB-deprojected ILC maps provide the highest SNR, while the 90 GHz map provides the best SNR across the board for the radio-clean samples.}\label{fig:yaps}
    \end{center}
\end{figure*}

\section{Results: Optical depth measurements}\label{sec:taucomparisons}

In this section, we present the optical depth measurements for each sample derived from both the tSZ and kSZ analyses. We present tSZ results derived from both the complete samples as well as the radio-clean samples discussed in Section \ref{sec:radio}. All kSZ optical depths presented here are derived from the full samples and thus are not directly comparable to the radio-clean tSZ optical depths. However, the null tests performed in the pairwise kSZ analyses \cite{hadzhiyska25,gong26} on randomized datasets give results consistent with zero signal, which implies this contamination does not bias the kSZ measurements. 

Our final tSZ Compton-$y$ measurements, taking into account the systematics discussed in Section \ref{sec:results}, are presented in Section \ref{sec:tsz} (Figure \ref{fig:yaps}). In Section \ref{sec:kSZmeas} we present the kSZ-derived optical depths from the companion papers. In Section \ref{sec:ytau} we develop $\bar y-\bar {\tau}$ scaling relations for each sample using a combination of simulations-based measurements and the real data. We use these scaling relations to convert our $\bar{y}_{\mathrm{AP}}$ values to optical depths, which we then compare to the values reported in the kSZ companion papers (Figure \ref{fig:taucomps}).

\subsection{tSZ measurements}\label{sec:tsz}

The final stacked, corrected tSZ Compton-$y$ measurements are shown in Figure \ref{fig:yaps} and tabulated in Appendix \ref{sec:yaps} Table \ref{tab:ytable}. We provide measurements for both the full samples, to complement the kSZ measurements, and for the radio-clean samples.

The LRG and BGS samples see the greatest shift in Compton-$y$ between the full and cleaned samples, with the magnitude of the increase varying with frequency and galaxy sample cut. The LRG sample sees the largest increase ($>30\sigma$) in the 90 GHz measurements, while the BGS sample sees a shift upwards of $20\sigma$ in both the 90 GHz and CIB-deprojected ILC measurements. The shift is less pronounced in the measurements of the \texttt{eROMaPPer} sample, which aligns with the more subtle changes observed in the stacked submaps in Figure \ref{fig:radio_cleaned_stacks}, however all of the clean ($R=R_{\mathrm{AP}}$) measurements are shifted at least $1\sigma$ higher than the original measurements, indicating this is a non-negligible source of contamination.

The observed difference between 90 and 150 GHz measurements of the radio-clean LRG and BGS samples is likely due to unremoved dust contamination. The slight decrease in signal observed in the radio-clean \texttt{eROMaPPer} $\lambda>20$ bin is likely due to the significantly decreased sample size: the $\lambda>5, \lambda>10$, and $\lambda>20$ bins lose 31, 37, and 45\% of sources when the radio source cut is applied, bringing the total number of sources down to approximately 150,000 ($\lambda>5$). 

Using the full samples, we obtain the best signal-to-noise with the 90 GHz and CIB-deprojected maps. For the LRG sample, we measure the tSZ signal of the L36 (full catalog) bin with $18\sigma$ significance using the CIB-deprojected ILC map, and surpass $20\sigma$ significance for the higher mass cumulative bins. The SNR of the disjoint bins is lower, though we achieve $>5\sigma$ significance for the L79 and L98 bins. The SNR jumps substantially when radio-contaminated sources are removed, particularly for the lower mass bins which maintain a high number of sources after the cut. For the L36 bin, we obtain SNR $>28$ using the CIB-deprojected ILC map and SNR $>38$ using the 90 GHz map. For the BGS sample, we achieve SNR of 13.5 for the $\log_{10} (M_*/M_{\odot})_*>10$ bin (full catalog) using the 90 GHz map, and that number jumps to 27 following removal of contaminated sources. With the radio-clean samples, we obtain SNR $>5$ for all samples and maps. We achieve the highest overall SNR with the \texttt{eROMaPPer} sample, approaching $40\sigma$ significance for the radio-clean $\lambda >5$ (full catalog) bin with the 90 GHz map. These SNR numbers are reported as including statistical uncertainties but without propagating uncertainties related to dust and CIB removal at either the ILC map-making level or the post-processing level described in this work.  

\subsection{kSZ measurements}
\label{sec:kSZmeas}

The optical depths derived from pairwise kSZ measurements of each  sample are presented in \cite{gong26, hadzhiyska25, hsu26}. In \cite{gong26}, the ACT DR6 ILC, 150 GHz, and combined 90+150+220 GHz (``ftot'') maps are used to compute the pairwise kSZ signal for the LRG sample. For the full sample (L36 bin; 913,286 galaxies) using the ILC map, a best-fit kSZ-derived optical depth of $\bar{\tau}_{AP} = 0.46 \pm 0.05 (\times 10^{-4})$ with $\chi^2_{\text{min}} = 18$ for 17 degrees of freedom and a signal-to-noise ratio of 9.3 was found. The SNR of this measurement is over 70\% greater than that reported in the previous analysis \cite{calafut21} and is the most significant pairwise kSZ measurement to date. Using the ILC map, SNR $> 5$ is achieved for all but the most luminous cumulative bin (L98). The SNR of the disjoint measurements is smaller, reaching a maximum of 4.2 for the L60D bin. 

For the BGS sample \cite{hadzhiyska25} a best-fit kSZ-derived optical depth of $\bar{\tau} = 7.19 \pm 1.37 \times 10^{-5}$ with $\chi^2_{\text{null}} = 42.98$ for 12 degrees of freedom and SNR $=5.24$ was found for the $\log_{10} (M_*/M_{\odot})_*>11$ sample using a 3.3 arcminute aperture and the ILC map. The kSZ analysis was conducted using a range of aperture sizes and 2.7 arcminutes was found to be best suited to the full sample. Using the fiducial aperture, $\bar{\tau} = 5.36 \pm 1.06 \times 10^{-5}$ was found for the $\log_{10} (M_*/M_{\odot})_*>11$ bin with $\chi^2_{\text{null}} = 38.73$ and SNR $=5.08$. For the full sample ($\log_{10} (M_*/M_{\odot})_*>10$; 1,610,381 galaxies), a best-fit $\bar{\tau} = 1.23 \pm 0.65 \times 10^{-5}$ with $\chi^2_{\text{null}} = 22.12$ was found with SNR $=1.90$.

Hsu et al. \cite{hsu26} presents the first pairwise kSZ measurement of \texttt{eROMaPPer} clusters using entirely spectroscopic redshifts, finding a best fit $\bar{\tau} = 0.63 \pm 0.10 (\times 10^{-4})$ with $6\sigma$ significance for the $\lambda > 5$ bin using the ILC map.

\subsection{$\bar{y}-\bar{\tau}$ scaling relations}\label{sec:ytau}
One of the aims of this paper is to use the combination of tSZ Compton-$y$ and kSZ $\bar{\tau}$ measurements to formulate a relationship between the two quantities that would enable the extraction of $\bar{\tau}$ from future tSZ measurements alone. Previous work using hydrodynamical simulations has developed a power-law $\bar y-\bar \tau$ scaling relation of the form

\begin{equation}\label{eq:tau}
    \ln\bar{\tau} = \ln\tau_0 + m\ln\frac{\bar{y}}{\bar y_0},
\end{equation}

\noindent where $\bar{\tau}$ is the average optical depth of the halo, $m$ is the power-law slope, and $\tau_0$ is the fit intercept \cite{battaglia16}. With tSZ and kSZ measurements of the same independent (disjoint) samples, we can fit this equation to the data and extract a best fit $\ln \tau_0$ and $m$, which we can then use to convert the tSZ Compton-$y$ measurements to an estimate of optical depth. Alternatively, one may use simulations-based measurements to extract the best-fit parameters. In this work, we employ different tactics for each of the three samples. For all samples, we use $\bar y_0 = 10^{-7}$.

\begin{figure*}[ht!]
\begin{center}
\hspace*{-0.3cm}
\includegraphics[width=15.6cm]{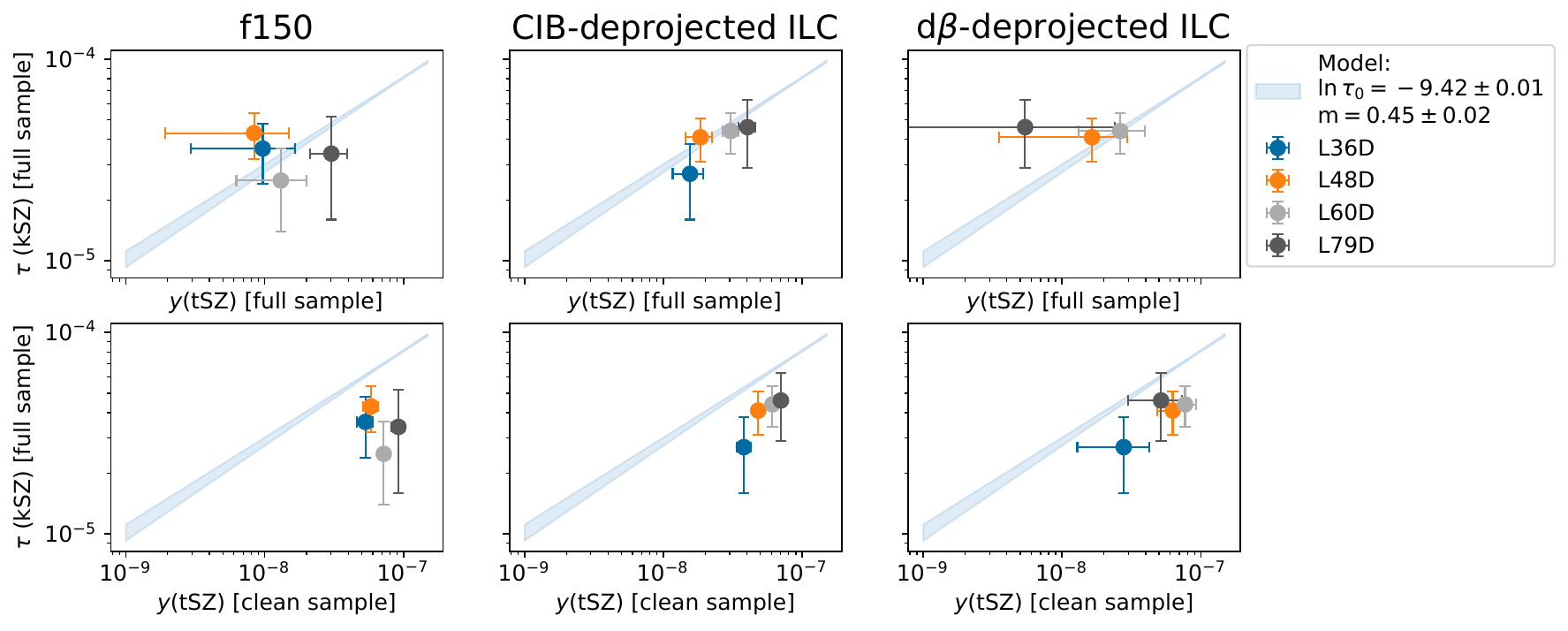}
\caption{kSZ-derived optical depths plotted versus tSZ-derived CIB-deprojected Compton-$y$ measurements for the disjoint LRG binned samples in the aperture for the full samples [top] and radio-clean samples [bottom] along with the mean and 1$\sigma$ uncertainty for the model $y-\tau$ scaling relation (Eq.~\ref{eq:tau}) derived from the simulated sample [light blue]. }
\label{fig:lrgytau}
\end{center}
\end{figure*}

For the LRG sample, we use the method outlined in \cite{gong24, gong25, soergel18} to develop a $\bar{y}-\bar{\tau}$ scaling relation using a simulated SZ map and sample, which is calibrated to match the characteristics of the full LRG sample. Using this methodology, we find best-fit $\ln\tau_0 = -9.42$ and $m = 0.45$ with systematic uncertainties of 0.01 and 0.02, respectively. We plot this model alongside the tSZ and kSZ measurements in Figure \ref{fig:lrgytau}. While $\bar{y}$ and $\bar{\tau}$ both appear to trend upwards with increasing luminosity in the CIB-deprojected ILC measurements, there is no distinct trend observed in either the 150 GHz or the d$\beta$-deprojected ILC measurements. Although the increased $\bar{y}$ of the radio-clean samples shifts the points away from the model curve, we see some improvement in the overall trend of the measurements. The 150 GHz $\bar{y}$ radio-clean measurements increase with increasing luminosity (though there is no trend in kSZ $\bar\tau$). We obtain positive Compton-$y$ for the L36D bin with the d$\beta$-deprojected ILC map, however, the L79D point is lower in $\bar{y}$ than both the L48D and L60D bins.

\begin{figure*}[ht!]
\begin{center}
\hspace*{-0.3cm}
\includegraphics[width=14.6cm]{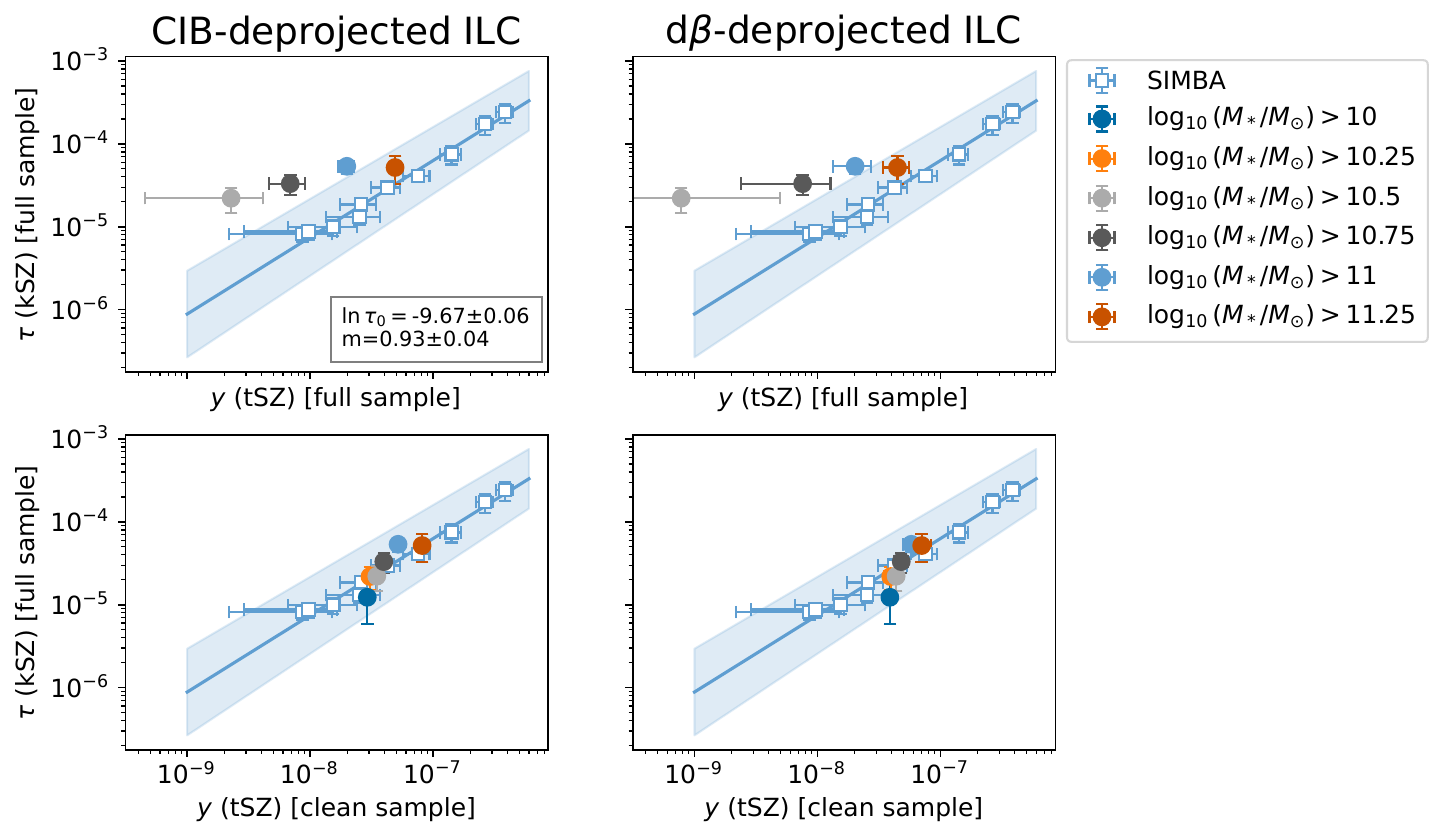}
\caption{kSZ-derived optical depths plotted versus tSZ-derived Compton-$y$ aperture photometry measurements for the full samples and [top] and radio-cleaned samples [bottom] for the BGS sample [filled circles] using the CIB- and d$\beta$-deprojected ILC maps. We also include measurements derived from a series of BGS-like disjoint samples generated with SIMBA using an ILC-like beam [light blue squares]; described in Appendix B of \cite{hadzhiyska25}) to which we fit Equation \ref{eq:tau} [light blue line and shaded region]. The SIMBA sample spans a halo mass range of $12.5 < \log_{10} M_h < 13.8$ $(M_{\odot}/h)$. Data points for the full sample $\log_{10} (M_*/M_{\odot})>10$ and $\log_{10} (M_*/M_{\odot})>10.25$ bins, which have negative values of Compton-$y$, are not shown.}
\label{fig:bgsytau}
\end{center}
\end{figure*}

We plot $\bar{y}-\bar{\tau}$ for the BGS sample in Figure \ref{fig:bgsytau}. We note the BGS bins are all cumulative (i.e., not independent) and therefore not ideal for extracting fit parameters. We instead use SIMBA \cite{dave19} to generate $\bar{y}$ and $\bar{\tau}$ measurements for disjoint binned BGS-like samples using an ILC-like beam. The simulated sample, described in detail in Hadzhiyska et al. 2025 Appendix B \cite{hadzhiyska25}, consists of sources with halo mass between 12.5 and $13.8 M_{\odot}/h$, binned into 10 disjoint samples (each with a thickness of $d\log M = 0.4$). The same aperture photometry methodology used for the real data is used with the simulation to extract $\bar{y}$ and $\bar{\tau}$, which are plotted alongside the data in Figure \ref{fig:bgsytau}. We fit Equation \ref{eq:tau} to the simulated $\bar{y}-\bar{\tau}$ measurements and find $\ln\tau_0 = -9.67 \pm 0.06$ and $m = 0.93 \pm 0.04$.  We find the measurements derived from the full BGS sample fall slightly outside of the model projection using both the CIB-deprojected and d$\beta$-deprojected ILC maps. Using the Compton-$y$ measurements obtained from the radio-clean samples, we find the data lie well within the error range of the SIMBA-derived model relation.

\begin{figure}[ht!]
\begin{center}
\hspace*{-0.3cm}
\includegraphics[width=8.6cm]{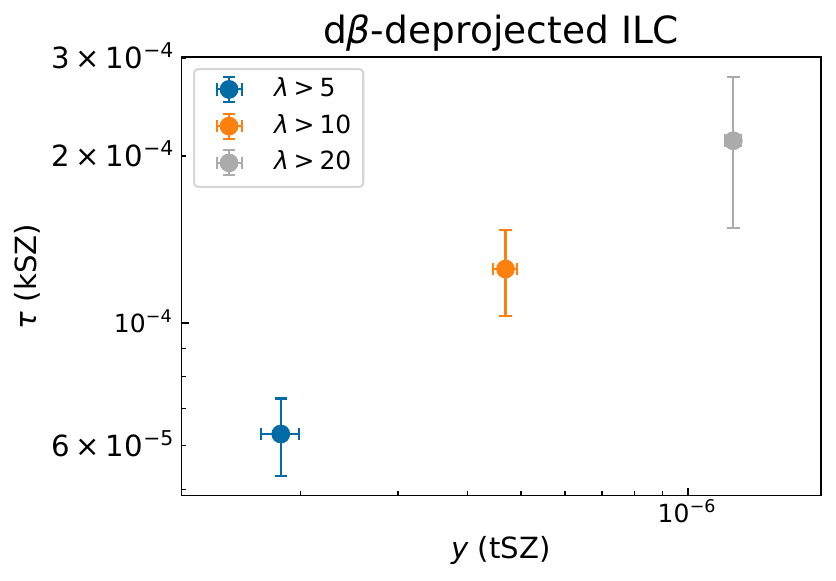}
\caption{kSZ-derived optical depths versus tSZ-derived Compton-$y$ aperture photometry measurements for the cumulative \texttt{eROMaPPer} samples with the d$\beta$-deprojected ILC map using the full sample also used to obtain the kSZ-derived $\tau$ values.}
\label{fig:optytau}
\end{center}
\end{figure}

We plot $\bar y-\bar\tau$ for the \texttt{eROMaPPer} samples in Figure \ref{fig:optytau}. While there is a very obvious trend in $\bar y-\bar\tau$ with increasing lower richness cut, like the BGS samples, these samples are cumulative. To facilitate a fit to $\bar y-\bar\tau$, we split the \texttt{eROMaPPer} sample into two disjoint bins which are described in Table \ref{tab:opt_disjoint}. Using the tSZ and kSZ measurements for these disjoint bins along with the measurements for the highest mass ($\lambda>20$) bin, we fit Equation \ref{eq:tau} to the d$\beta$-deprojected data and find $\ln\tau_0=-9.62 \pm 0.06$, $m=0.46 \pm 0.03$ for the full sample and $\ln\tau_0=-10.05 \pm 0.04$, $m=0.66 \pm 0.02$ for the radio-clean sample. These fits are shown in Figure \ref{fig:optytau_disjoint}. The $5 < \lambda < 10$ and $10 < \lambda < 20$ tSZ measurements increase after removing contaminated sources, resulting in a slightly steeper fit to the data. Notably, the Compton-$y$ error for the lowest richness sample shrinks considerably with the removal of radio-contaminated sources. We find no significant difference in fit parameters if we instead fit to the CIB-deprojected measurements. These are the first empirical $\bar y-\bar\tau$ fits to stacked tSZ and pairwise kSZ measurements to date.

\begin{table}[]
    \centering
    \setlength{\tabcolsep}{2.5pt}
    \renewcommand{\arraystretch}{1.25}
    \begin{tabular}{|c|c|c|c|c|c|}
    \cline{5-6}
    \multicolumn{4}{c|}{} &  CIB-deproj. & d$\beta$-deproj \\ \hline
    Bin & N & $\langle z \rangle$ & $\langle \lambda_{\mathrm{DES}} \rangle$ & $\bar{y}_{\mathrm{AP}}/10^{-7}$ & $\bar{y}_{\mathrm{AP}}/10^{-7}$ \\ \hline
    $5<\lambda<10$  & 140,251 & 0.48 &  6.89 & $0.20 \pm 0.06$ & $0.21 \pm 0.12$ \\
    (clean) & 100,465 & 0.48 & 6.86 & $0.68 \pm 0.06$ & $0.72 \pm 0.14$ \\\hline
    $10<\lambda<20$ & 58,127 & 0.48 &  13.63 & $1.77 \pm 0.08$ & $1.85 \pm 0.20$ \\
    (clean) & 38,519 & 0.49 & 13.55 & $2.43 \pm 0.10$ & $2.46 \pm 0.23$ \\\hline
    
    \end{tabular}
    \caption{Disjoint binned \texttt{eROMaPPer} samples. For all samples, we see consistency in the Compton-$y$ measurements with both deprojected ILC maps, and a significant increase in Compton-$y$ in the radio-clean samples. These disjoint bins are used along with the $\lambda >20$ bin to construct a $\bar{y}-\bar{\tau}$ scaling relation (Eq. \ref{eq:tau}) to convert the Compton-$y$ values shown here to optical depth.}
    \label{tab:opt_disjoint}
\end{table}

\begin{figure}[ht!]
\begin{center}
\hspace*{-0.3cm}
\includegraphics[width=8.6cm]{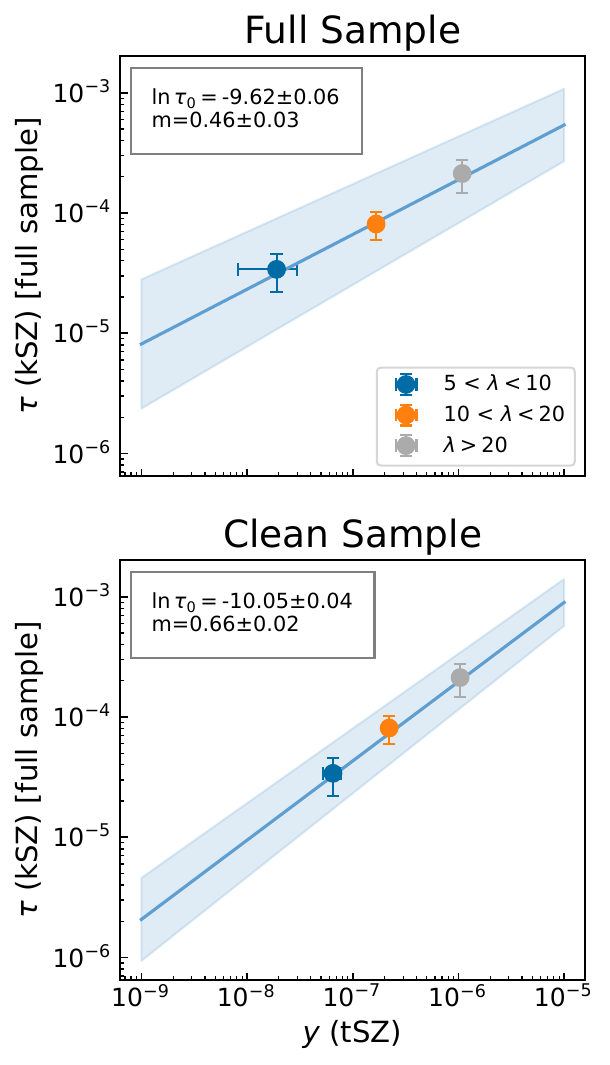}
\caption{kSZ-derived optical depths plotted versus tSZ-derived Compton-$y$ aperture photometry measurements for the disjoint \texttt{eROMaPPer} samples obtained with the d$\beta$-deprojected ILC map for the full samples [top] and radio-clean sample [bottom]. The blue line and shading show the $\bar{y}-\bar{\tau}$ fit and $1\sigma$ uncertainty.} 
\label{fig:optytau_disjoint}
\end{center}
\end{figure}

For all samples, we use the aforementioned values of $\ln\tau_0$ and $m$ (compiled in Table \ref{tab:taufit}) in Equation \ref{eq:tau} to convert our average Compton-$y$ measurements for the disjoint binned samples to optical depth $\bar{\tau}$. We employ a Monte Carlo routine to propagate the errors on our model parameters into our tSZ-derived optical depth estimates. We tabulate the tSZ and kSZ optical depths in Table \ref{lrgtautable} and plot all of the optical depths together in Figure \ref{fig:taucomps}. 

\begin{table}[]
    \centering
    \renewcommand{\arraystretch}{1.25}
    \setlength{\tabcolsep}{6pt}

    \begin{tabular}{|c|c|c|}\hline
    Sample & $\ln \tau_0$ & $m$ \\\hline
    LRG & $-9.42\pm 0.01$ &  $0.45\pm0.02$ \\ \hline
    BGS & $-9.67\pm 0.06$& $0.93\pm0.04$ \\ \hline
    \texttt{eROMaPPer} (full sample) & $-9.62\pm 0.06$ & $0.46\pm0.03$ \\ \hline
    \texttt{eROMaPPer} (clean sample) & $-10.05\pm0.04$ & $0.66\pm0.02$ \\ \hline
    
    \end{tabular}
    \caption{Fit parameters for the $\bar y-\bar \tau$ scaling relations derived for each sample.}
    \label{tab:taufit}
\end{table}

\begin{table*}[]
\begin{center}

\hspace*{-1.5cm}
\begin{tabular}{|c|c|c|c||c|c|c|c|c|}
\cline{2 -9}
\multicolumn{1}{c|}{\multirow{2}{*}{}} & \multicolumn{3}{c||}{f150}  & \multicolumn{5}{c|}{ILC}\\
\hline
{} & $\bar{\tau}_{\rm tSZ}$ & $\sigma_{\rm sys.}$ & $\bar{\tau}_{\rm kSZ}$ [G25] & $\bar{\tau}_{\rm tSZ}$ (CIB-deproj.) & $\sigma_{\rm sys.}$  & $\bar{\tau}_{\rm tSZ}$ (d$\beta$-deproj.) & $\sigma_{\rm sys.}$  &$\bar{\tau}_{\rm kSZ}$ [G25] \\
Bin & $[10^{-5}]$ & $[10^{-5}]$ & $[10^{-5}]$ & $[10^{-5}]$ & $[10^{-5}]$ & $[10^{-5}]$ & $[10^{-5}]$ & $[10^{-5}]$ \\
\hline 
L36D  & $2.84 \pm 0.76$ & 0.13 & $3.6 \pm 1.2$ & $3.51 \pm 0.37$ & 0.13 & -- & -- & $2.7 \pm 1.1$ \\
L36D (clean) & $6.10 \pm 0.34$ & 0.10 &  & $5.25 \pm 0.26$ & 0.11 & $4.56 \pm 0.97$ & 0.12 &  \\ \hline
L48D  & $2.66 \pm 0.78$ & 0.14 & $4.3 \pm 1.1$ & $3.79 \pm 0.35$ & 0.13 & $3.60 \pm 1.07$ & 0.14 & $4.1 \pm 1.0$ \\
L48D (clean) & $6.36 \pm 0.33$ & 0.09 &  & $5.84 \pm 0.23$ & 0.10 & $6.57 \pm 0.65$ & 0.09 & \\\hline
L60D  & $3.25 \pm 0.68$ & 0.14 & $2.5 \pm 1.1$ & $4.76 \pm 0.27$ & 0.12 & $4.44 \pm 0.89$ & 0.13 & $4.4 \pm 1.0$ \\
L60D (clean) & $7.00 \pm 0.29$ & 0.08 & & $6.49 \pm 0.21$ & 0.09 & $7.21 \pm 0.60$ & 0.08 &  \\\hline
L79D  & $4.73 \pm 0.58$ & 0.12 & $3.4 \pm 1.8$ & $5.39 \pm 0.33$ & 0.11 & $2.18 \pm 2.09$ & 0.13 & $4.6 \pm 1.7$ \\
L79D (clean) & $7.81 \pm 0.36$ & 0.08 & & $6.92 \pm 0.28$ & 0.09 & $6.03 \pm 1.03$ & 0.10 &  \\
\hline
\end{tabular}
\caption{Optical depth measurements derived from the tSZ and pairwise kSZ measurements for the LRG disjoint binned samples. We present the tSZ-derived optical depths ($\bar{\tau}_{\text{tSZ}}$) with statistical errors for both the f150 and CIB- and d$\beta$-deprojected ILC maps. Systematic errors ($\sigma_{\text{sys}}$), which are subdominant to the statistical errors, are also provided. We note that the d$\beta$-deprojected ILC Compton-$y$ measurement for the full L36D bin is negative, thus we are unable to compute $\bar{\tau}$ for this sample.} \label{lrgtautable}
\end{center}
\end{table*}

For both the 150 GHz and ILC measurements of the full LRG sample, we obtain $\bar{\tau}$ values that agree within the error with the  kSZ $\bar{\tau}$. Removing the radio-contaminated sources from the sample leads to significantly higher $\bar{\tau}$, though for any given bin the f150, CIB-deprojected, and d$\beta$-deprojected values are consistent with each other. Additional work is needed to examine the impact of radio contamination on the kSZ measurements of this sample and the calibrated $\bar{y}-\bar{\tau}$ relationship.

 With the exception of the $\log_{10} (M_*/M_{\odot})_*10$ and $\log_{10} (M_*/M_{\odot})_*11$ bins, we find good agreement between the kSZ-derived $\bar{\tau}$ values and the tSZ-derived $\bar{\tau}$ measurements from the radio-clean BGS samples using either deprojected ILC map. The full sample tSZ optical depths are substantially lower, though consistent with each other for any given bin. All $\bar  \tau$ values for this sample are tabulated in Table \ref{bgstautable}. As expected, we also find good agreement between the tSZ-derived and kSZ-derived measurements for all bins of the \texttt{eROMaPPer} sample, which are tabulated in Table \ref{opttautable}. While we do not presently have a way to independently validate this result, further study of this sample with next-generation CMB data will provide an important consistency check.

\begin{figure*}
    \begin{center}
        \includegraphics[width=17.6cm]{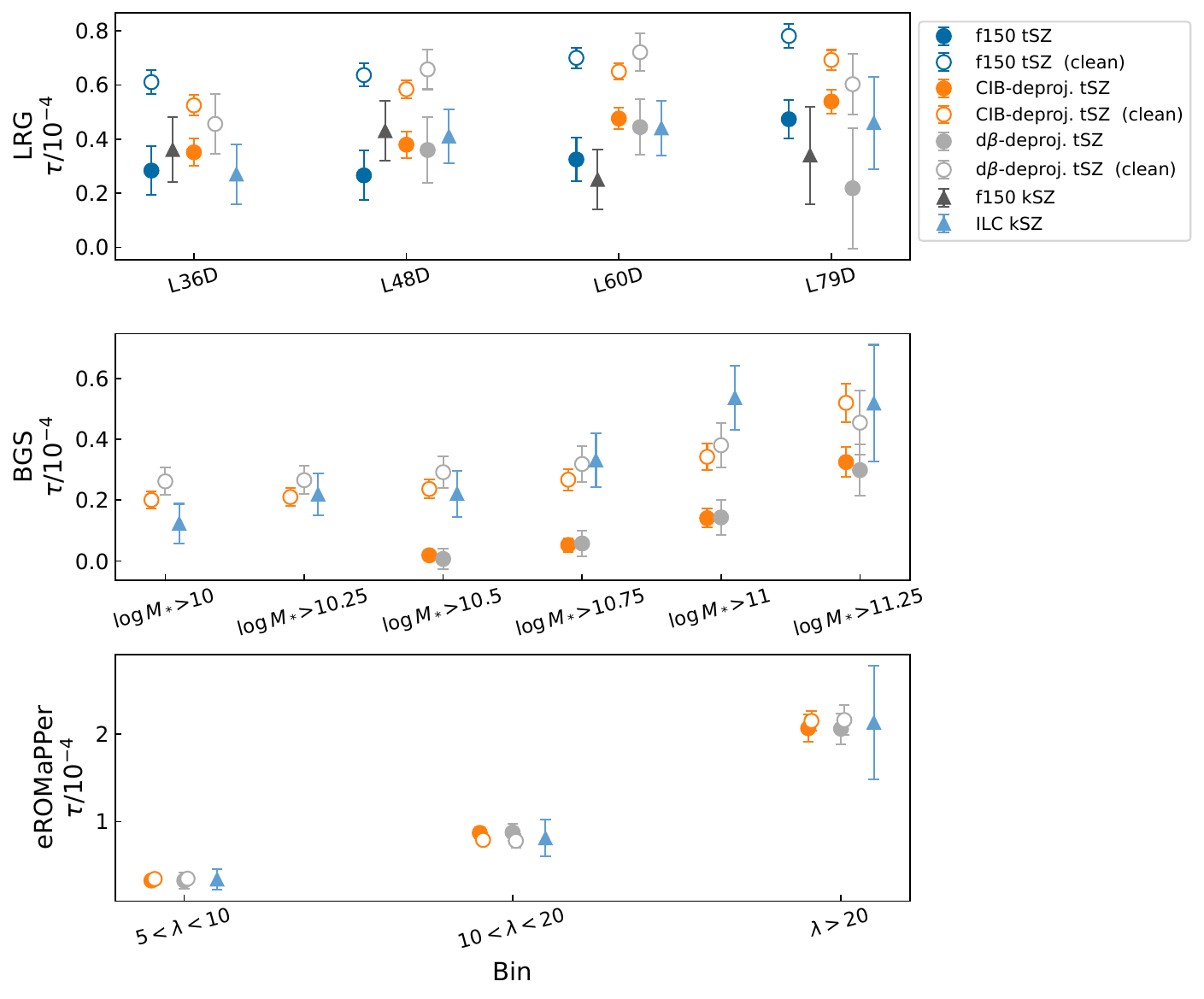}
        \caption{Optical depth comparisons for each sample derived from the tSZ measurements using the full [filled circles] and radio-cleaned [empty circles] samples and kSZ measurements [triangles]. The tSZ optical depths for the LRG and BGS samples are obtained using simulations-based scaling relations. For the LRG sample, we see agreement between the full-sample tSZ values and the kSZ-derived values. For the BGS sample, we see agreement between the radio-clean tSZ values and the kSZ-derived values. 
        We see remarkable consistency in the optical depths for the \texttt{eROMaPPer} sample, which are obtained using an empirically-derived $\bar y-\bar \tau$ scaling relation.}\label{fig:taucomps}
    \end{center}
\end{figure*}

\begin{table*}[]
\begin{center}

\hspace*{-1.5cm}
\begin{tabular}{|c|c|c||c|c||c|}

\cline{2 -6}
\multicolumn{1}{c|}{\multirow{2}{*}{}} & \multicolumn{5}{c|}{ILC} \\
\hline
{} & $\bar{\tau}_{\rm tSZ}$ (CIB-deproj.)& $\sigma_{\rm sys.}$ & $\bar{\tau}_{\rm tSZ}$ (d$\beta$-deproj.) & $\sigma_{\rm sys.}$ &$\bar{\tau}_{\rm kSZ}$ [H25] \\
Bin & $[10^{-5}]$ & $[10^{-5}]$ & $[10^{-5}]$ & $[10^{-5}]$ & $[10^{-5}]$ \\
\hline 

$M>10$  & -- & -- & -- & -- & $1.2 \pm 0.7$ \\
$M>10$ (clean) & $2.01 \pm 0.12$ & 0.15 & $2.62 \pm 0.26$ & 0.18 & \\\hline
$M>10.25$  & -- & -- & -- & -- & $2.2 \pm 0.7$ \\
$M>10.25$ (clean) & $2.10 \pm 0.13$ & 0.16 & $2.66 \pm 0.28$ & 0.18 &  \\\hline
$M>10.5$  & $0.19 \pm 0.14$ & 0.03 & $0.07 \pm 0.32$ & 0.01 & $2.2 \pm 0.8$ \\
$M>10.5$ (clean) & $2.37 \pm 0.14$ & 0.17 & $2.92 \pm 0.32$ & 0.19 &  \\\hline
$M>10.75$  & $0.53 \pm 0.16$ & 0.06 & $0.58 \pm 0.36$ & 0.07 & $3.3 \pm 0.9$ \\
$M>10.75$ (clean) & $2.67 \pm 0.17$ & 0.19 & $3.19 \pm 0.38$ & 0.21 &  \\\hline
$M>11$  & $1.41 \pm 0.19$ & 0.12 & $1.43 \pm 0.44$ & 0.12 & $5.4 \pm 1.1$ \\
$M>11$ (clean) & $3.42 \pm 0.22$ & 0.22 & $3.80 \pm 0.50$ & 0.24 &  \\\hline
$M>11.25$  & $3.25 \pm 0.27$ & 0.21 & $2.99 \pm 0.65$ & 0.20 & $5.2 \pm 1.9$ \\
$M>11.25$ (clean) & $5.19 \pm 0.32$ & 0.31 & $4.54 \pm 0.77$ & 0.27 &  \\ \hline

\end{tabular}
\caption{Optical depth measurements derived from the tSZ and pairwise kSZ measurements for the BGS binned samples. We present the tSZ-derived optical depths ($\bar{\tau}_{\text{tSZ}}$) for both the full and radio-clean samples with statistical errors derived from the CIB-deprojected ILC maps. Systematic errors ($\sigma_{\text{sys}}$) are also provided. Compton-y measurements for the two lowest-mass samples are negative when radio-contaminated sources are included, thus there is no corresponding $\bar \tau$ for these samples.}\label{bgstautable}
\end{center}
\end{table*}

The $\bar y-\bar\tau$ scaling relation parameters presented in Table \ref{tab:taufit} depend on the characteristics of the sample, such as mass range and redshift. We note the errors on the fit parameters include systematic uncertainties arising from the fit as well as statistical uncertainties from the measurements, but do not encompass the uncertainty arising from the tSZ systematics. In Table \ref{tab:taufit}, we see variation in both $\ln \tau_0$ and $m$ from sample to sample. For comparison, \cite{battaglia16} finds $\ln\tau_0$ of -6.47 with $m=0.49$ for a simulated sample of clusters of mass $M_{500}=10^{14}-10^{15}M_{\odot}$ at $z=0.5$.  These fit values are also sensitive to galaxy formation and feedback properties that are currently poorly constrained and vary widely in simulations, such as active galactic nuclei (AGN) feedback, radiative cooling, star formation, galactic winds, supernova feedback, and cosmic ray physics \cite{springel03, jubelgas08, battaglia10, battaglia16}. These fits therefore provide a key opportunity to derive insights into the properties and formation histories of these halo populations, and provide valuable checks on simulation models in the case of empirical relationships from measurements alone, which we leave to future study. 

\begin{table}[]
\begin{center}

\hspace*{-1.5cm}
\begin{tabular}{|c|c|c||c|}

\cline{2 -4}
\multicolumn{1}{c|}{\multirow{2}{*}{}} & \multicolumn{3}{c|}{ILC} \\
\hline
{} & $\bar{\tau}_{\rm tSZ}$ & $\sigma_{\rm sys.}$ & $\bar{\tau}_{\rm kSZ}$ [H26] \\
Bin & $[10^{-4}]$ & $[10^{-4}]$ & $[10^{-4}]$ \\
\hline 
$5<\lambda<10$  & $0.33 \pm 0.07$ & 0.02 & $0.34 \pm 0.12$ \\
$5<\lambda<10$ (clean) & $0.35 \pm 0.04$ & 0.01 &  \\ \hline
$10<\lambda<20$  & $0.88 \pm 0.04$ & 0.05 & $0.81 \pm 0.21$ \\
$10<\lambda<20$ (clean) & $0.78 \pm 0.05$ & 0.03 & \\\hline
$\lambda>20$  & $2.06 \pm 0.03$ & 0.14 & $2.13 \pm 0.65$ \\
$\lambda>20$ (clean) & $2.16 \pm 0.06$ & 0.11 &  \\ \hline

\end{tabular}
\caption{Optical depth measurements derived from the tSZ and pairwise kSZ measurements for the \texttt{eROMaPPer} binned samples. We present the tSZ-derived optical depths ($\bar{\tau}_{\text{tSZ}}$) with statistical errors for the d$\beta$-deprojected ILC maps. Systematic errors ($\sigma_{\text{sys}}$) are also provided.}\label{opttautable}
\end{center}
\end{table}

\section{Conclusion}\label{sec:conclusion}

In this work we have used several ACT DR6 maps to measure the stacked thermal Sunayev-Zel'dovich signal associated with halos from three tracer samples: Luminous red galaxies from the DESI DR1 LRG catalog, low-redshift galaxies from the DESI Bright Galaxy Survey, and \texttt{eROMaPPer} galaxy clusters selected using the \texttt{eROMaPPer} cluster finder. Using disk-ring aperture photometry, and accounting for various sources of contamination, we have measured stacked signals with signal-to-noise of over 38 for the LRG sample, 27 for the BGS sample, and 39 for the \texttt{eROMaPPer} sample using the ACT DR6+\textit{Planck} 90 GHz co-added map. Drawing on the kSZ pairwise-derived optical depth estimates in the companion papers \cite{gong26, hadzhiyska25, hsu26}, this  work presents the first joint comparison of thermal and kinematic SZ cluster optical depth estimates for these three galaxy  tracer samples.

This work presents a substantial exploration of systematic considerations when disentangling secondary anisotropies in multi-frequency stacked tSZ measurements, a key step forward as the field enters a systematics-dominated regime. We have explored different methods for correcting our aperture photometry measurements to account for contamination from dust and the cosmic infrared background. We have generated radial profiles using numerous component-separated ILC maps \cite{coulton24} and find the measured signal at large scales ($R\gtrsim 4'$) can vary significantly depending on the method of CIB deprojection and the assumed CIB SED parameters. In comparing radial profiles generated from each of the ILC maps (Fig.~\ref{fig:ilc_deproj_profiles}), we find the CIB-deprojected ILC map offers a cleaner measurement on scales near and below the aperture size, where the dust emission dominates in the fiducial maps, but gives a substantially decreased signal at larger radii. In contrast, using the d$\beta$-deprojected ILC maps results in a significantly decreased signal at smaller scales compared to that obtained with the CIB-deprojected maps, contrary to what would be expected for a decontaminated measurement. Additionally, the d$\beta$-deprojected maps are much noisier, resulting in a decrease in SNR of $68\%$ for the radio-clean LRG sample, $41\%$ for the BGS sample, and $31\%$ for the \texttt{eROMaPPer} sample relative to the CIB-deprojected measurements for the default aperture sizes of $2.1',2.7'$ and $2.4'$ respectively.

We find the aperture photometry measurement for the single-frequency ACT DR6+\textit{Planck} coadded maps centered for each of the galaxy samples are contaminated on small scales by dust emission (Figure \ref{fig:f220_ims_and_profiles}). This contamination completely overwhelms the tSZ signal in the LRG and BGS samples, but is subdominant in the \texttt{eROMaPPer} sample (Figure \ref{fig:raw_profiles}). The \texttt{eROMaPPer} sample consists of higher mass clusters, resulting in a stronger overall tSZ signal despite the dust emission. Using a ring-ring aperture photometry filter to mask this central contamination \cite{liu25}, we find good agreement in measured signal in 90 and 150 GHz for all samples (Figure \ref{fig:rr_profiles}), indicating little large-scale CIB contamination is present. By rescaling the signal measured at 220 GHz, we are able to correct the full sample LRG and BGS measurements for the centralized dust contamination in the 150 GHz measurements following the method presented in \cite{liu25} (Figure \ref{fig:deproj_sf_profiles}).

We have also explored additional contamination from nearby radio sources, which, like dust and CIB contamination, biases our tSZ measurements low. By cross-matching our catalogs with the NRAO VLA Sky Survey catalog of radio sources, we find $\sim10$\% of our target sources are within $1'$ of a known radio source, and between 20 and 30\% of the sources in each sample have a radio source within the aperture photometry disk radius. We find this contamination significantly impacts our results (Figure \ref{fig:radio_cleaned_stacks}), with 90 GHz aperture photometry measurements of the radio-clean sources shifted upwards of $30\sigma$, $20\sigma$, and $7\sigma$ from the baseline measurements for the LRG, BGS, and \texttt{eROMaPPer} samples, respectively. Radio contamination is expected to have little impact on the pairwise kSZ measurements, but exploration of this is left to future study.

We have used a forward-modeling approach to develop average electron pressure profiles for each of our samples using the halo mass and redshift distributions. As described in Section \ref{sec:beamcorr}, we convolve these model profiles with the effective beams to develop fractional corrections to account for the different beam size of each of the ACT map products relative to 150 GHz. We have attempted to extract fits of the one-halo and two-halo contribution from these profiles, but due to the aforementioned variation in Compton-$y$ at large scales with the different map products, we are unable to obtain reliable estimates for the two-halo effect, which biases our measurements high. Well-constrained measurements at large radii, which may be obtained in future analyses with improved handling of contaminants, are necessary to break the degeneracy in the one- and two-halo terms of the forward model. Previous analyses \cite{vavagiakis21} have quoted two-halo contributions of $2-10\%$. Even assuming a worst case ($10\%$) impact, the takeaways from our analysis would not change. 

For each sample, we have computed $\bar y-\bar\tau$ scaling relations using either a simulated sample or the real data to determine the best-fit $\ln\tau_0$ and $m$ values for Eq.~\ref{eq:tau}, which we then use to convert our Compton-$y$ measurements to optical depths (Fig.~\ref{fig:taucomps}). For the LRG sample, we use a simulations-derived scaling relation formulated from a simulated map and sample which is calibrated to the characteristics of the full LRG sample. Using the 150 GHz and deprojected ILC maps, we obtain optical depths for the full LRG samples that agree within $1\sigma$ with the kSZ-derived values with the exception of the L36D bin using the d$\beta$-deprojected ILC map, which gives a negative $\bar{y_{\mathrm{AP}}}$. The optical depths derived from the radio-clean measurements are significantly higher, though the scaling relation is not well-matched to the data. With additional study into the effects of radio contamination on the pairwise kSZ signal and improved handling of tSZ contaminants, it may be possible to obtain an empirically-derived scaling relationship for this sample.

For the BGS sample, we use a $\bar y-\bar\tau$ scaling relation derived from SIMBA measurements of a BGS-like sample.  For nearly all of the bins, we find good agreement between the radio-clean tSZ-derived optical depths and the kSZ-derived optical depths from the full samples using the deprojected ILC maps, but obtain systematically lower $\bar{\tau}$ from the full sample tSZ measurements. For the \texttt{eROMaPPer} sample, we present the first scaling relation to be derived directly from the SZ measurements of \texttt{eROMaPPer} galaxies.
We fit the $\bar y-\bar\tau$ scaling relation directly to the tSZ and kSZ measurements. For all bins, both the full sample and radio-clean tSZ optical depths align with the kSZ-derived values using the deprojected ILC maps. 

We find the fit parameters $\ln\tau_0$ and $m$ vary from sample to sample (Table \ref{tab:taufit}), and differ from previous simulations-based values \cite{battaglia16}, which showed modest differences in fit parameters for different redshifts, aperture sizes, and feedback models. At present, it is difficult to discern whether the differences we see in the best-fit parameters are a result of differing sample properties (e.g. mass, redshift) or residual systematics. Future analyses with better constraints on the tSZ systematics discussed in this paper will offer additional insight into the universality of the $\bar y-\bar\tau$ scaling relationship. As these fit parameters are related to the intrinsic tSZ signal of the sample, we expect them to remain consistent across CMB surveys. Future analysis of the samples presented here with new CMB data from the Simons Observatory (SO) \cite{so19} will help validate this methodology.

We have shown in this work that the combination of high-resolution maps from ACT and expansive source catalogs from DESI in tandem with complementary VLA catalogs allows us to extract measurements with relatively low noise. However, a better understanding of foregrounds, especially dust and CIB, is necessary to untangle the true tSZ signal from the measurements. Previous analyses \cite{vavagiakis21} have used higher frequency data from \textit{Herschel} to model the dust spectral energy distribution of the target galaxies and infer the dust level at 90 and 150 GHz. In this analysis, we found that less than $5\%$ of our sources lie in the \textit{Herschel} footprint and thus elected to pursue the correction method of \cite{liu25} instead. In future studies, cross-correlation with submillimeter data over larger areas of the sky will be possible with data from the CCAT Observatory's Fred Young Submillimeter Telescope \cite{ccat23}, which is soon to be commissioned atop Cerro Chajnantor in the Atacama Desert of Chile. CCAT's Prime-Cam instrument will survey the sky in 5 frequency bands from 280-850 GHz, allowing for unique windows into thermal dust contamination. The survey footprint of CCAT will match that of SO, the newest CMB observatory which has recently begun observations at the former ACT site. SO's Large Aperture Telescope \cite{so25} will provide wide-survey CMB maps with exquisite angular resolution and depth, allowing for higher significance measurements. 

Combining the expanding galaxy catalogs expected from the full DESI survey and its proposed extension with data from SO and CCAT will enable improved SZ measurements through powerful new cross-correlation analyses. This will enable more precise tSZ and kSZ measurements over a wider range of samples, with a better understanding of foregrounds and systematics. This will allow us to precisely probe the baryonic content of galaxies and clusters and improve our understanding of the relationship between $\bar y$ and $\bar \tau$. As we have shown in this work, with careful consideration of foregrounds, it is indeed possible to extract an empirical $y-\tau$ scaling relation directly from the SZ measurements. With these scaling relations, we will be able to use the tSZ-derived optical depth to break the degeneracy between optical depth and velocity in the kSZ measurements. With the degeneracy broken, one may then extract velocity information from the kSZ measurements and provide constraints on cosmology. 

\section{Acknowledgements}
The authors would like to thank R. H. Liu and W. Coulton for helpful discussions that aided the analysis presented in this paper. 

JEM and EMV are supported by the Department of Energy grant DE-SC0010007. The work of YG and RB is supported by NSF grant AST-2206088, NASA grant 22-ROMAN11-0011 and NASA grant 12-EUCLID12-0004. YH and DG were in part funded by the Deutsche Forschungsgemeinschaft (DFG, German Research Foundation) under Germany's Excellence Strategy – EXC-2094/2 – 390783311. E. Bulbul acknowledges financial support from the European Research Council (ERC) Consolidator Grant under the European Union’s Horizon 2020 research and innovation program (grant agreement CoG DarkQuest No 101002585). TM acknowledges support from the Agencia Estatal de Investigaci\'on (AEI) and the Ministerio de Ciencia, Innovaci\'on y Universidades (MICIU) Grant ATRAE2024-154740 funded by MICIU/AEI//10.13039/501100011033. TM is also partly supported by the Spanish program Unidad de Excelencia María de Maeztu CEX2020-001058-M, financed by MCIN/AEI//10.13039/501100011033, and by the MaX-CSIC Excellence Award MaX4-SOMMA-ICE.

Support for ACT was through the U.S.~National Science Foundation through awards AST-0408698, AST-0965625, and AST-1440226 for the ACT project, as well as awards PHY-0355328, PHY-0855887 and PHY-1214379. Funding was also provided by Princeton University, the University of Pennsylvania, and a Canada Foundation for Innovation (CFI) award to UBC. ACT operated in the Parque Astron\'omico Atacama in northern Chile under the auspices of the Agencia Nacional de Investigaci\'on y Desarrollo (ANID). The development of multichroic detectors and lenses was supported by NASA grants NNX13AE56G and NNX14AB58G. Detector research at NIST was supported by the NIST Innovations in Measurement Science program. Computing for ACT was performed using the Princeton Research Computing resources at Princeton University, the National Energy Research Scientific Computing Center (NERSC), and the Niagara supercomputer at the SciNet HPC Consortium. SciNet is funded by the CFI under the auspices of Compute Canada, the Government of Ontario, the Ontario Research Fund–Research Excellence, and the University of Toronto. We thank the Republic of Chile for hosting ACT in the northern Atacama, and the local indigenous Licanantay communities whom we follow in observing and learning from the night sky.

This research used resources of the National Energy Research Scientific Computing Center (NERSC), a U.S. Department of Energy Office of Science User Facility located at Lawrence Berkeley National Laboratory, operated under Contract No. DE-AC02-05CH11231 using NERSC award HEP-ERCAPmp107

This material is based upon work supported by the U.S. Department of Energy (DOE), Office of Science, Office of High-Energy Physics, under Contract No. DE–AC02–05CH11231, and by the National Energy Research Scientific Computing Center, a DOE Office of Science User Facility under the same contract. Additional support for DESI was provided by the U.S. National Science Foundation (NSF), Division of Astronomical Sciences under Contract No. AST-0950945 to the NSF’s National Optical-Infrared Astronomy Research Laboratory; the Science and Technology Facilities Council of the United Kingdom; the Gordon and Betty Moore Foundation; the Heising-Simons Foundation; the French Alternative Energies and Atomic Energy Commission (CEA); the Secretariat of Science, Humanities, Technology and Innovation (SECIHTI) of Mexico; the Ministry of Science, Innovation and Universities of Spain (MICIU/AEI/10.13039/501100011033), and by the DESI Member Institutions: \url{https://www.desi.lbl.gov/collaborating-institutions}. 

The DESI Legacy Imaging Surveys consist of three individual and complementary projects: the Dark Energy Camera Legacy Survey (DECaLS), the Beijing-Arizona Sky Survey (BASS), and the Mayall z-band Legacy Survey (MzLS). DECaLS, BASS and MzLS together include data obtained, respectively, at the Blanco telescope, Cerro Tololo Inter-American Observatory, NSF’s NOIRLab; the Bok telescope, Steward Observatory, University of Arizona; and the Mayall telescope, Kitt Peak National Observatory, NOIRLab. NOIRLab is operated by the Association of Universities for Research in Astronomy (AURA) under a cooperative agreement with the National Science Foundation. Pipeline processing and analyses of the data were supported by NOIRLab and the Lawrence Berkeley National Laboratory. Legacy Surveys also uses data products from the Near-Earth Object Wide-field Infrared Survey Explorer (NEOWISE), a project of the Jet Propulsion Laboratory/California Institute of Technology, funded by the National Aeronautics and Space Administration. Legacy Surveys was supported by: the Director, Office of Science, Office of High Energy Physics of the U.S. Department of Energy; the National Energy Research Scientific Computing Center, a DOE Office of Science User Facility; the U.S. National Science Foundation, Division of Astronomical Sciences; the National Astronomical Observatories of China, the Chinese Academy of Sciences and the Chinese National Natural Science Foundation. LBNL is managed by the Regents of the University of California under contract to the U.S. Department of Energy. The complete acknowledgments can be found at \url{https://www.legacysurvey.org/}.

Any opinions, findings, and conclusions or recommendations expressed in this material are those of the author(s) and do not necessarily reflect the views of the U. S. National Science Foundation, the U. S. Department of Energy, or any of the listed funding agencies.

The authors are honored to be permitted to conduct scientific research on I'oligam Du'ag (Kitt Peak), a mountain with particular significance to the Tohono O’odham Nation.

We also acknowledge the use of the National Energy Research Scientific Computing Center (NERSC) computational resources and platform, through an allocation provided to the DESI collaboration, in this work.

Some of the computing for this project was performed on the Duke Compute Cluster. We thank Duke University Research Computing for providing resources and support.

\section*{DATA AVAILABILITY}
The data corresponding to the figures in this paper are available at https://doi.org/10.5281/zenodo.21795818.

\appendix
\section{Null and gradient test}\label{sec:nullgrad}
A noticeable gradient exists in the stacked single-frequency submaps for all samples that likely arises from residual point source contamination. Although we apply a point source mask to our catalogs, point sources fainter than our cutoff can still be seen in the stacked single-frequency maps. Since our average aperture photometry measurements are computed from the individual source measurements and not the stacked images, we don't expect this gradient to bias our signal. In Figure \ref{fig:rotated_im} we show that applying a random rotation to each submap prior to stacking results in a much smoother background, though the average pixel values differ $<1\%$ from the non-rotated image. Furthermore, we find no significant difference between radial profiles computed from the original and rotated stacked images. 

A null test was conducted by stacking $2\times10^6$ randomly distributed coordinates within the ACT footprint on the 150 GHz map. This sample was processed using the same pipeline and jackknife estimation procedure used in the analysis of the samples. This yielded a Compton-$y$ value consistent with zero within 1$\sigma$, indicating no statistically significant signal introduced by the map or analysis methodology.

\begin{figure*}
\begin{center}
\includegraphics[width=17.6cm]{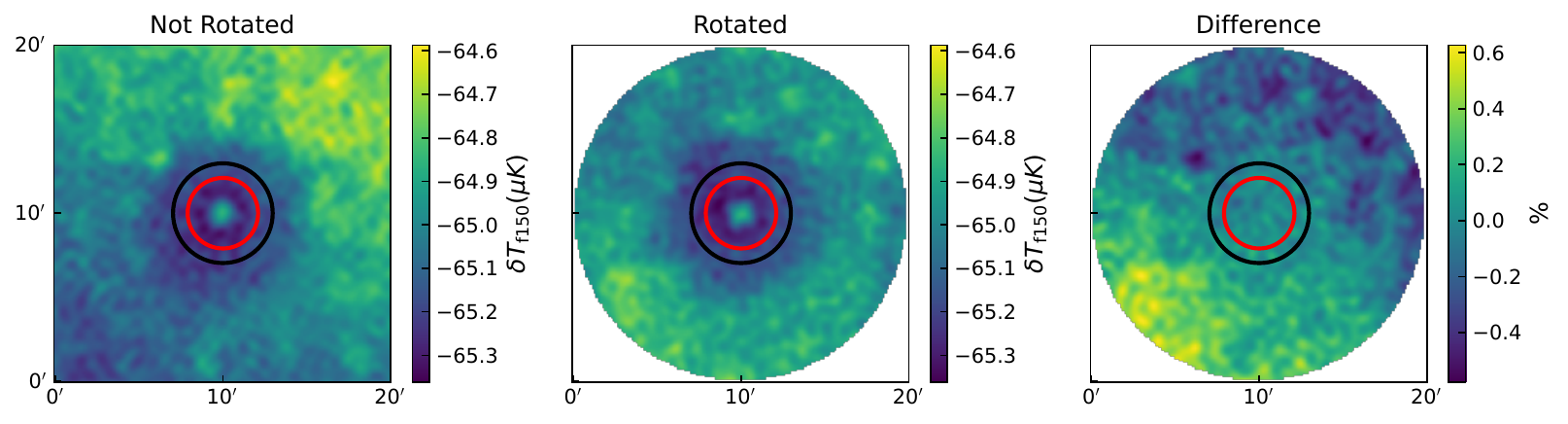}
\caption{The original [left] stacked f150 image for the full LRG (L36) sample is shown against the sample stacked after each submap undergoes a random rotation [center] and the difference between the two stacked maps [right]. We restrict the stack to areas with uniform pixel overlap across all submaps, thus the resulting image is a circle.}
\label{fig:rotated_im}
\end{center}
\end{figure*}

\section{Binned profiles}\label{sec:binnedprofs}
In Figure \ref{fig:raw_profiles_binned} we show the binned radial profiles for each full sample as computed from the f090 and f150 maps. For all LRG and BGS binned samples, we see a substantial suppression of signal at 150 GHz relative to 90 GHz at small scales, indicating significant contamination. At larger radii, the profiles are more consistent, indicating large-scale contamination from the CIB is likely not a dominant foreground.  

Figure \ref{fig:ilc_deproj_profiles_binned} shows the ILC radial profiles for each of the binned samples. For all samples, we see the CIB-deprojected ILC map results in an increase in signal (relative to the fiducial profile) on small ($R<2'$) scales and a decrease in signal on large ($R>3-4'$) scales. The d$\beta$-deprojected maps give more inconsistent results. In some LRG bins (e.g. L48D, L60D), the d$\beta$-deprojected profiles are largely consistent with the fiducial profiles on small scales. In other bins (e.g. all cumulative bins, L36D, L79D), the d$\beta$-deprojected profiles are somewhat aligned with the CIB-deprojected profiles on the very smallest scales, but then show a dip in signal around the aperture size before falling more in line with the fiducial profiles. For the BGS samples, the d$\beta$-deprojected profiles are most closely aligned with the fiducial profiles, but on the smallest scales they are significantly decreased relative to the fiducial profiles, particularly for the higher mass bins. We see very minor differences in the three profiles for all of the \texttt{eROMaPPer} samples, which indicates any dust or CIB contamination present may be subdominant to the tSZ signal.

\section{ILC profiles}\label{sec:compare_ilc}
In Figures \ref{fig:yprofiles_compare_beta} and \ref{fig:ilc_deproj_profile_temp_comparison} we show the radial profiles computed from the deprojected ILC maps using different assumed values of spectral index $\beta$ and CIB temperature, respectively.  In Figure \ref{fig:yprofiles_compare_beta} we  examine the stability of the deprojected radial profiles to changes in assumed spectral index $\beta$ (using $T_{\mathrm{CIB}}=10.7$K). For the CIB-deprojected map, we see that choosing a smaller value of $\beta$ consistently leads to larger shifts in signal relative to the fiducial measurement at smaller scales. We do not find the profiles for our samples to be particularly sensitive to the choice of $\beta$; all profiles for a given sample and deprojection type are consistent within the error at all radii regardless of the assumed $\beta$. 

In Figure \ref{fig:ilc_deproj_profile_temp_comparison} we show the radial profiles for each sample computed from the deprojected ILC maps with both the fiducial choice of $T_{\mathrm{CIB}}=10.7$ K and a higher $T_{\mathrm{CIB}}=24$ K (with $\beta=1.6$). We see for the CIB-deprojected maps, the choice of $T_{\mathrm{CIB}}$ has only a modest impact on the measured signal at scales below the aperture size while at larger scales, using a larger value of $T_{\mathrm{CIB}}$ in the CIB-deprojected map leads to significantly higher values of Compton-$y$ that are more consistent with the fiducial profile. For all samples, we find the choice of $T_{\mathrm{CIB}}$ does not significantly impact the signal measured from the d$\beta$-deprojected map at any radius. 

\section{Compton-$y$ aperture photometry values}\label{sec:yaps}

In Table \ref{tab:ytable} we present the final corrected Compton-$y$ aperture photometry values ($\bar{y_{\mathrm{AP}}}$ for each binned sample and map. We provide values for both the complete samples as well as the radio-cleaned samples as described in Section \ref{sec:radio}. Select binned measurements from Table \ref{tab:ytable} are plotted in the main text in Figure \ref{fig:yaps}.

\begin{figure*}[ht!]
\begin{center}
\hspace*{-0.2cm}
\includegraphics[width=14.5cm]{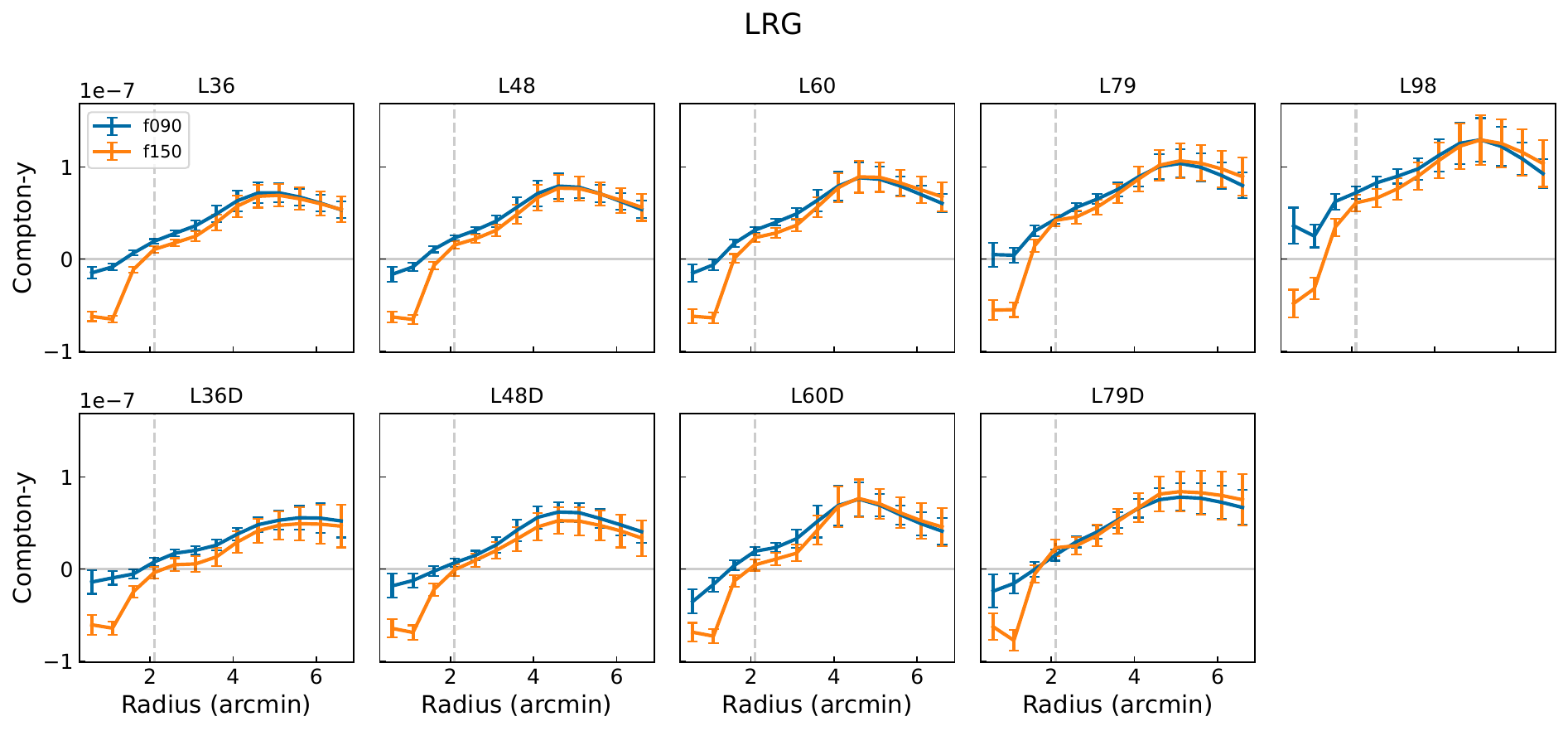}
\includegraphics[width=9.5cm]{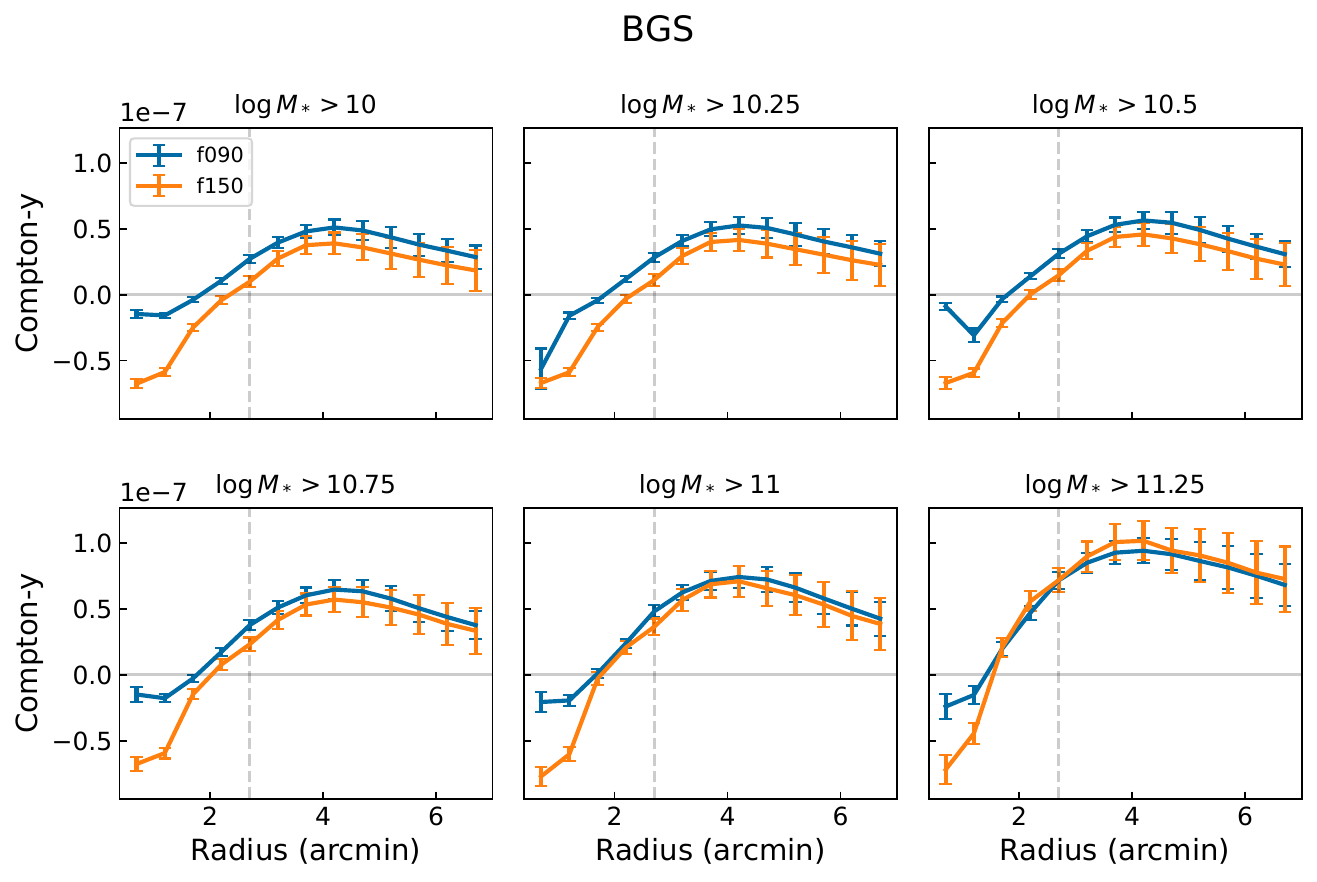}
\includegraphics[width=9.5cm]{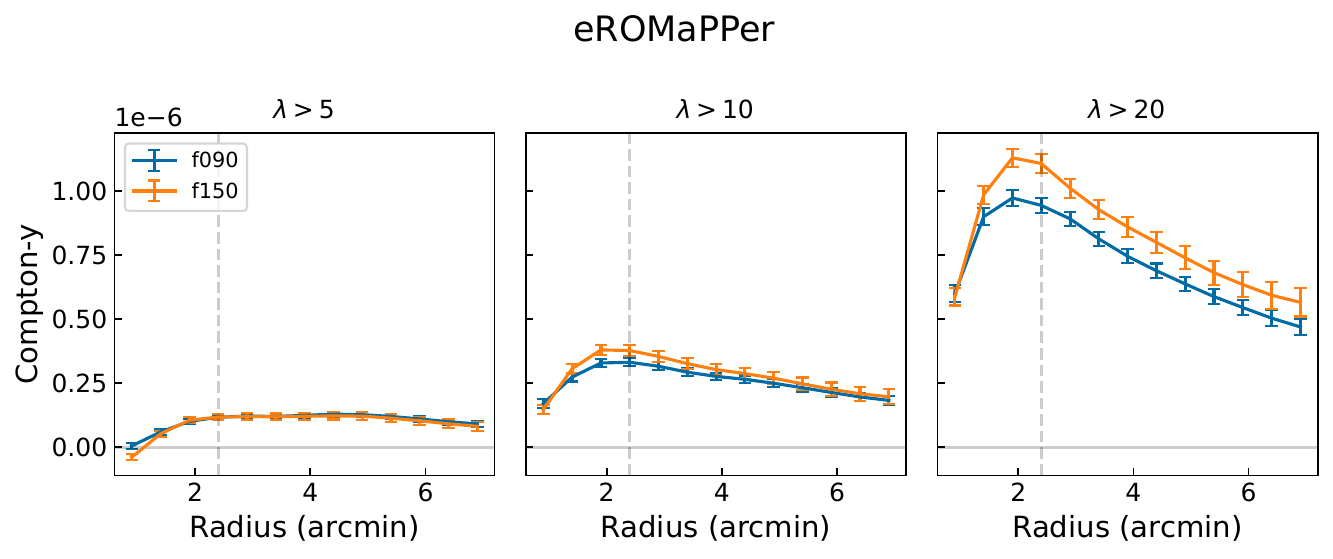}
\caption{An extended version of Figure \ref{fig:raw_profiles}, showing the raw profiles for the binned samples calculated using the f090 [blue] and f150 [orange] maps, in units of Compton-$y$. 90 GHz profiles have been corrected to account for the larger beam sizes relative to 150 GHz as described in Section \ref{sec:beamcorr}. \textit{Rows 1-2:} LRG cumulative [top row] and disjoint [bottom row] profiles. \textit{Rows 3-4:} BGS profiles. \textit{Row 5:} \texttt{eROMaPPer} profiles.}
\label{fig:raw_profiles_binned}
\end{center}
\end{figure*}

\begin{figure*}[ht!]
\begin{center}
\hspace*{-0.2cm}
\includegraphics[width=14.6cm]{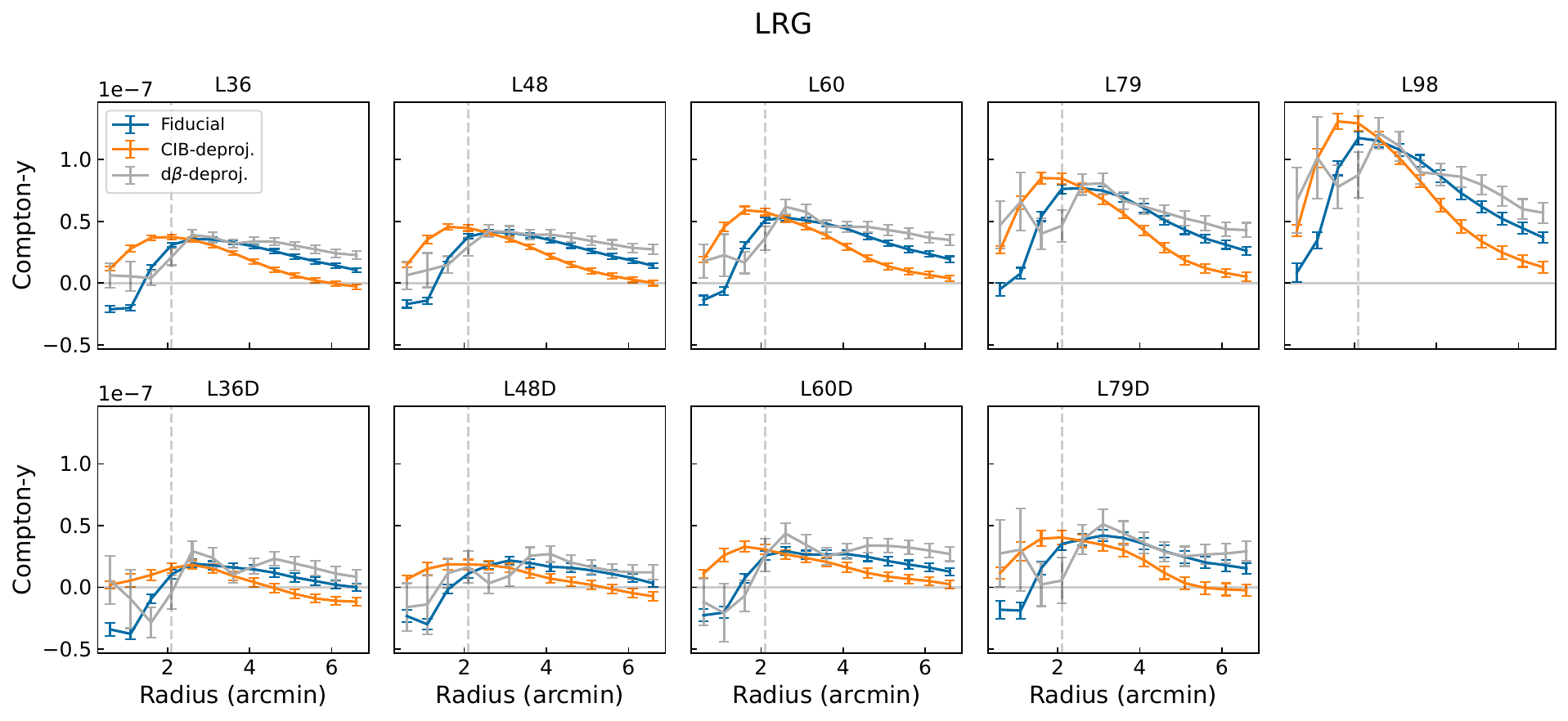}
\includegraphics[width=9.6cm]{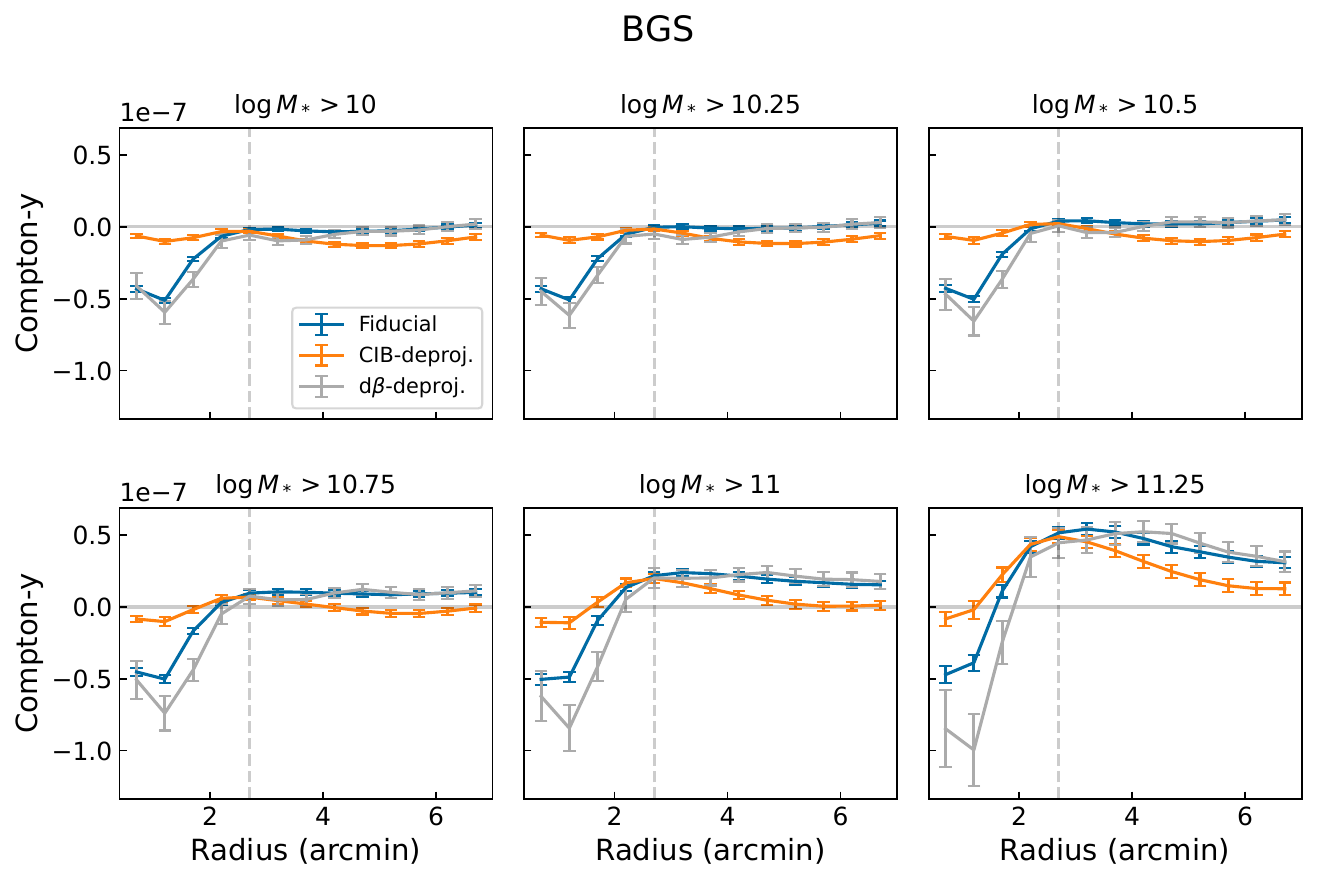}
\includegraphics[width=9.6cm]{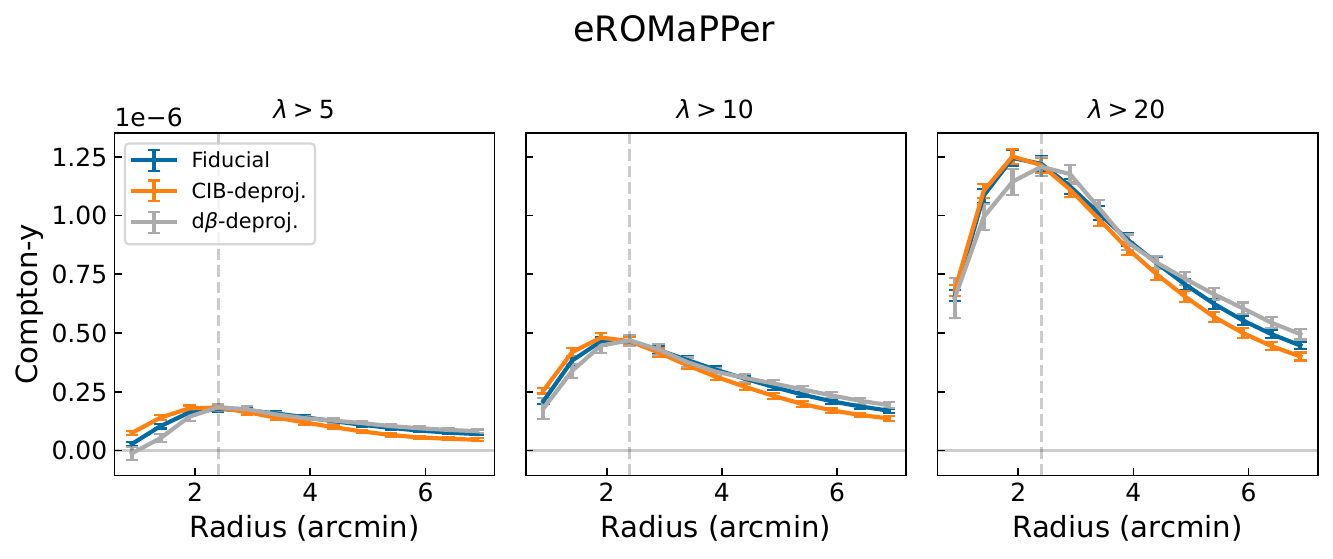}
\caption{An expanded verision of Figure \ref{fig:ilc_deproj_profiles} showing profiles for the binned samples computed from the ILC maps. For the CIB and d$\beta$-deprojected maps, we plot the profiles computed from the $\beta = 1.6$, $T_{\mathrm{CIB}}=10.7$ K maps. Lines are used to guide the eye and do not indicate a fit to the data. The aperture size is indicated by the gray dashed vertical line. \textit{Rows 1-2:} LRG cumulative [top row] and disjoint [bottom row] profiles. \textit{Rows 3-4:} BGS profiles. \textit{Row 5: } Profiles for the \texttt{eROMaPPer} samples.}
\label{fig:ilc_deproj_profiles_binned}
\end{center}
\end{figure*}

\begin{figure*}
\begin{center}
\includegraphics[width=17.6cm]{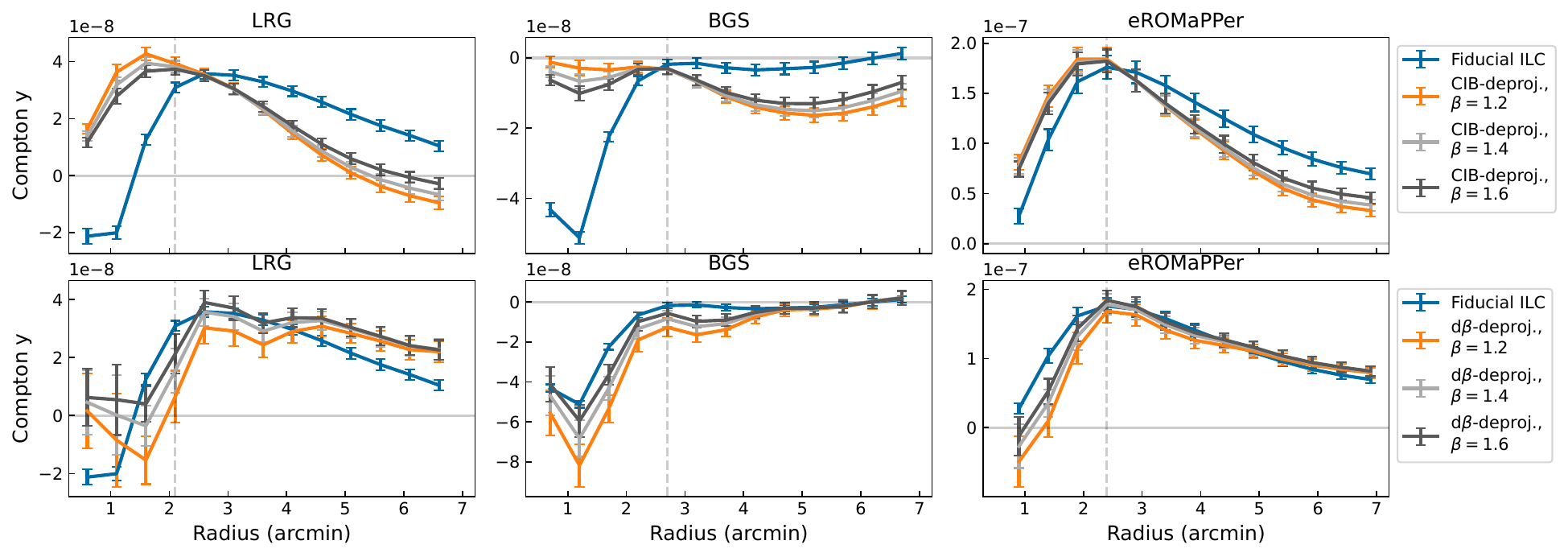}
\caption{Evaluating the impact of assumed $\beta$ on the CIB-deprojected [top row] and d$\beta$-deprojected [bottom row] radial profiles for the LRG [left], BGS [middle], and \texttt{eROMaPPer} [right] samples (with $T_{\mathrm{CIB}} = 10.7$K).} 
\label{fig:yprofiles_compare_beta}
\end{center}
\end{figure*}

\begin{figure*}[ht!]
\begin{center}
\hspace*{-0.2cm}
\includegraphics[width=17.6cm]{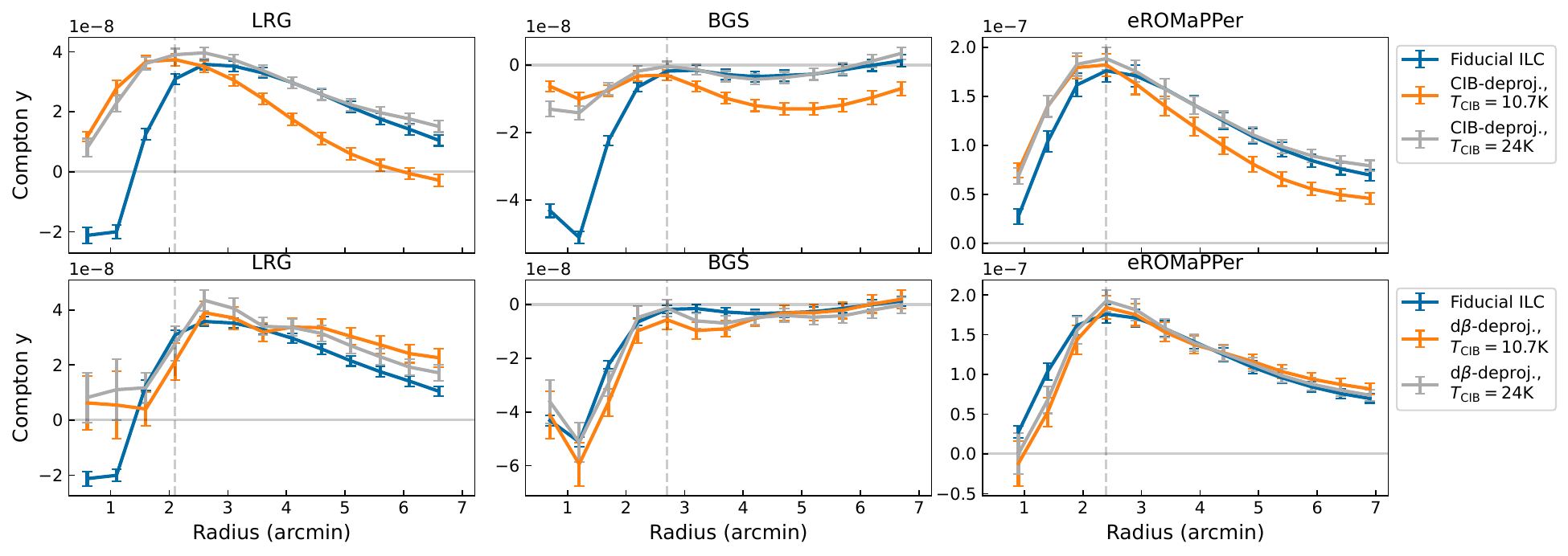}
\caption{Evaluating the impact of assumed CIB temperature $T_{\mathrm{CIB}}$ on the CIB-deprojected and d$\beta$-deprojected radial profiles for the three samples (with $\beta=1.6$). We see a significant difference in CIB-deprojected signal at large scales, with higher $T_{\mathrm{CIB}}$ leading to increased Compton-$y$.}
\label{fig:ilc_deproj_profile_temp_comparison}
\end{center}
\end{figure*}


\begin{table*}
\centering
\setlength\extrarowheight{1.5pt}
\begin{tabular}{|c|c|c||c|c||c|c||c|c||c|c|}
\hline
Bin
& \multicolumn{2}{c||}{f090}
& \multicolumn{2}{c||}{f150}
& \multicolumn{2}{c||}{ILC (fiducial)}
& \multicolumn{2}{c||}{ILC (CIB-deproj.)}
& \multicolumn{2}{c|}{ILC (d$\beta$-deproj.)} \\
\cline{2-11}
& $\bar{y}_{\rm AP}/10^{-7}$ & SNR
& $\bar{y}_{\rm AP}/10^{-7}$ & SNR
& $\bar{y}_{\rm AP}/10^{-7}$ & SNR
& $\bar{y}_{\rm AP}/10^{-7}$ & SNR
& $\bar{y}_{\rm AP}/10^{-7}$ & SNR \\
\hline

\multicolumn{11}{|c|}{\textbf{LRG}}\\
\hline
L36   & $0.19 \pm 0.02$  & 8.9  & $0.21 \pm 0.03$  & 6.5  & 0.32 $\pm$ 0.02  & 17.5  & 0.39 $\pm$ 0.02  & 18.3  & 0.22 $\pm$ 0.07  & 3.1 \\
L36  (clean)  & $0.96 \pm 0.03$  & 38.4  & $0.73 \pm 0.04$  & 19.5  & 0.62 $\pm$ 0.02  & 30.2  & 0.68 $\pm$ 0.02  & 28.5  & 0.70 $\pm$ 0.08  & 9.1 \\ \hline
L48   & $0.23 \pm 0.03$  & 9.2  & $0.25 \pm 0.04$  & 6.7  & 0.39 $\pm$ 0.02  & 18.6  & 0.47 $\pm$ 0.02  & 18.8  & 0.31 $\pm$ 0.08  & 3.9 \\
L48  (clean)  & $1.03 \pm 0.03$  & 36.3  & $0.79 \pm 0.04$  & 18.7  & 0.70 $\pm$ 0.02  & 29.4  & 0.77 $\pm$ 0.03  & 27.4  & 0.83 $\pm$ 0.09  & 9.6 \\\hline
L60   & $0.32 \pm 0.03$  & 10.1  & $0.33 \pm 0.05$  & 7.2  & 0.53 $\pm$ 0.03  & 20.3  & 0.60 $\pm$ 0.03  & 19.9  & 0.38 $\pm$ 0.10  & 3.9 \\
L60  (clean)  & $1.16 \pm 0.03$  & 34.5  & $0.90 \pm 0.05$  & 17.6  & 0.85 $\pm$ 0.03  & 28.5  & 0.91 $\pm$ 0.03  & 26.7  & 0.93 $\pm$ 0.11  & 8.6 \\\hline
L79   & $0.43 \pm 0.05$  & 9.6  & $0.54 \pm 0.06$  & 8.3  & 0.80 $\pm$ 0.04  & 20.9  & 0.89 $\pm$ 0.04  & 20.4  & 0.48 $\pm$ 0.14  & 3.6 \\
L79  (clean)  & $1.32 \pm 0.05$  & 29.1  & $1.08 \pm 0.07$  & 15.4  & 1.11 $\pm$ 0.04  & 26.3  & 1.19 $\pm$ 0.05  & 24.4  & 1.06 $\pm$ 0.16  & 6.8 \\\hline
L98   & $0.72 \pm 0.07$  & 10.8  & $0.79 \pm 0.10$  & 8.2  & 1.23 $\pm$ 0.05  & 22.6  & 1.35 $\pm$ 0.06  & 21.8  & 0.92 $\pm$ 0.19  & 4.8 \\
L98  (clean)  & $1.68 \pm 0.06$  & 26.2  & $1.25 \pm 0.10$  & 12.4  & 1.53 $\pm$ 0.06  & 24.8  & 1.67 $\pm$ 0.07  & 24.1  & 1.59 $\pm$ 0.22  & 7.3 \\\hline
L36D   & $0.08 \pm 0.04$  & 1.9  & $0.10 \pm 0.07$  & 1.4  & 0.11 $\pm$ 0.04  & 3.1  & 0.16 $\pm$ 0.04  & 4.0  & -0.04 $\pm$ 0.14  & -0.3 \\
L36D  (clean)  & $0.73 \pm 0.04$  & 16.8  & $0.53 \pm 0.07$  & 7.8  & 0.37 $\pm$ 0.04  & 9.5  & 0.40 $\pm$ 0.05  & 8.7  & 0.29 $\pm$ 0.16  & 1.9 \\\hline
L48D   & $0.07 \pm 0.04$  & 1.6  & $0.08 \pm 0.06$  & 1.3  & 0.12 $\pm$ 0.03  & 3.3  & 0.19 $\pm$ 0.04  & 4.6  & 0.17 $\pm$ 0.13  & 1.3 \\
L48D  (clean)  & $0.81 \pm 0.04$  & 18.4  & $0.58 \pm 0.07$  & 8.4  & 0.43 $\pm$ 0.04  & 11.0  & 0.50 $\pm$ 0.05  & 11.1  & 0.65 $\pm$ 0.15  & 4.3 \\\hline
L60D   & $0.20 \pm 0.04$  & 4.5  & $0.13 \pm 0.07$  & 1.9  & 0.27 $\pm$ 0.04  & 7.4  & 0.32 $\pm$ 0.04  & 7.7  & 0.27 $\pm$ 0.14  & 2.0 \\
L60D  (clean)  & $0.99 \pm 0.05$  & 21.1  & $0.72 \pm 0.07$  & 10.6  & 0.59 $\pm$ 0.04  & 14.8  & 0.64 $\pm$ 0.05  & 13.9  & 0.80 $\pm$ 0.16  & 5.1 \\\hline
L79D   & $0.15 \pm 0.06$  & 2.5  & $0.30 \pm 0.09$  & 3.4  & 0.37 $\pm$ 0.05  & 7.1  & 0.42 $\pm$ 0.06  & 7.1  & 0.06 $\pm$ 0.19  & 0.3 \\
L79D  (clean)  & $0.98 \pm 0.06$  & 16.0  & $0.92 \pm 0.10$  & 9.5  & 0.71 $\pm$ 0.06  & 12.4  & 0.74 $\pm$ 0.07  & 10.8  & 0.54 $\pm$ 0.23  & 2.4 \\

\hline

\multicolumn{11}{|c|}{\textbf{BGS}}\\
\hline
$M>10$   & $0.28 \pm 0.02$  & 13.5  & $0.19 \pm 0.03$  & 5.9  & -0.02 $\pm$ 0.01  & -1.3  & -0.03 $\pm$ 0.02  & -2.0  & -0.06 $\pm$ 0.04  & -1.6 \\
$M>10$  (clean)  & $0.92 \pm 0.03$  & 27.0  & $0.72 \pm 0.05$  & 15.3  & 0.28 $\pm$ 0.02  & 16.1  & 0.30 $\pm$ 0.02  & 15.5  & 0.40 $\pm$ 0.04  & 9.1 \\\hline
$M>10.25$   & $0.29 \pm 0.02$  & 14.2  & $0.21 \pm 0.03$  & 6.4  & -0.00 $\pm$ 0.02  & -0.1  & -0.01 $\pm$ 0.02  & -0.9  & -0.05 $\pm$ 0.04  & -1.3 \\
$M>10.25$  (clean)  & $0.94 \pm 0.04$  & 26.2  & $0.73 \pm 0.05$  & 14.8  & 0.29 $\pm$ 0.02  & 15.8  & 0.31 $\pm$ 0.02  & 15.3  & 0.40 $\pm$ 0.05  & 8.8 \\\hline
$M>10.5$   & $0.31 \pm 0.02$  & 13.6  & $0.24 \pm 0.04$  & 6.5  & 0.04 $\pm$ 0.02  & 2.5  & 0.02 $\pm$ 0.02  & 1.2  & 0.01 $\pm$ 0.04  & 0.2 \\
$M>10.5$  (clean)  & $0.96 \pm 0.04$  & 24.7  & $0.76 \pm 0.05$  & 14.3  & 0.34 $\pm$ 0.02  & 16.3  & 0.35 $\pm$ 0.02  & 15.4  & 0.44 $\pm$ 0.05  & 8.3 \\\hline
$M>10.75$   & $0.38 \pm 0.03$  & 13.4  & $0.29 \pm 0.04$  & 6.8  & 0.10 $\pm$ 0.02  & 4.7  & 0.07 $\pm$ 0.02  & 3.1  & 0.08 $\pm$ 0.05  & 1.5 \\
$M>10.75$  (clean)  & $1.04 \pm 0.05$  & 22.8  & $0.83 \pm 0.06$  & 13.5  & 0.40 $\pm$ 0.03  & 15.8  & 0.41 $\pm$ 0.03  & 14.6  & 0.49 $\pm$ 0.06  & 7.8 \\\hline
$M>11$   & $0.48 \pm 0.04$  & 12.9  & $0.41 \pm 0.06$  & 7.5  & 0.22 $\pm$ 0.03  & 8.4  & 0.20 $\pm$ 0.03  & 6.9  & 0.21 $\pm$ 0.07  & 3.0 \\
$M>11$  (clean)  & $1.13 \pm 0.06$  & 19.8  & $0.94 \pm 0.08$  & 12.3  & 0.51 $\pm$ 0.03  & 15.6  & 0.53 $\pm$ 0.04  & 14.4  & 0.59 $\pm$ 0.09  & 7.0 \\\hline
$M>11.25$   & $0.72 \pm 0.06$  & 12.2  & $0.74 \pm 0.08$  & 8.9  & 0.53 $\pm$ 0.04  & 12.8  & 0.50 $\pm$ 0.05  & 10.9  & 0.46 $\pm$ 0.11  & 4.2 \\
$M>11.25$  (clean)  & $1.39 \pm 0.07$  & 18.9  & $1.27 \pm 0.11$  & 11.5  & 0.82 $\pm$ 0.05  & 16.4  & 0.83 $\pm$ 0.06  & 14.9  & 0.72 $\pm$ 0.13  & 5.4 \\\hline

\multicolumn{11}{|c|}{\textbf{\texttt{eROMaPPer}}}\\
\hline
$\lambda>5$ & $1.17\pm0.10$ & 11.8 & $1.18\pm0.13$ & 9.4 & $1.82\pm0.12$ & 15.1 & $1.88\pm0.12$ & 15.7 & $1.91\pm0.15$ & 12.7 \\
$\lambda>5$ (clean) & $1.97\pm0.05$ & 39.4 & $1.99\pm0.11$ & 17.4 & $1.97\pm0.10$ & 20.3 & $2.05\pm0.10$ & 21.0 & $2.11\pm0.15$ & 14.4 \\\hline
$\lambda>10$ & $3.32\pm0.16$ & 21.1 & $3.78\pm0.20$ & 18.8 & $4.78\pm0.19$ & 25.1 & $4.81\pm0.19$ & 25.4 & $4.84\pm0.24$ & 20.2 \\
$\lambda>10$ (clean) & $4.03\pm0.09$ & 43.4 & $4.06\pm0.14$ & 28.3 & $4.67\pm0.16$ & 29.1 & $4.72\pm0.16$ & 29.2 & $4.81\pm0.25$ & 19.4 \\\hline
$\lambda>20$ & $9.45\pm0.28$ & 33.3 & $11.08\pm0.36$ & 30.4 & $12.61\pm0.34$ & 37.4 & $12.55\pm0.33$ & 37.5 & $12.48\pm0.43$ & 29.3 \\
$\lambda>20$ (clean) & $9.66\pm0.29$ & 33.7 & $10.54\pm0.39$ & 27.3 & $11.70\pm0.33$ & 36.0 & $11.62\pm0.33$ & 35.4 & $11.91\pm0.50$ & 24.0 \\\hline

\end{tabular}
\caption{Compton-$y$ aperture photometry values for each binned sample and map calculated as described in Section \ref{sec:methods}. The BGS bin labels have been shortened from $\log_{10} (M_*/M_{\odot})_* >$ to $M>$ for brevity. For the full samples, the 90 GHz and CIB-deprojected ILC maps provide the highest SNR, while the 90 GHz map provides the best SNR across the board for the radio-clean samples.}
\label{tab:ytable}
\end{table*}

\section{Optical depth measurements}
We present our tSZ-derived optical depth measurements alongside the kSZ-derived values reported in \cite{gong26, hadzhiyska25, hsu26} in Tables \ref{lrgtautable}-\ref{opttautable}.

\clearpage
\bibliographystyle{apsrev}
\bibliography{Moore26}

@string{PRL={Phys. Rev. Lett.}}

@ARTICLE{abdul25,
       author = {{Abdul Karim}, M. and {Aguilar}, J. and {Ahlen}, S. and {Alam}, S. and {Allen}, L. and {Prieto}, C. Allende and {Alves}, O. and {Anand}, A. and {Andrade}, U. and {Armengaud}, E. and {Aviles}, A. and {Bailey}, S. and {Baltay}, C. and {Bansal}, P. and {Bault}, A. and {Behera}, J. and {BenZvi}, S. and {Bianchi}, D. and {Blake}, C. and {Brieden}, S. and {Brodzeller}, A. and {Brooks}, D. and {Buckley-Geer}, E. and {Burtin}, E. and {Calderon}, R. and {Canning}, R. and {Rosell}, A. Carnero and {Carrilho}, P. and {Casas}, L. and {Castander}, F.~J. and {Charles}, M. and {Chaussidon}, E. and {Chaves-Montero}, J. and {Chebat}, D. and {Chen}, X. and {Claybaugh}, T. and {Cole}, S. and {Cooper}, A.~P. and {Cuceu}, A. and {Dawson}, K.~S. and {de la Macorra}, A. and {de Mattia}, A. and {Deiosso}, N. and {Della Costa}, J. and {Demina}, R. and {Dey}, A. and {Dey}, B. and {Ding}, Z. and {Doel}, P. and {Edelstein}, J. and {Eisenstein}, D.~J. and {Elbers}, W. and {Fagrelius}, P. and {Fanning}, K. and {Fern{\'a}ndez-Garc{\'\i}a}, E. and {Ferraro}, S. and {Font-Ribera}, A. and {Forero-Romero}, J.~E. and {Frenk}, C.~S. and {Garcia-Quintero}, C. and {Garrison}, L.~H. and {Gazta{\~n}aga}, E. and {Gil-Mar{\'\i}n}, H. and {Gontcho A Gontcho}, S. and {Gonzalez}, D. and {Gonzalez-Morales}, A.~X. and {Gordon}, C. and {Green}, D. and {Gutierrez}, G. and {Guy}, J. and {Hadzhiyska}, B. and {Hahn}, C. and {He}, S. and {Herbold}, M. and {Herrera-Alcantar}, H.~K. and {Ho}, M.-F. and {Honscheid}, K. and {Howlett}, C. and {Huterer}, D. and {Ishak}, M. and {Juneau}, S. and {Kamble}, N.~V. and {Kara{\c{c}}ayl{\i}}, N.~G. and {Kehoe}, R. and {Kent}, S. and {Kim}, A.~G. and {Kirkby}, D. and {Kisner}, T. and {Koposov}, S.~E. and {Kremin}, A. and {Krolewski}, A. and {Lahav}, O. and {Lamman}, C. and {Landriau}, M. and {Lang}, D. and {Lasker}, J. and {Le Goff}, J.~M. and {Le Guillou}, L. and {Leauthaud}, A. and {Levi}, M.~E. and {Li}, Q. and {Li}, T.~S. and {Lodha}, K. and {Lokken}, M. and {Lozano-Rodr{\'\i}guez}, F. and {Magneville}, C. and {Manera}, M. and {Martini}, P. and {Matthewson}, W.~L. and {Meisner}, A. and {Mena-Fern{\'a}ndez}, J. and {Menegas}, A. and {Mergulh{\~a}o}, T. and {Miquel}, R. and {Moustakas}, J. and {Mu{\~n}oz-Guti{\'e}rrez}, A. and {Mu{\~n}oz-Santos}, D. and {Myers}, A.~D. and {Nadathur}, S. and {Naidoo}, K. and {Napolitano}, L. and {Newman}, J.~A. and {Niz}, G. and {Noriega}, H.~E. and {Paillas}, E. and {Palanque-Delabrouille}, N. and {Pan}, J. and {Peacock}, J.~A. and {Pellejero Ibanez}, M. and {Percival}, W.~J. and {P{\'e}rez-Fern{\'a}ndez}, A. and {P{\'e}rez-R{\`a}fols}, I. and {Pieri}, M.~M. and {Poppett}, C. and {Prada}, F. and {Rabinowitz}, D. and {Raichoor}, A. and {Ram{\'\i}rez-P{\'e}rez}, C. and {Rashkovetskyi}, M. and {Ravoux}, C. and {Rich}, J. and {Rocher}, A. and {Rockosi}, C. and {Rohlf}, J. and {Rom{\'a}n-Herrera}, J.~O. and {Ross}, A.~J. and {Rossi}, G. and {Ruggeri}, R. and {Ruhlmann-Kleider}, V. and {Samushia}, L. and {Sanchez}, E. and {Sanders}, N. and {Schlegel}, D. and {Schubnell}, M. and {Seo}, H. and {Shafieloo}, A. and {Sharples}, R. and {Silber}, J. and {Sinigaglia}, F. and {Sprayberry}, D. and {Tan}, T. and {Tarl{\'e}}, G. and {Taylor}, P. and {Turner}, W. and {Ure{\~n}a-L{\'o}pez}, L.~A. and {Vaisakh}, R. and {Valdes}, F. and {Valogiannis}, G. and {Vargas-Maga{\~n}a}, M. and {Verde}, L. and {Walther}, M. and {Weaver}, B.~A. and {Weinberg}, D.~H. and {White}, M. and {Wolfson}, M. and {Y{\`e}che}, C. and {Yu}, J. and {Zaborowski}, E.~A. and {Zarrouk}, P. and {Zhai}, Z. and {Zhang}, H. and {Zhao}, C. and {Zhao}, G.~B. and {Zhou}, R. and {Zou}, H. and {DESI Collaboration}},
        title = "{DESI DR2 results. II. Measurements of baryon acoustic oscillations and cosmological constraints}",
      journal = {\prd},
         year = 2025,
        month = oct,
       volume = {112},
       number = {8},
          eid = {083515},
        pages = {083515},
          doi = {10.1103/tr6y-kpc6},
archivePrefix = {arXiv},
       eprint = {2503.14738},
 primaryClass = {astro-ph.CO},
       adsurl = {https://ui.adsabs.harvard.edu/abs/2025PhRvD.112h3515A}
}

@ARTICLE{act26,
       author = {{Aguena}, M. and {Aiola}, S. and {Allam}, S. and {Andrade-Oliveira}, F. and {Bacon}, D. and {Bahcall}, N. and {Battaglia}, N. and {Battistelli}, E.~S. and {Bocquet}, S. and {Bolliet}, B. and {Bond}, J.~R. and {Brooks}, D. and {Calabrese}, E. and {Carretero}, J. and {Choi}, S.~K. and {da Costa}, L.~N. and {Costanzi}, M. and {Coulton}, W. and {Davis}, T.~M. and {Desai}, S. and {Devlin}, M.~J. and {Dicker}, S. and {Doel}, P. and {Duivenvoorden}, A.~J. and {Dunkley}, J. and {Ferraro}, S. and {Flaugher}, B. and {Frieman}, J. and {Gallardo}, P.~A. and {Gatti}, M. and {Gaztanaga}, E. and {Gill}, A.~S. and {Golec}, J.~E. and {Gruen}, D. and {Gruendl}, R.~A. and {Halpern}, M. and {Hasselfield}, M. and {Hill}, J.~C. and {Hilton}, M. and {Hincks}, A.~D. and {Hinton}, S.~R. and {Hollowood}, D.~L. and {Honscheid}, K. and {Hubmayr}, J. and {Huffenberger}, K.~M. and {Hughes}, J.~P. and {James}, D.~J. and {Klein}, M. and {Knowles}, K. and {Koopman}, B.~J. and {Kosowsky}, A. and {Lahav}, O. and {Lee}, E. and {Lin}, Y. and {Lokken}, M. and {Madhavacheril}, M.~S. and {Malag{\'o}n}, A.~A. Plazas and {Marrewijk}, J. v. and {Marshall}, J.~L. and {McMahon}, J. and {Mena-Fern{\'a}ndez}, J. and {Miquel}, R. and {Miyatake}, H. and {Mohr}, J.~J. and {Moodley}, K. and {Mroczkowski}, T. and {Naess}, S. and {Nati}, F. and {Nicola}, A. and {Niemack}, M.~D. and {Ogando}, R.~L.~C. and {Oguri}, M. and {Orlowski-Scherer}, J. and {Page}, L.~A. and {Partridge}, B. and {da Silva Pereira}, M.~E. and {Porredon}, A. and {Qu}, F.~J. and {Ragavan}, D.~C. and {Guachalla}, B. Ried and {Romer}, A.~K. and {Rosell}, A. Carnero and {Rykoff}, E.~S. and {Samuroff}, S. and {Sanchez}, E. and {Sevilla-Noarbe}, I. and {Sierra}, C. and {Sif{\'o}n}, C. and {Smith}, M. and {Staggs}, S.~T. and {Suchyta}, E. and {Swanson}, M.~E.~C. and {Tucker}, D.~L. and {Vargas}, C. and {Vavagiakis}, E.~M. and {De Vicente}, J. and {Weaverdyck}, N. and {Weller}, J. and {Wollack}, E.~J. and {Zubeldia}, I.},
        title = "{The Atacama Cosmology Telescope: DR6 Sunyaev-Zel'dovich Selected Galaxy Clusters Catalog}",
      journal = {The Open Journal of Astrophysics},
         year = 2026,
        month = jan,
       volume = {9},
        pages = {55863},
          doi = {10.33232/001c.155863},
archivePrefix = {arXiv},
       eprint = {2507.21459},
 primaryClass = {astro-ph.CO},
       adsurl = {https://ui.adsabs.harvard.edu/abs/2026OJAp....955863A}
}

@ARTICLE{amodeo21,
       author = {{Amodeo}, Stefania and {Battaglia}, Nicholas and {Schaan}, Emmanuel and {Ferraro}, Simone and {Moser}, Emily and {Aiola}, Simone and {Austermann}, Jason E. and {Beall}, James A. and {Bean}, Rachel and {Becker}, Daniel T. and {Bond}, Richard J. and {Calabrese}, Erminia and {Calafut}, Victoria and {Choi}, Steve K. and {Denison}, Edward V. and {Devlin}, Mark and {Duff}, Shannon M. and {Duivenvoorden}, Adriaan J. and {Dunkley}, Jo and {D{\"u}nner}, Rolando and {Gallardo}, Patricio A. and {Hall}, Kirsten R. and {Han}, Dongwon and {Hill}, J. Colin and {Hilton}, Gene C. and {Hilton}, Matt and {Hlo{\v{z}}ek}, Ren{\'e}e and {Hubmayr}, Johannes and {Huffenberger}, Kevin M. and {Hughes}, John P. and {Koopman}, Brian J. and {MacInnis}, Amanda and {McMahon}, Jeff and {Madhavacheril}, Mathew S. and {Moodley}, Kavilan and {Mroczkowski}, Tony and {Naess}, Sigurd and {Nati}, Federico and {Newburgh}, Laura B. and {Niemack}, Michael D. and {Page}, Lyman A. and {Partridge}, Bruce and {Schillaci}, Alessandro and {Sehgal}, Neelima and {Sif{\'o}n}, Crist{\'o}bal and {Spergel}, David N. and {Staggs}, Suzanne and {Storer}, Emilie R. and {Ullom}, Joel N. and {Vale}, Leila R. and {van Engelen}, Alexander and {Van Lanen}, Jeff and {Vavagiakis}, Eve M. and {Wollack}, Edward J. and {Xu}, Zhilei},
        title = "{Atacama Cosmology Telescope: Modeling the gas thermodynamics in BOSS CMASS galaxies from kinematic and thermal Sunyaev-Zel'dovich measurements}",
      journal = {\prd},
         year = 2021,
        month = mar,
       volume = {103},
       number = {6},
          eid = {063514},
        pages = {063514},
          doi = {10.1103/PhysRevD.103.063514},
archivePrefix = {arXiv},
       eprint = {2009.05558},
 primaryClass = {astro-ph.CO},
       adsurl = {https://ui.adsabs.harvard.edu/abs/2021PhRvD.103f3514A}
}

@ARTICLE{battaglia10,
       author = {{Battaglia}, N. and {Bond}, J.~R. and {Pfrommer}, C. and {Sievers}, J.~L. and {Sijacki}, D.},
        title = "{Simulations of the Sunyaev-Zel'dovich Power Spectrum with Active Galactic Nucleus Feedback}",
      journal = {\apj},
         year = 2010,
        month = dec,
       volume = {725},
       number = {1},
        pages = {91-99},
          doi = {10.1088/0004-637X/725/1/91},
archivePrefix = {arXiv},
       eprint = {1003.4256},
 primaryClass = {astro-ph.CO},
       adsurl = {https://ui.adsabs.harvard.edu/abs/2010ApJ...725...91B}
}

@ARTICLE{battaglia12,
   author = {{Battaglia}, N. and {Bond}, J.~R. and {Pfrommer}, C. and {Sievers}, J.~L.
	},
    title = "{On the Cluster Physics of Sunyaev-Zel'dovich and X-Ray Surveys. II. Deconstructing the Thermal SZ Power Spectrum}",
  journal = {\apj},
archivePrefix = "arXiv",
   eprint = {arXiv:1109.3711},
     year = 2012,
    month = oct,
   volume = 758,
      eid = {75},
    pages = {75},
      doi = {10.1088/0004-637X/758/2/75},
   adsurl = {http://adsabs.harvard.edu/abs/2012ApJ...758...75B}
}

@ARTICLE{battaglia16,
   author = {{Battaglia}, N.},
    title = "{The tau of galaxy clusters}",
  journal = {JCAP},
archivePrefix = "arXiv",
   eprint = {arXiv:1607.02442},
     year = 2016,
    month = aug,
   volume = 8,
      eid = {058},
    pages = {058},
      doi = {10.1088/1475-7516/2016/08/058},
   adsurl = {http://adsabs.harvard.edu/abs/2016JCAP...08..058B}
}

@ARTICLE{battaglia17,
       author = {{Battaglia}, Nicholas and {Ferraro}, Simone and {Schaan}, Emmanuel and {Spergel}, David N.},
        title = "{Future constraints on halo thermodynamics from combined Sunyaev-Zel'dovich measurements}",
      journal = {\jcap},
         year = 2017,
        month = nov,
       volume = {2017},
       number = {11},
          eid = {040},
        pages = {040},
          doi = {10.1088/1475-7516/2017/11/040},
archivePrefix = {arXiv},
       eprint = {1705.05881},
 primaryClass = {astro-ph.CO},
       adsurl = {https://ui.adsabs.harvard.edu/abs/2017JCAP...11..040B}
}

@ARTICLE{battaglia26,
       author = {{Battaglia}, Nicholas and {Melin}, Jean-Baptiste and {Hill}, J. Colin and {Bartlett}, James},
        title = "{Thermal Sunyaev-Zel'dovich Measurements of Locally Bright Galaxies with ACT DR6: Radio Source Contamination and Excess Compton-y Signal}",
      journal = {arXiv e-prints},
         year = 2026,
        month = jul,
          eid = {arXiv:2607.08721},
        pages = {arXiv:2607.08721},
          doi = {10.48550/arXiv.2607.08721},
archivePrefix = {arXiv},
       eprint = {2607.08721},
 primaryClass = {astro-ph.GA},
       adsurl = {https://ui.adsabs.harvard.edu/abs/2026arXiv260708721B}
}

@ARTICLE{bhattacharya08,
       author = {{Bhattacharya}, Suman and {Kosowsky}, Arthur},
        title = "{Dark energy constraints from galaxy cluster peculiar velocities}",
      journal = {\prd},
         year = 2008,
        month = apr,
       volume = {77},
       number = {8},
          eid = {083004},
        pages = {083004},
          doi = {10.1103/PhysRevD.77.083004},
archivePrefix = {arXiv},
       eprint = {0712.0034},
 primaryClass = {astro-ph},
       adsurl = {https://ui.adsabs.harvard.edu/abs/2008PhRvD..77h3004B}
}

@ARTICLE{calafut21,
       author = {{Calafut}, V. and {Gallardo}, P.~A. and {Vavagiakis}, E.~M. and {Amodeo}, S. and {Aiola}, S. and {Austermann}, J.~E. and {Battaglia}, N. and {Battistelli}, E.~S. and {Beall}, J.~A. and {Bean}, R. and {Bond}, J.~R. and {Calabrese}, E. and {Choi}, S.~K. and {Cothard}, N.~F. and {Devlin}, M.~J. and {Duell}, C.~J. and {Duff}, S.~M. and {Duivenvoorden}, A.~J. and {Dunkley}, J. and {Dunner}, R. and {Ferraro}, S. and {Guan}, Y. and {Hill}, J.~C. and {Hilton}, G.~C. and {Hilton}, M. and {Hlo{\v{z}}ek}, R. and {Huber}, Z.~B. and {Hubmayr}, J. and {Huffenberger}, K.~M. and {Hughes}, J.~P. and {Koopman}, B.~J. and {Kosowsky}, A. and {Li}, Y. and {Lokken}, M. and {Madhavacheril}, M. and {McMahon}, J. and {Moodley}, K. and {Naess}, S. and {Nati}, F. and {Newburgh}, L.~B. and {Niemack}, M.~D. and {Page}, L.~A. and {Partridge}, B. and {Schaan}, E. and {Schillaci}, A. and {Sif{\'o}n}, C. and {Spergel}, D.~N. and {Staggs}, S.~T. and {Ullom}, J.~N. and {Vale}, L.~R. and {Van Engelen}, A. and {Van Lanen}, J. and {Wollack}, E.~J. and {Xu}, Z.},
        title = "{The Atacama Cosmology Telescope: Detection of the pairwise kinematic Sunyaev-Zel'dovich effect with SDSS DR15 galaxies}",
      journal = {\prd},
         year = 2021,
        month = aug,
       volume = {104},
       number = {4},
          eid = {043502},
        pages = {043502},
          doi = {10.1103/PhysRevD.104.043502},
archivePrefix = {arXiv},
       eprint = {2101.08374},
 primaryClass = {astro-ph.CO},
       adsurl = {https://ui.adsabs.harvard.edu/abs/2021PhRvD.104d3502C}
}

@ARTICLE{carlstrom11,
       author = {{Carlstrom}, J.~E. and {Ade}, P.~A.~R. and {Aird}, K.~A. and {Benson}, B.~A. and {Bleem}, L.~E. and {Busetti}, S. and {Chang}, C.~L. and {Chauvin}, E. and {Cho}, H.-M. and {Crawford}, T.~M. and {Crites}, A.~T. and {Dobbs}, M.~A. and {Halverson}, N.~W. and {Heimsath}, S. and {Holzapfel}, W.~L. and {Hrubes}, J.~D. and {Joy}, M. and {Keisler}, R. and {Lanting}, T.~M. and {Lee}, A.~T. and {Leitch}, E.~M. and {Leong}, J. and {Lu}, W. and {Lueker}, M. and {Luong-Van}, D. and {McMahon}, J.~J. and {Mehl}, J. and {Meyer}, S.~S. and {Mohr}, J.~J. and {Montroy}, T.~E. and {Padin}, S. and {Plagge}, T. and {Pryke}, C. and {Ruhl}, J.~E. and {Schaffer}, K.~K. and {Schwan}, D. and {Shirokoff}, E. and {Spieler}, H.~G. and {Staniszewski}, Z. and {Stark}, A.~A. and {Tucker}, C. and {Vanderlinde}, K. and {Vieira}, J.~D. and {Williamson}, R.},
        title = "{The 10 Meter South Pole Telescope}",
      journal = {\pasp},
         year = 2011,
        month = may,
       volume = {123},
       number = {903},
        pages = {568},
          doi = {10.1086/659879},
archivePrefix = {arXiv},
       eprint = {0907.4445},
 primaryClass = {astro-ph.IM},
       adsurl = {https://ui.adsabs.harvard.edu/abs/2011PASP..123..568C}
}

@ARTICLE{ccat23,
       author = {{CCAT-Prime Collaboration} and {Aravena}, Manuel and {Austermann}, Jason E. and {Basu}, Kaustuv and {Battaglia}, Nicholas and {Beringue}, Benjamin and {Bertoldi}, Frank and {Bigiel}, Frank and {Bond}, J. Richard and {Breysse}, Patrick C. and {Broughton}, Colton and {Bustos}, Ricardo and {Chapman}, Scott C. and {Charmetant}, Maude and {Choi}, Steve K. and {Chung}, Dongwoo T. and {Clark}, Susan E. and {Cothard}, Nicholas F. and {Crites}, Abigail T. and {Dev}, Ankur and {Douglas}, Kaela and {Duell}, Cody J. and {D{\"u}nner}, Rolando and {Ebina}, Haruki and {Erler}, Jens and {Fich}, Michel and {Fissel}, Laura M. and {Foreman}, Simon and {Freundt}, R.~G. and {Gallardo}, Patricio A. and {Gao}, Jiansong and {Garc{\'\i}a}, Pablo and {Giovanelli}, Riccardo and {Golec}, Joseph E. and {Groppi}, Christopher E. and {Haynes}, Martha P. and {Henke}, Douglas and {Hensley}, Brandon and {Herter}, Terry and {Higgins}, Ronan and {Hlo{\v{z}}ek}, Ren{\'e}e and {Huber}, Anthony and {Huber}, Zachary and {Hubmayr}, Johannes and {Jackson}, Rebecca and {Johnstone}, Douglas and {Karoumpis}, Christos and {Keating}, Laura C. and {Komatsu}, Eiichiro and {Li}, Yaqiong and {Magnelli}, Benjamin and {Matthews}, Brenda C. and {Mauskopf}, Philip D. and {McMahon}, Jeffrey J. and {Meerburg}, P. Daniel and {Meyers}, Joel and {Muralidhara}, Vyoma and {Murray}, Norman W. and {Niemack}, Michael D. and {Nikola}, Thomas and {Okada}, Yoko and {Puddu}, Roberto and {Riechers}, Dominik A. and {Rosolowsky}, Erik and {Rossi}, Kayla and {Rotermund}, Kaja and {Roy}, Anirban and {Sadavoy}, Sarah I. and {Schaaf}, Reinhold and {Schilke}, Peter and {Scott}, Douglas and {Simon}, Robert and {Sinclair}, Adrian K. and {Sivakoff}, Gregory R. and {Stacey}, Gordon J. and {Stutz}, Amelia M. and {Stutzki}, Juergen and {Tahani}, Mehrnoosh and {Thanjavur}, Karun and {Timmermann}, Ralf A. and {Ullom}, Joel N. and {van Engelen}, Alexander and {Vavagiakis}, Eve M. and {Vissers}, Michael R. and {Wheeler}, Jordan D. and {White}, Simon D.~M. and {Zhu}, Yijie and {Zou}, Bugao},
        title = "{CCAT-prime Collaboration: Science Goals and Forecasts with Prime-Cam on the Fred Young Submillimeter Telescope}",
      journal = {\apjs},
         year = 2023,
        month = jan,
       volume = {264},
       number = {1},
          eid = {7},
        pages = {7},
          doi = {10.3847/1538-4365/ac9838},
archivePrefix = {arXiv},
       eprint = {2107.10364},
 primaryClass = {astro-ph.CO},
       adsurl = {https://ui.adsabs.harvard.edu/abs/2023ApJS..264....7C}
}

@ARTICLE{cooray98,
       author = {{Cooray}, Asantha R. and {Grego}, Laura and {Holzapfel}, William L. and {Joy}, Marshall and {Carlstrom}, John E.},
        title = "{Radio Sources in Galaxy Clusters at 28.5 GHz}",
      journal = {\aj},
         year = 1998,
        month = apr,
       volume = {115},
       number = {4},
        pages = {1388-1399},
          doi = {10.1086/300310},
archivePrefix = {arXiv},
       eprint = {astro-ph/9711218},
 primaryClass = {astro-ph},
       adsurl = {https://ui.adsabs.harvard.edu/abs/1998AJ....115.1388C}
}

@ARTICLE{condon98,
       author = {{Condon}, J.~J. and {Cotton}, W.~D. and {Greisen}, E.~W. and {Yin}, Q.~F. and {Perley}, R.~A. and {Taylor}, G.~B. and {Broderick}, J.~J.},
        title = "{The NRAO VLA Sky Survey}",
      journal = {\aj},
         year = 1998,
        month = may,
       volume = {115},
       number = {5},
        pages = {1693-1716},
          doi = {10.1086/300337},
       adsurl = {https://ui.adsabs.harvard.edu/abs/1998AJ....115.1693C}
}

@ARTICLE{coulton24,
       author = {{Coulton}, William and {Madhavacheril}, Mathew S. and {Duivenvoorden}, Adriaan J. and {Hill}, J. Colin and {Abril-Cabezas}, Irene and {Ade}, Peter A.~R. and {Aiola}, Simone and {Alford}, Tommy and {Amiri}, Mandana and {Amodeo}, Stefania and {An}, Rui and {Atkins}, Zachary and {Austermann}, Jason E. and {Battaglia}, Nicholas and {Battistelli}, Elia Stefano and {Beall}, James A. and {Bean}, Rachel and {Beringue}, Benjamin and {Bhandarkar}, Tanay and {Biermann}, Emily and {Bolliet}, Boris and {Bond}, J. Richard and {Cai}, Hongbo and {Calabrese}, Erminia and {Calafut}, Victoria and {Capalbo}, Valentina and {Carrero}, Felipe and {Chesmore}, Grace E. and {Cho}, Hsiao-mei and {Choi}, Steve K. and {Clark}, Susan E. and {Rosado}, Rodrigo C{\'o}rdova and {Cothard}, Nicholas F. and {Coughlin}, Kevin and {Crowley}, Kevin T. and {Devlin}, Mark J. and {Dicker}, Simon and {Doze}, Peter and {Duell}, Cody J. and {Duff}, Shannon M. and {Dunkley}, Jo and {D{\"u}nner}, Rolando and {Fanfani}, Valentina and {Fankhanel}, Max and {Farren}, Gerrit and {Ferraro}, Simone and {Freundt}, Rodrigo and {Fuzia}, Brittany and {Gallardo}, Patricio A. and {Garrido}, Xavier and {Givans}, Jahmour and {Gluscevic}, Vera and {Golec}, Joseph E. and {Guan}, Yilun and {Halpern}, Mark and {Han}, Dongwon and {Hasselfield}, Matthew and {Healy}, Erin and {Henderson}, Shawn and {Hensley}, Brandon and {Herv{\'\i}as-Caimapo}, Carlos and {Hilton}, Gene C. and {Hilton}, Matt and {Hincks}, Adam D. and {Hlo{\v{z}}ek}, Ren{\'e}e and {Ho}, Shuay-Pwu Patty and {Huber}, Zachary B. and {Hubmayr}, Johannes and {Huffenberger}, Kevin M. and {Hughes}, John P. and {Irwin}, Kent and {Isopi}, Giovanni and {Jense}, Hidde T. and {Keller}, Ben and {Kim}, Joshua and {Knowles}, Kenda and {Koopman}, Brian J. and {Kosowsky}, Arthur and {Kramer}, Darby and {Kusiak}, Aleksandra and {La Posta}, Adrien and {Lakey}, Victoria and {Lee}, Eunseong and {Li}, Zack and {Li}, Yaqiong and {Limon}, Michele and {Lokken}, Martine and {Louis}, Thibaut and {Lungu}, Marius and {MacCrann}, Niall and {MacInnis}, Amanda and {Maldonado}, Diego and {Maldonado}, Felipe and {Mallaby-Kay}, Maya and {Marques}, Gabriela A. and {van Marrewijk}, Joshiwa and {McCarthy}, Fiona and {McMahon}, Jeff and {Mehta}, Yogesh and {Menanteau}, Felipe and {Moodley}, Kavilan and {Morris}, Thomas W. and {Mroczkowski}, Tony and {Naess}, Sigurd and {Namikawa}, Toshiya and {Nati}, Federico and {Newburgh}, Laura and {Nicola}, Andrina and {Niemack}, Michael D. and {Nolta}, Michael R. and {Orlowski-Scherer}, John and {Page}, Lyman A. and {Pandey}, Shivam and {Partridge}, Bruce and {Prince}, Heather and {Puddu}, Roberto and {Qu}, Frank J. and {Radiconi}, Federico and {Robertson}, Naomi and {Rojas}, Felipe and {Sakuma}, Tai and {Salatino}, Maria and {Schaan}, Emmanuel and {Schmitt}, Benjamin L. and {Sehgal}, Neelima and {Shaikh}, Shabbir and {Sherwin}, Blake D. and {Sierra}, Carlos and {Sievers}, Jon and {Sif{\'o}n}, Crist{\'o}bal and {Simon}, Sara and {Sonka}, Rita and {Spergel}, David N. and {Staggs}, Suzanne T. and {Storer}, Emilie and {Switzer}, Eric R. and {Tampier}, Niklas and {Thornton}, Robert and {Trac}, Hy and {Treu}, Jesse and {Tucker}, Carole and {Ullom}, Joel and {Vale}, Leila R. and {Van Engelen}, Alexander and {Van Lanen}, Jeff and {Vargas}, Cristian and {Vavagiakis}, Eve M. and {Wagoner}, Kasey and {Wang}, Yuhan and {Wenzl}, Lukas and {Wollack}, Edward J. and {Xu}, Zhilei and {Zago}, Fernando and {Zheng}, Kaiwen},
        title = "{Atacama Cosmology Telescope: High-resolution component-separated maps across one third of the sky}",
      journal = {\prd},
         year = 2024,
        month = mar,
       volume = {109},
       number = {6},
          eid = {063530},
        pages = {063530},
          doi = {10.1103/PhysRevD.109.063530},
archivePrefix = {arXiv},
       eprint = {2307.01258},
 primaryClass = {astro-ph.CO},
       adsurl = {https://ui.adsabs.harvard.edu/abs/2024PhRvD.109f3530C}
}

@ARTICLE{coulton26,
       author = {{Coulton}, William R. and {Duivenvoorden}, Adriaan J. and {Atkins}, Zachary and {Battaglia}, Nicholas and {Battistelli}, Elia Stefano and {Bond}, J. Richard and {Cai}, Hongbo and {Calabrese}, Erminia and {Choi}, Steve K. and {Crowley}, Kevin T. and {Devlin}, Mark J. and {Dunkley}, Jo and {Ferraro}, Simone and {Guan}, Yilun and {Herv{\'\i}as-Caimapo}, Carlos and {Hill}, J. Colin and {Hilton}, Matt and {Hincks}, Adam D. and {Kosowsky}, Arthur and {Madhavacheril}, Mathew S. and {van Marrewijk}, Joshiwa and {McCarthy}, Fiona and {Moodley}, Kavilan and {Mroczkowski}, Tony and {Niemack}, Michael D. and {Page}, Lyman A. and {Partridge}, Bruce and {Schaan}, Emmanuel and {Sehgal}, Neelima and {Sherwin}, Blake D. and {Sif{\'o}n}, Crist{\'o}bal and {Spergel}, David N. and {Staggs}, Suzanne T. and {Van Engelen}, Alexander and {Vavagiakis}, Eve M. and {Wollack}, Edward J.},
        title = "{Atacama Cosmology Telescope: A measurement of galaxy cluster temperatures through relativistic corrections to the thermal Sunyaev-Zeldovich effect}",
      journal = {\prd},
         year = 2026,
        month = feb,
       volume = {113},
       number = {4},
          eid = {043520},
        pages = {043520},
          doi = {10.1103/n7p5-pc66},
archivePrefix = {arXiv},
       eprint = {2410.19046},
 primaryClass = {astro-ph.CO},
       adsurl = {https://ui.adsabs.harvard.edu/abs/2026PhRvD.113d3520C}
}

@ARTICLE{dave19,
       author = {{Dav{\'e}}, Romeel and {Angl{\'e}s-Alc{\'a}zar}, Daniel and {Narayanan}, Desika and {Li}, Qi and {Rafieferantsoa}, Mika H. and {Appleby}, Sarah},
        title = "{SIMBA: Cosmological simulations with black hole growth and feedback}",
      journal = {\mnras},
         year = 2019,
        month = jun,
       volume = {486},
       number = {2},
        pages = {2827-2849},
          doi = {10.1093/mnras/stz937},
archivePrefix = {arXiv},
       eprint = {1901.10203},
 primaryClass = {astro-ph.GA},
       adsurl = {https://ui.adsabs.harvard.edu/abs/2019MNRAS.486.2827D}
}

@ARTICLE{dedeo05,
       author = {{DeDeo}, Simon and {Spergel}, David N. and {Trac}, Hy},
        title = "{The kinetic Sunyaev-Zel'dovitch effect as a dark energy probe}",
      journal = {arXiv e-prints},
         year = 2005,
        month = nov,
          eid = {astro-ph/0511060},
        pages = {astro-ph/0511060},
          doi = {10.48550/arXiv.astro-ph/0511060},
archivePrefix = {arXiv},
       eprint = {astro-ph/0511060},
 primaryClass = {astro-ph},
       adsurl = {https://ui.adsabs.harvard.edu/abs/2005astro.ph.11060D}
}

@ARTICLE{desi16.1,
       author = {{DESI Collaboration} and {Aghamousa}, Amir and {Aguilar}, Jessica and {Ahlen}, Steve and {Alam}, Shadab and {Allen}, Lori E. and {Allende Prieto}, Carlos and {Annis}, James and {Bailey}, Stephen and {Balland}, Christophe and {Ballester}, Otger and {Baltay}, Charles and {Beaufore}, Lucas and {Bebek}, Chris and {Beers}, Timothy C. and {Bell}, Eric F. and {Bernal}, Jos{\'e} Luis and {Besuner}, Robert and {Beutler}, Florian and {Blake}, Chris and {Bleuler}, Hannes and {Blomqvist}, Michael and {Blum}, Robert and {Bolton}, Adam S. and {Briceno}, Cesar and {Brooks}, David and {Brownstein}, Joel R. and {Buckley-Geer}, Elizabeth and {Burden}, Angela and {Burtin}, Etienne and {Busca}, Nicolas G. and {Cahn}, Robert N. and {Cai}, Yan-Chuan and {Cardiel-Sas}, Laia and {Carlberg}, Raymond G. and {Carton}, Pierre-Henri and {Casas}, Ricard and {Castander}, Francisco J. and {Cervantes-Cota}, Jorge L. and {Claybaugh}, Todd M. and {Close}, Madeline and {Coker}, Carl T. and {Cole}, Shaun and {Comparat}, Johan and {Cooper}, Andrew P. and {Cousinou}, M. -C. and {Crocce}, Martin and {Cuby}, Jean-Gabriel and {Cunningham}, Daniel P. and {Davis}, Tamara M. and {Dawson}, Kyle S. and {de la Macorra}, Axel and {De Vicente}, Juan and {Delubac}, Timoth{\'e}e and {Derwent}, Mark and {Dey}, Arjun and {Dhungana}, Govinda and {Ding}, Zhejie and {Doel}, Peter and {Duan}, Yutong T. and {Ealet}, Anne and {Edelstein}, Jerry and {Eftekharzadeh}, Sarah and {Eisenstein}, Daniel J. and {Elliott}, Ann and {Escoffier}, St{\'e}phanie and {Evatt}, Matthew and {Fagrelius}, Parker and {Fan}, Xiaohui and {Fanning}, Kevin and {Farahi}, Arya and {Farihi}, Jay and {Favole}, Ginevra and {Feng}, Yu and {Fernandez}, Enrique and {Findlay}, Joseph R. and {Finkbeiner}, Douglas P. and {Fitzpatrick}, Michael J. and {Flaugher}, Brenna and {Flender}, Samuel and {Font-Ribera}, Andreu and {Forero-Romero}, Jaime E. and {Fosalba}, Pablo and {Frenk}, Carlos S. and {Fumagalli}, Michele and {Gaensicke}, Boris T. and {Gallo}, Giuseppe and {Garcia-Bellido}, Juan and {Gaztanaga}, Enrique and {Pietro Gentile Fusillo}, Nicola and {Gerard}, Terry and {Gershkovich}, Irena and {Giannantonio}, Tommaso and {Gillet}, Denis and {Gonzalez-de-Rivera}, Guillermo and {Gonzalez-Perez}, Violeta and {Gott}, Shelby and {Graur}, Or and {Gutierrez}, Gaston and {Guy}, Julien and {Habib}, Salman and {Heetderks}, Henry and {Heetderks}, Ian and {Heitmann}, Katrin and {Hellwing}, Wojciech A. and {Herrera}, David A. and {Ho}, Shirley and {Holland}, Stephen and {Honscheid}, Klaus and {Huff}, Eric and {Hutchinson}, Timothy A. and {Huterer}, Dragan and {Hwang}, Ho Seong and {Illa Laguna}, Joseph Maria and {Ishikawa}, Yuzo and {Jacobs}, Dianna and {Jeffrey}, Niall and {Jelinsky}, Patrick and {Jennings}, Elise and {Jiang}, Linhua and {Jimenez}, Jorge and {Johnson}, Jennifer and {Joyce}, Richard and {Jullo}, Eric and {Juneau}, St{\'e}phanie and {Kama}, Sami and {Karcher}, Armin and {Karkar}, Sonia and {Kehoe}, Robert and {Kennamer}, Noble and {Kent}, Stephen and {Kilbinger}, Martin and {Kim}, Alex G. and {Kirkby}, David and {Kisner}, Theodore and {Kitanidis}, Ellie and {Kneib}, Jean-Paul and {Koposov}, Sergey and {Kovacs}, Eve and {Koyama}, Kazuya and {Kremin}, Anthony and {Kron}, Richard and {Kronig}, Luzius and {Kueter-Young}, Andrea and {Lacey}, Cedric G. and {Lafever}, Robin and {Lahav}, Ofer and {Lambert}, Andrew and {Lampton}, Michael and {Landriau}, Martin and {Lang}, Dustin and {Lauer}, Tod R. and {Le Goff}, Jean-Marc and {Le Guillou}, Laurent and {Le Van Suu}, Auguste and {Lee}, Jae Hyeon and {Lee}, Su-Jeong and {Leitner}, Daniela and {Lesser}, Michael and {Levi}, Michael E. and {L'Huillier}, Benjamin and {Li}, Baojiu and {Liang}, Ming and {Lin}, Huan and {Linder}, Eric and {Loebman}, Sarah R. and {Luki{\'c}}, Zarija and {Ma}, Jun and {MacCrann}, Niall and {Magneville}, Christophe and {Makarem}, Laleh and {Manera}, Marc and {Manser}, Christopher J. and {Marshall}, Robert and {Martini}, Paul and {Massey}, Richard and {Matheson}, Thomas and {McCauley}, Jeremy and {McDonald}, Patrick and {McGreer}, Ian D. and {Meisner}, Aaron and {Metcalfe}, Nigel and {Miller}, Timothy N. and {Miquel}, Ramon and {Moustakas}, John and {Myers}, Adam and {Naik}, Milind and {Newman}, Jeffrey A. and {Nichol}, Robert C. and {Nicola}, Andrina and {Nicolati da Costa}, Luiz and {Nie}, Jundan and {Niz}, Gustavo and {Norberg}, Peder and {Nord}, Brian and {Norman}, Dara and {Nugent}, Peter and {O'Brien}, Thomas and {Oh}, Minji and {Olsen}, Knut A.~G.},
        title = "{The DESI Experiment Part I: Science,Targeting, and Survey Design}",
      journal = {arXiv e-prints},
         year = 2016,
        month = oct,
          eid = {arXiv:1611.00036},
        pages = {arXiv:1611.00036},
          doi = {10.48550/arXiv.1611.00036},
archivePrefix = {arXiv},
       eprint = {1611.00036},
 primaryClass = {astro-ph.IM},
       adsurl = {https://ui.adsabs.harvard.edu/abs/2016arXiv161100036D}
}

@ARTICLE{desi16.2,
       author = {{DESI Collaboration} and {Aghamousa}, Amir and {Aguilar}, Jessica and {Ahlen}, Steve and {Alam}, Shadab and {Allen}, Lori E. and {Allende Prieto}, Carlos and {Annis}, James and {Bailey}, Stephen and {Balland}, Christophe and {Ballester}, Otger and {Baltay}, Charles and {Beaufore}, Lucas and {Bebek}, Chris and {Beers}, Timothy C. and {Bell}, Eric F. and {Bernal}, Jos{\'e} Luis and {Besuner}, Robert and {Beutler}, Florian and {Blake}, Chris and {Bleuler}, Hannes and {Blomqvist}, Michael and {Blum}, Robert and {Bolton}, Adam S. and {Briceno}, Cesar and {Brooks}, David and {Brownstein}, Joel R. and {Buckley-Geer}, Elizabeth and {Burden}, Angela and {Burtin}, Etienne and {Busca}, Nicolas G. and {Cahn}, Robert N. and {Cai}, Yan-Chuan and {Cardiel-Sas}, Laia and {Carlberg}, Raymond G. and {Carton}, Pierre-Henri and {Casas}, Ricard and {Castander}, Francisco J. and {Cervantes-Cota}, Jorge L. and {Claybaugh}, Todd M. and {Close}, Madeline and {Coker}, Carl T. and {Cole}, Shaun and {Comparat}, Johan and {Cooper}, Andrew P. and {Cousinou}, M. -C. and {Crocce}, Martin and {Cuby}, Jean-Gabriel and {Cunningham}, Daniel P. and {Davis}, Tamara M. and {Dawson}, Kyle S. and {de la Macorra}, Axel and {De Vicente}, Juan and {Delubac}, Timoth{\'e}e and {Derwent}, Mark and {Dey}, Arjun and {Dhungana}, Govinda and {Ding}, Zhejie and {Doel}, Peter and {Duan}, Yutong T. and {Ealet}, Anne and {Edelstein}, Jerry and {Eftekharzadeh}, Sarah and {Eisenstein}, Daniel J. and {Elliott}, Ann and {Escoffier}, St{\'e}phanie and {Evatt}, Matthew and {Fagrelius}, Parker and {Fan}, Xiaohui and {Fanning}, Kevin and {Farahi}, Arya and {Farihi}, Jay and {Favole}, Ginevra and {Feng}, Yu and {Fernandez}, Enrique and {Findlay}, Joseph R. and {Finkbeiner}, Douglas P. and {Fitzpatrick}, Michael J. and {Flaugher}, Brenna and {Flender}, Samuel and {Font-Ribera}, Andreu and {Forero-Romero}, Jaime E. and {Fosalba}, Pablo and {Frenk}, Carlos S. and {Fumagalli}, Michele and {Gaensicke}, Boris T. and {Gallo}, Giuseppe and {Garcia-Bellido}, Juan and {Gaztanaga}, Enrique and {Pietro Gentile Fusillo}, Nicola and {Gerard}, Terry and {Gershkovich}, Irena and {Giannantonio}, Tommaso and {Gillet}, Denis and {Gonzalez-de-Rivera}, Guillermo and {Gonzalez-Perez}, Violeta and {Gott}, Shelby and {Graur}, Or and {Gutierrez}, Gaston and {Guy}, Julien and {Habib}, Salman and {Heetderks}, Henry and {Heetderks}, Ian and {Heitmann}, Katrin and {Hellwing}, Wojciech A. and {Herrera}, David A. and {Ho}, Shirley and {Holland}, Stephen and {Honscheid}, Klaus and {Huff}, Eric and {Hutchinson}, Timothy A. and {Huterer}, Dragan and {Hwang}, Ho Seong and {Illa Laguna}, Joseph Maria and {Ishikawa}, Yuzo and {Jacobs}, Dianna and {Jeffrey}, Niall and {Jelinsky}, Patrick and {Jennings}, Elise and {Jiang}, Linhua and {Jimenez}, Jorge and {Johnson}, Jennifer and {Joyce}, Richard and {Jullo}, Eric and {Juneau}, St{\'e}phanie and {Kama}, Sami and {Karcher}, Armin and {Karkar}, Sonia and {Kehoe}, Robert and {Kennamer}, Noble and {Kent}, Stephen and {Kilbinger}, Martin and {Kim}, Alex G. and {Kirkby}, David and {Kisner}, Theodore and {Kitanidis}, Ellie and {Kneib}, Jean-Paul and {Koposov}, Sergey and {Kovacs}, Eve and {Koyama}, Kazuya and {Kremin}, Anthony and {Kron}, Richard and {Kronig}, Luzius and {Kueter-Young}, Andrea and {Lacey}, Cedric G. and {Lafever}, Robin and {Lahav}, Ofer and {Lambert}, Andrew and {Lampton}, Michael and {Landriau}, Martin and {Lang}, Dustin and {Lauer}, Tod R. and {Le Goff}, Jean-Marc and {Le Guillou}, Laurent and {Le Van Suu}, Auguste and {Lee}, Jae Hyeon and {Lee}, Su-Jeong and {Leitner}, Daniela and {Lesser}, Michael and {Levi}, Michael E. and {L'Huillier}, Benjamin and {Li}, Baojiu and {Liang}, Ming and {Lin}, Huan and {Linder}, Eric and {Loebman}, Sarah R. and {Luki{\'c}}, Zarija and {Ma}, Jun and {MacCrann}, Niall and {Magneville}, Christophe and {Makarem}, Laleh and {Manera}, Marc and {Manser}, Christopher J. and {Marshall}, Robert and {Martini}, Paul and {Massey}, Richard and {Matheson}, Thomas and {McCauley}, Jeremy and {McDonald}, Patrick and {McGreer}, Ian D. and {Meisner}, Aaron and {Metcalfe}, Nigel and {Miller}, Timothy N. and {Miquel}, Ramon and {Moustakas}, John and {Myers}, Adam and {Naik}, Milind and {Newman}, Jeffrey A. and {Nichol}, Robert C. and {Nicola}, Andrina and {Nicolati da Costa}, Luiz and {Nie}, Jundan and {Niz}, Gustavo and {Norberg}, Peder and {Nord}, Brian and {Norman}, Dara and {Nugent}, Peter and {O'Brien}, Thomas and {Oh}, Minji and {Olsen}, Knut A.~G.},
        title = "{The DESI Experiment Part II: Instrument Design}",
      journal = {arXiv e-prints},
         year = 2016,
        month = oct,
          eid = {arXiv:1611.00037},
        pages = {arXiv:1611.00037},
          doi = {10.48550/arXiv.1611.00037},
archivePrefix = {arXiv},
       eprint = {1611.00037},
 primaryClass = {astro-ph.IM},
       adsurl = {https://ui.adsabs.harvard.edu/abs/2016arXiv161100037D}
}

@ARTICLE{desi22,
       author = {{DESI Collaboration} and {Abareshi}, B. and {Aguilar}, J. and {Ahlen}, S. and {Alam}, Shadab and {Alexander}, David M. and {Alfarsy}, R. and {Allen}, L. and {Allende Prieto}, C. and {Alves}, O. and {Ameel}, J. and {Armengaud}, E. and {Asorey}, J. and {Aviles}, Alejandro and {Bailey}, S. and {Balaguera-Antol{\'\i}nez}, A. and {Ballester}, O. and {Baltay}, C. and {Bault}, A. and {Beltran}, S.~F. and {Benavides}, B. and {BenZvi}, S. and {Berti}, A. and {Besuner}, R. and {Beutler}, Florian and {Bianchi}, D. and {Blake}, C. and {Blanc}, P. and {Blum}, R. and {Bolton}, A. and {Bose}, S. and {Bramall}, D. and {Brieden}, S. and {Brodzeller}, A. and {Brooks}, D. and {Brownewell}, C. and {Buckley-Geer}, E. and {Cahn}, R.~N. and {Cai}, Z. and {Canning}, R. and {Capasso}, R. and {Carnero Rosell}, A. and {Carton}, P. and {Casas}, R. and {Castander}, F.~J. and {Cervantes-Cota}, J.~L. and {Chabanier}, S. and {Chaussidon}, E. and {Chuang}, C. and {Circosta}, C. and {Cole}, S. and {Cooper}, A.~P. and {da Costa}, L. and {Cousinou}, M. -C. and {Cuceu}, A. and {Davis}, T.~M. and {Dawson}, K. and {de la Cruz-Noriega}, R. and {de la Macorra}, A. and {de Mattia}, A. and {Della Costa}, J. and {Demmer}, P. and {Derwent}, M. and {Dey}, A. and {Dey}, B. and {Dhungana}, G. and {Ding}, Z. and {Dobson}, C. and {Doel}, P. and {Donald-McCann}, J. and {Donaldson}, J. and {Douglass}, K. and {Duan}, Y. and {Dunlop}, P. and {Edelstein}, J. and {Eftekharzadeh}, S. and {Eisenstein}, D.~J. and {Enriquez-Vargas}, M. and {Escoffier}, S. and {Evatt}, M. and {Fagrelius}, P. and {Fan}, X. and {Fanning}, K. and {Fawcett}, V.~A. and {Ferraro}, S. and {Ereza}, J. and {Flaugher}, B. and {Font-Ribera}, A. and {Forero-Romero}, J.~E. and {Frenk}, C.~S. and {Fromenteau}, S. and {G{\"a}nsicke}, B.~T. and {Garcia-Quintero}, C. and {Garrison}, L. and {Gazta{\~n}aga}, E. and {Gerardi}, F. and {Gil-Mar{\'\i}n}, H. and {Gontcho a Gontcho}, S. and {Gonzalez-Morales}, Alma X. and {Gonzalez-de-Rivera}, G. and {Gonzalez-Perez}, V. and {Gordon}, C. and {Graur}, O. and {Green}, D. and {Grove}, C. and {Gruen}, D. and {Gutierrez}, G. and {Guy}, J. and {Hahn}, C. and {Harris}, S. and {Herrera}, D. and {Herrera-Alcantar}, Hiram K. and {Honscheid}, K. and {Howlett}, C. and {Huterer}, D. and {Ir{\v{s}}i{\v{c}}}, V. and {Ishak}, M. and {Jelinsky}, P. and {Jiang}, L. and {Jimenez}, J. and {Jing}, Y.~P. and {Joyce}, R. and {Jullo}, E. and {Juneau}, S. and {Kara{\c{c}}ayl{\i}}, N.~G. and {Karamanis}, M. and {Karcher}, A. and {Karim}, T. and {Kehoe}, R. and {Kent}, S. and {Kirkby}, D. and {Kisner}, T. and {Kitaura}, F. and {Koposov}, S.~E. and {Kov{\'a}cs}, A. and {Kremin}, A. and {Krolewski}, Alex and {L'Huillier}, B. and {Lahav}, O. and {Lambert}, A. and {Lamman}, C. and {Lan}, Ting-Wen and {Landriau}, M. and {Lane}, S. and {Lang}, D. and {Lange}, J.~U. and {Lasker}, J. and {Le Guillou}, L. and {Leauthaud}, A. and {Le Van Suu}, A. and {Levi}, Michael E. and {Li}, T.~S. and {Magneville}, C. and {Manera}, M. and {Manser}, Christopher J. and {Marshall}, B. and {Martini}, Paul and {McCollam}, W. and {McDonald}, P. and {Meisner}, Aaron M. and {Mena-Fern{\'a}ndez}, J. and {Meneses-Rizo}, J. and {Mezcua}, M. and {Miller}, T. and {Miquel}, R. and {Montero-Camacho}, P. and {Moon}, J. and {Moustakas}, J. and {Mueller}, E. and {Mu{\~n}oz-Guti{\'e}rrez}, Andrea and {Myers}, Adam D. and {Nadathur}, S. and {Najita}, J. and {Napolitano}, L. and {Neilsen}, E. and {Newman}, Jeffrey A. and {Nie}, J.~D. and {Ning}, Y. and {Niz}, G. and {Norberg}, P. and {Noriega}, Hern{\'a}n E. and {O'Brien}, T. and {Obuljen}, A. and {Palanque-Delabrouille}, N. and {Palmese}, A. and {Zhiwei}, P. and {Pappalardo}, D. and {PENG}, X. and {Percival}, W.~J. and {Perruchot}, S. and {Pogge}, R. and {Poppett}, C. and {Porredon}, A. and {Prada}, F. and {Prochaska}, J. and {Pucha}, R. and {P{\'e}rez-Fern{\'a}ndez}, A. and {P{\'e}rez-R{\`a}fols}, I. and {Rabinowitz}, D. and {Raichoor}, A. and {Ramirez-Solano}, S. and {Ram{\'\i}rez-P{\'e}rez}, C{\'e}sar and {Ravoux}, C. and {Reil}, K. and {Rezaie}, M. and {Rocher}, A. and {Rockosi}, C. and {Roe}, N.~A. and {Roodman}, A. and {Ross}, A.~J. and {Rossi}, G. and {Ruggeri}, R. and {Ruhlmann-Kleider}, V. and {Sabiu}, C.~G. and {Safonova}, S. and {Said}, K. and {Saintonge}, A. and {Salas Catonga}, Javier and {Samushia}, L. and {Sanchez}, E. and {Saulder}, C. and {Schaan}, E. and {Schlafly}, E. and {Schlegel}, D. and {Schmoll}, J. and {Scholte}, D. and {Schubnell}, M. and {Secroun}, A. and {Seo}, H. and {Serrano}, S. and {Sharples}, Ray M. and {Sholl}, Michael J. and {Silber}, Joseph Harry and {Silva}, D.~R. and {Sirk}, M. and {Siudek}, M. and {Smith}, A. and {Sprayberry}, D. and {Staten}, R. and {Stupak}, B. and {Tan}, T. and {Tarl{\'e}}, Gregory and {Tie}, Suk Sien and {Tojeiro}, R. and {Ure{\~n}a-L{\'o}pez}, L.~A. and {Valdes}, F. and {Valenzuela}, O. and {Valluri}, M. and {Vargas-Maga{\~n}a}, M. and {Verde}, L. and {Walther}, M. and {Wang}, B. and {Wang}, M.~S. and {Weaver}, B.~A. and {Weaverdyck}, C. and {Wechsler}, R. and {Wilson}, Michael J. and {Yang}, J. and {Yu}, Y. and {Yuan}, S. and {Y{\`e}che}, Christophe and {Zhang}, H. and {Zhang}, K. and {Zhao}, Cheng and {Zhou}, Rongpu and {Zhou}, Zhimin and {Zou}, H. and {Zou}, J. and {Zou}, S. and {Zu}, Y. and {DESI Collaboration}},
        title = "{Overview of the Instrumentation for the Dark Energy Spectroscopic Instrument}",
      journal = {\aj},
         year = 2022,
        month = nov,
       volume = {164},
       number = {5},
          eid = {207},
        pages = {207},
          doi = {10.3847/1538-3881/ac882b},
archivePrefix = {arXiv},
       eprint = {2205.10939},
 primaryClass = {astro-ph.IM},
       adsurl = {https://ui.adsabs.harvard.edu/abs/2022AJ....164..207A}
}

@ARTICLE{desi24,
       author = {{DESI Collaboration} and {Adame}, A.~G. and {Aguilar}, J. and {Ahlen}, S. and {Alam}, S. and {Alexander}, D.~M. and {Allende Prieto}, C. and {Alvarez}, M. and {Alves}, O. and {Anand}, A. and {Andrade}, U. and {Armengaud}, E. and {Avila}, S. and {Aviles}, A. and {Awan}, H. and {Bahr-Kalus}, B. and {Bailey}, S. and {Baltay}, C. and {Bault}, A. and {Behera}, J. and {BenZvi}, S. and {Beutler}, F. and {Bianchi}, D. and {Blake}, C. and {Blum}, R. and {Bonici}, M. and {Brieden}, S. and {Brodzeller}, A. and {Brooks}, D. and {Buckley-Geer}, E. and {Burtin}, E. and {Calderon}, R. and {Canning}, R. and {Carnero Rosell}, A. and {Cereskaite}, R. and {Cervantes-Cota}, J.~L. and {Chabanier}, S. and {Chaussidon}, E. and {Chaves-Montero}, J. and {Chebat}, D. and {Chen}, S. and {Chen}, X. and {Claybaugh}, T. and {Cole}, S. and {Cuceu}, A. and {Davis}, T.~M. and {Dawson}, K. and {de la Macorra}, A. and {de Mattia}, A. and {Deiosso}, N. and {Dey}, A. and {Dey}, B. and {Ding}, Z. and {Doel}, P. and {Edelstein}, J. and {Eftekharzadeh}, S. and {Eisenstein}, D.~J. and {Elbers}, W. and {Elliott}, A. and {Fagrelius}, P. and {Fanning}, K. and {Ferraro}, S. and {Ereza}, J. and {Findlay}, N. and {Flaugher}, B. and {Font-Ribera}, A. and {Forero-S{\'a}nchez}, D. and {Forero-Romero}, J.~E. and {Frenk}, C.~S. and {Garcia-Quintero}, C. and {Garrison}, L.~H. and {Gazta{\~n}aga}, E. and {Gil-Mar{\'\i}n}, H. and {Gontcho}, S. Gontcho A. and {Gonzalez-Morales}, A.~X. and {Gonzalez-Perez}, V. and {Gordon}, C. and {Green}, D. and {Gruen}, D. and {Gsponer}, R. and {Gutierrez}, G. and {Guy}, J. and {Hadzhiyska}, B. and {Hahn}, C. and {Hanif}, M.~M.~S. and {Herrera-Alcantar}, H.~K. and {Honscheid}, K. and {Howlett}, C. and {Huterer}, D. and {Ir{\v{s}}i{\v{c}}}, V. and {Ishak}, M. and {Joyce}, R. and {Juneau}, S. and {Kara{\c{c}}ayl{\i}}, N.~G. and {Kehoe}, R. and {Kent}, S. and {Kirkby}, D. and {Kong}, H. and {Koposov}, S.~E. and {Kremin}, A. and {Krolewski}, A. and {Lahav}, O. and {Lai}, Y. and {Lan}, T.-W. and {Landriau}, M. and {Lang}, D. and {Lasker}, J. and {Le Goff}, J.~M. and {Le Guillou}, L. and {Leauthaud}, A. and {Levi}, M.~E. and {Li}, T.~S. and {Lodha}, K. and {Magneville}, C. and {Manera}, M. and {Margala}, D. and {Martini}, P. and {Matthewson}, W. and {Maus}, M. and {McDonald}, P. and {Medina-Varela}, L. and {Meisner}, A. and {Mena-Fern{\'a}ndez}, J. and {Miquel}, R. and {Moon}, J. and {Moore}, S. and {Moustakas}, J. and {Mudur}, N. and {Mueller}, E. and {Mu{\~n}oz-Guti{\'e}rrez}, A. and {Myers}, A.~D. and {Nadathur}, S. and {Napolitano}, L. and {Neveux}, R. and {Newman}, J.~A. and {Nguyen}, N.~M. and {Nie}, J. and {Niz}, G. and {Noriega}, H.~E. and {Padmanabhan}, N. and {Paillas}, E. and {Palanque-Delabrouille}, N. and {Pan}, J. and {Penmetsa}, S. and {Percival}, W.~J. and {Pieri}, M.~M. and {Pinon}, M. and {Poppett}, C. and {Porredon}, A. and {Prada}, F. and {P{\'e}rez-Fern{\'a}ndez}, A. and {P{\'e}rez-R{\`a}fols}, I. and {Rabinowitz}, D. and {Raichoor}, A. and {Ram{\'\i}rez-P{\'e}rez}, C. and {Ramirez-Solano}, S. and {Rashkovetskyi}, M. and {Ravoux}, C. and {Rezaie}, M. and {Rich}, J. and {Rocher}, A. and {Rockosi}, C. and {Roe}, N.~A. and {Rosado-Marin}, A. and {Ross}, A.~J. and {Rossi}, G. and {Ruggeri}, R. and {Ruhlmann-Kleider}, V. and {Samushia}, L. and {Sanchez}, E. and {Saulder}, C. and {Schlafly}, E.~F. and {Schlegel}, D. and {Schubnell}, M. and {Seo}, H. and {Shafieloo}, A. and {Sharples}, R. and {Silber}, J. and {Slosar}, A. and {Smith}, A. and {Sprayberry}, D. and {Tan}, T. and {Tarl{\'e}}, G. and {Taylor}, P. and {Trusov}, S. and {Vaisakh}, R. and {Valcin}, D. and {Valdes}, F. and {Valogiannis}, G. and {Vargas-Maga{\~n}a}, M. and {Verde}, L. and {Walther}, M. and {Wang}, B. and {Wang}, M.~S. and {Weaver}, B.~A. and {Weaverdyck}, N. and {Wechsler}, R.~H. and {Weinberg}, D.~H. and {White}, M. and {Wilson}, M.~J. and {Yi}, L.},
        title = "{DESI 2024 VII: cosmological constraints from the full-shape modeling of clustering measurements}",
      journal = {\jcap},
         year = 2025,
        month = jul,
       volume = {2025},
       number = {7},
          eid = {028},
        pages = {028},
          doi = {10.1088/1475-7516/2025/07/028},
archivePrefix = {arXiv},
       eprint = {2411.12022},
 primaryClass = {astro-ph.CO},
       adsurl = {https://ui.adsabs.harvard.edu/abs/2025JCAP...07..028A}
}

@ARTICLE{desi25,
       author = {{DESI Collaboration} and {Abdul Karim}, M. and {Adame}, A.~G. and {Aguado}, D. and {Aguilar}, J. and {Ahlen}, S. and {Alam}, S. and {Aldering}, G. and {Alexander}, D.~M. and {Alfarsy}, R. and {Allen}, L. and {Allende Prieto}, C. and {Alves}, O. and {Anand}, A. and {Andrade}, U. and {Armengaud}, E. and {Avila}, S. and {Aviles}, A. and {Awan}, H. and {Bailey}, S. and {Baleato Lizancos}, A. and {Ballester}, O. and {Bault}, A. and {Bautista}, J. and {Bean}, R. and {Behera}, J. and {BenZvi}, S. and {Beraldo e Silva}, L. and {Bermejo-Climent}, J.~R. and {Beutler}, F. and {Bianchi}, D. and {Blake}, C. and {Blum}, R. and {Bolton}, A.~S. and {Bonici}, M. and {Brieden}, S. and {Brodzeller}, A. and {Brooks}, D. and {Buckley-Geer}, E. and {Burtin}, E. and {Bystr{\"o}m}, A. and {Canning}, R. and {Carnero Rosell}, A. and {Carr}, A. and {Carrilho}, P. and {Casas}, L. and {Castander}, F.~J. and {Cereskaite}, R. and {Cervantes-Cota}, J.~L. and {Chaussidon}, E. and {Chaves-Montero}, J. and {Chen}, S. and {Chen}, X. and {Circosta}, C. and {Claybaugh}, T. and {Cole}, S. and {Cooper}, A.~P. and {Cousinou}, M.-C. and {Cuceu}, A. and {Davis}, T.~M. and {Dawson}, K.~S. and {de Belsunce}, R. and {de la Cruz}, R. and {de la Macorra}, A. and {de Mattia}, A. and {Deiosso}, N. and {Della Costa}, J. and {Demina}, R. and {Demirbozan}, U. and {DeRose}, J. and {Dey}, A. and {Dey}, B. and {Ding}, J. and {Ding}, Z. and {Doel}, P. and {Douglass}, K. and {Dowicz}, M. and {Ebina}, H. and {Edelstein}, J. and {Eisenstein}, D.~J. and {Elbers}, W. and {Emas}, N. and {Escoffier}, S. and {Fagrelius}, P. and {Fan}, X. and {Fanning}, K. and {Favole}, G. and {Fawcett}, V.~A. and {Fern{\'a}ndez-Garc{\'\i}a}, E. and {Ferraro}, S. and {Findlay}, N. and {Font-Ribera}, A. and {Forero-Romero}, J.~E. and {Forero-S{\'a}nchez}, D. and {Frenk}, C.~S. and {G{\"a}nsicke}, B.~T. and {Galbany}, L. and {Garc{\'\i}a-Bellido}, J. and {Garcia-Quintero}, C. and {Garrison}, L.~H. and {Gazta{\~n}aga}, E. and {Gil-Mar{\'\i}n}, H. and {Gloudemans}, A. and {Gnedin}, O.~Y. and {Gontcho A Gontcho}, S. and {Gonzalez}, D. and {Gonzalez-Morales}, A.~X. and {Gonzalez-Perez}, V. and {Gordon}, C. and {Graur}, O. and {Green}, D. and {Gruen}, D. and {Gsponer}, R. and {Guandalin}, C. and {Gutierrez}, G. and {Guy}, J. and {Hahn}, C. and {Han}, J.~J. and {Han}, J. and {He}, S. and {Herrera-Alcantar}, H.~K. and {Heydenreich}, S. and {Honscheid}, K. and {Hou}, J. and {Howlett}, C. and {Huterer}, D. and {Ir{\v{s}}i{\v{c}}}, V. and {Ishak}, M. and {Jacques}, A. and {Jiang}, L. and {Jimenez}, J. and {Jing}, Y.~P. and {Joachimi}, B. and {Joudaki}, S. and {Joyce}, R. and {Jullo}, E. and {Juneau}, S. and {Kara{\c{c}}ayl{\i}}, N.~G. and {Karim}, T. and {Kehoe}, R. and {Kent}, S. and {Khederlarian}, A. and {Kirkby}, D. and {Kisner}, T. and {Kitaura}, F.-S. and {Kizhuprakkat}, N. and {Kong}, H. and {Koposov}, S.~E. and {Kremin}, A. and {Krolewski}, A. and {Lahav}, O. and {Lai}, Y. and {Lamman}, C. and {Lan}, T.-W. and {Landriau}, M. and {Lang}, D. and {Lange}, J.~U. and {Lasker}, J. and {Le Goff}, J.~M. and {Le Guillou}, L. and {Leauthaud}, A. and {Levi}, M.~E. and {Li}, S. and {Li}, T.~S. and {Liu}, W. and {Lodha}, K. and {Lokken}, M. and {Luo}, Y. and {Magneville}, C. and {Manera}, M. and {Manser}, C.~J. and {Margala}, D. and {Martini}, P. and {Maus}, M. and {McCullough}, J. and {McDonald}, P. and {Medina}, G.~E. and {Medina-Varela}, L. and {Meisner}, A. and {Mena-Fern{\'a}ndez}, J. and {Menegas}, A. and {Meneses-Rizo}, J. and {Mezcua}, M. and {Miquel}, R. and {Montero-Camacho}, P. and {Moon}, J. and {Moustakas}, J. and {Mu{\~n}oz-Guti{\'e}rrez}, A. and {Mu noz-Santos}, D. and {Myers}, A.~D. and {Myles}, J. and {Nadathur}, S. and {Najita}, J. and {Napolitano}, L. and {Newman}, J.~A. and {Nikakhtar}, F. and {Nikutta}, R. and {Niz}, G. and {Noriega}, H.~E. and {Nugent}, P.},
        title = "{Data Release 1 of the Dark Energy Spectroscopic Instrument}",
      journal = {\aj},
         year = 2026,
        month = may,
       volume = {171},
       number = {5},
          eid = {285},
        pages = {285},
          doi = {10.3847/1538-3881/ae4c43},
archivePrefix = {arXiv},
       eprint = {2503.14745},
 primaryClass = {astro-ph.CO},
       adsurl = {https://ui.adsabs.harvard.edu/abs/2026AJ....171..285D}
}

@ARTICLE{desi25.2,
       author = {{DESI Collaboration} and {Adame}, A.~G. and {Aguilar}, J. and {Ahlen}, S. and {Alam}, S. and {Alexander}, D.~M. and {Allende Prieto}, C. and {Alvarez}, M. and {Alves}, O. and {Anand}, A. and {Andrade}, U. and {Armengaud}, E. and {Avila}, S. and {Aviles}, A. and {Awan}, H. and {Bahr-Kalus}, B. and {Bailey}, S. and {Baltay}, C. and {Bault}, A. and {Behera}, J. and {BenZvi}, S. and {Beutler}, F. and {Bianchi}, D. and {Blake}, C. and {Blum}, R. and {Bonici}, M. and {Brieden}, S. and {Brodzeller}, A. and {Brooks}, D. and {Buckley-Geer}, E. and {Burtin}, E. and {Calderon}, R. and {Canning}, R. and {Carnero Rosell}, A. and {Cereskaite}, R. and {Cervantes-Cota}, J.~L. and {Chabanier}, S. and {Chaussidon}, E. and {Chaves-Montero}, J. and {Chebat}, D. and {Chen}, S. and {Chen}, X. and {Claybaugh}, T. and {Cole}, S. and {Cuceu}, A. and {Davis}, T.~M. and {Dawson}, K. and {de la Macorra}, A. and {de Mattia}, A. and {Deiosso}, N. and {Dey}, A. and {Dey}, B. and {Ding}, Z. and {Doel}, P. and {Edelstein}, J. and {Eftekharzadeh}, S. and {Eisenstein}, D.~J. and {Elbers}, W. and {Elliott}, A. and {Fagrelius}, P. and {Fanning}, K. and {Ferraro}, S. and {Ereza}, J. and {Findlay}, N. and {Flaugher}, B. and {Font-Ribera}, A. and {Forero-S{\'a}nchez}, D. and {Forero-Romero}, J.~E. and {Frenk}, C.~S. and {Garcia-Quintero}, C. and {Garrison}, L.~H. and {Gazta{\~n}aga}, E. and {Gil-Mar{\'\i}n}, H. and {Gontcho}, S. Gontcho A. and {Gonzalez-Morales}, A.~X. and {Gonzalez-Perez}, V. and {Gordon}, C. and {Green}, D. and {Gruen}, D. and {Gsponer}, R. and {Gutierrez}, G. and {Guy}, J. and {Hadzhiyska}, B. and {Hahn}, C. and {Hanif}, M.~M.~S. and {Herrera-Alcantar}, H.~K. and {Honscheid}, K. and {Howlett}, C. and {Huterer}, D. and {Ir{\v{s}}i{\v{c}}}, V. and {Ishak}, M. and {Joyce}, R. and {Juneau}, S. and {Kara{\c{c}}ayl{\i}}, N.~G. and {Kehoe}, R. and {Kent}, S. and {Kirkby}, D. and {Kong}, H. and {Koposov}, S.~E. and {Kremin}, A. and {Krolewski}, A. and {Lahav}, O. and {Lai}, Y. and {Lan}, T. -W. and {Landriau}, M. and {Lang}, D. and {Lasker}, J. and {Le Goff}, J.~M. and {Le Guillou}, L. and {Leauthaud}, A. and {Levi}, M.~E. and {Li}, T.~S. and {Lodha}, K. and {Magneville}, C. and {Manera}, M. and {Margala}, D. and {Martini}, P. and {Matthewson}, W. and {Maus}, M. and {McDonald}, P. and {Medina-Varela}, L. and {Meisner}, A. and {Mena-Fern{\'a}ndez}, J. and {Miquel}, R. and {Moon}, J. and {Moore}, S. and {Moustakas}, J. and {Mudur}, N. and {Mueller}, E. and {Mu{\~n}oz-Guti{\'e}rrez}, A. and {Myers}, A.~D. and {Nadathur}, S. and {Napolitano}, L. and {Neveux}, R. and {Newman}, J.~A. and {Nguyen}, N.~M. and {Nie}, J. and {Niz}, G. and {Noriega}, H.~E. and {Padmanabhan}, N. and {Paillas}, E. and {Palanque-Delabrouille}, N. and {Pan}, J. and {Penmetsa}, S. and {Percival}, W.~J. and {Pieri}, M.~M. and {Pinon}, M. and {Poppett}, C. and {Porredon}, A. and {Prada}, F. and {P{\'e}rez-Fern{\'a}ndez}, A. and {P{\'e}rez-R{\`a}fols}, I. and {Rabinowitz}, D. and {Raichoor}, A. and {Ram{\'\i}rez-P{\'e}rez}, C. and {Ramirez-Solano}, S. and {Rashkovetskyi}, M. and {Ravoux}, C. and {Rezaie}, M. and {Rich}, J. and {Rocher}, A. and {Rockosi}, C. and {Roe}, N.~A. and {Rosado-Marin}, A. and {Ross}, A.~J. and {Rossi}, G. and {Ruggeri}, R. and {Ruhlmann-Kleider}, V. and {Samushia}, L. and {Sanchez}, E. and {Saulder}, C. and {Schlafly}, E.~F. and {Schlegel}, D. and {Schubnell}, M. and {Seo}, H. and {Shafieloo}, A. and {Sharples}, R. and {Silber}, J. and {Slosar}, A. and {Smith}, A. and {Sprayberry}, D. and {Tan}, T. and {Tarl{\'e}}, G. and {Taylor}, P. and {Trusov}, S. and {Vaisakh}, R. and {Valcin}, D. and {Valdes}, F. and {Valogiannis}, G. and {Vargas-Maga{\~n}a}, M. and {Verde}, L. and {Walther}, M. and {Wang}, B. and {Wang}, M.~S. and {Weaver}, B.~A. and {Weaverdyck}, N. and {Wechsler}, R.~H. and {Weinberg}, D.~H. and {White}, M. and {Wilson}, M.~J. and {Yi}, L.},
        title = "{DESI 2024 VII: cosmological constraints from the full-shape modeling of clustering measurements}",
      journal = {\jcap},
         year = 2025,
        month = jul,
       volume = {2025},
       number = {7},
          eid = {028},
        pages = {028},
          doi = {10.1088/1475-7516/2025/07/028},
archivePrefix = {arXiv},
       eprint = {2411.12022},
 primaryClass = {astro-ph.CO},
       adsurl = {https://ui.adsabs.harvard.edu/abs/2025JCAP...07..028A}
}

@ARTICLE{desi25.3,
  title = {DESI DR2 results. II. Measurements of baryon acoustic oscillations and cosmological constraints},
  author = {Abdul Karim, M. and Aguilar, J. and Ahlen, S. and Alam, S. and Allen, L. and Prieto, C. Allende and Alves, O. and Anand, A. and Andrade, U. and Armengaud, E. and Aviles, A. and Bailey, S. and Baltay, C. and Bansal, P. and Bault, A. and Behera, J. and BenZvi, S. and Bianchi, D. and Blake, C. and Brieden, S. and Brodzeller, A. and Brooks, D. and Buckley-Geer, E. and Burtin, E. and Calderon, R. and Canning, R. and Rosell, A. Carnero and Carrilho, P. and Casas, L. and Castander, F. J. and Charles, M. and Chaussidon, E. and Chaves-Montero, J. and Chebat, D. and Chen, X. and Claybaugh, T. and Cole, S. and Cooper, A. P. and Cuceu, A. and Dawson, K. S. and de la Macorra, A. and de Mattia, A. and Deiosso, N. and Della Costa, J. and Demina, R. and Dey, A. and Dey, B. and Ding, Z. and Doel, P. and Edelstein, J. and Eisenstein, D. J. and Elbers, W. and Fagrelius, P. and Fanning, K. and Fern\'andez-Garc\'{\i}a, E. and Ferraro, S. and Font-Ribera, A. and Forero-Romero, J. E. and Frenk, C. S. and Garcia-Quintero, C. and Garrison, L. H. and Gazta\~naga, E. and Gil-Mar\'{\i}n, H. and Gontcho, S. Gontcho A. and Gonzalez, D. and Gonzalez-Morales, A. X. and Gordon, C. and Green, D. and Gutierrez, G. and Guy, J. and Hadzhiyska, B. and Hahn, C. and He, S. and Herbold, M. and Herrera-Alcantar, H. K. and Ho, M.-F. and Honscheid, K. and Howlett, C. and Huterer, D. and Ishak, M. and Juneau, S. and Kamble, N. V. and Kara\ifmmode \mbox{\c{c}}\else \c{c}\fi{}ayl��, N. G. and Kehoe, R. and Kent, S. and Kim, A. G. and Kirkby, D. and Kisner, T. and Koposov, S. E. and Kremin, A. and Krolewski, A. and Lahav, O. and Lamman, C. and Landriau, M. and Lang, D. and Lasker, J. and Le Goff, J. M. and Le Guillou, L. and Leauthaud, A. and Levi, M. E. and Li, Q. and Li, T. S. and Lodha, K. and Lokken, M. and Lozano-Rodr\'{\i}guez, F. and Magneville, C. and Manera, M. and Martini, P. and Matthewson, W. L. and Meisner, A. and Mena-Fern\'andez, J. and Menegas, A. and Mergulh\~ao, T. and Miquel, R. and Moustakas, J. and Mu\~noz-Guti\'errez, A. and Mu\~noz-Santos, D. and Myers, A. D. and Nadathur, S. and Naidoo, K. and Napolitano, L. and Newman, J. A. and Niz, G. and Noriega, H. E. and Paillas, E. and Palanque-Delabrouille, N. and Pan, J. and Peacock, J. A. and Ibanez, M. P. and Percival, W. J. and P\'erez-Fern\'andez, A. and P\'erez-R\`afols, I. and Pieri, M. M. and Poppett, C. and Prada, F. and Rabinowitz, D. and Raichoor, A. and Ram\'{\i}rez-P\'erez, C. and Rashkovetskyi, M. and Ravoux, C. and Rich, J. and Rocher, A. and Rockosi, C. and Rohlf, J. and Rom\'an-Herrera, J. O. and Ross, A. J. and Rossi, G. and Ruggeri, R. and Ruhlmann-Kleider, V. and Samushia, L. and Sanchez, E. and Sanders, N. and Schlegel, D. and Schubnell, M. and Seo, H. and Shafieloo, A. and Sharples, R. and Silber, J. and Sinigaglia, F. and Sprayberry, D. and Tan, T. and Tarl\'e, G. and Taylor, P. and Turner, W. and Ure\~na-L\'opez, L. A. and Vaisakh, R. and Valdes, F. and Valogiannis, G. and Vargas-Maga\~na, M. and Verde, L. and Walther, M. and Weaver, B. A. and Weinberg, D. H. and White, M. and Wolfson, M. and Y\`eche, C. and Yu, J. and Zaborowski, E. A. and Zarrouk, P. and Zhai, Z. and Zhang, H. and Zhao, C. and Zhao, G. B. and Zhou, R. and Zou, H.},
  collaboration = {DESI Collaboration},
  journal = {Phys. Rev. D},
  volume = {112},
  issue = {8},
  pages = {083515},
  numpages = {40},
  year = {2025},
  month = {Oct},
  publisher = {American Physical Society},
  doi = {10.1103/tr6y-kpc6},
  url = {https://link.aps.org/doi/10.1103/tr6y-kpc6},
  archivePrefix = {arXiv},
  eprint = {2503.14738},
  primaryClass = {astro-ph.CO}
}

@ARTICLE{dey19,
       author = {{Dey}, Arjun and {Schlegel}, David J. and {Lang}, Dustin and {Blum}, Robert and {Burleigh}, Kaylan and {Fan}, Xiaohui and {Findlay}, Joseph R. and {Finkbeiner}, Doug and {Herrera}, David and {Juneau}, St{\'e}phanie and {Landriau}, Martin and {Levi}, Michael and {McGreer}, Ian and {Meisner}, Aaron and {Myers}, Adam D. and {Moustakas}, John and {Nugent}, Peter and {Patej}, Anna and {Schlafly}, Edward F. and {Walker}, Alistair R. and {Valdes}, Francisco and {Weaver}, Benjamin A. and {Y{\`e}che}, Christophe and {Zou}, Hu and {Zhou}, Xu and {Abareshi}, Behzad and {Abbott}, T.~M.~C. and {Abolfathi}, Bela and {Aguilera}, C. and {Alam}, Shadab and {Allen}, Lori and {Alvarez}, A. and {Annis}, James and {Ansarinejad}, Behzad and {Aubert}, Marie and {Beechert}, Jacqueline and {Bell}, Eric F. and {BenZvi}, Segev Y. and {Beutler}, Florian and {Bielby}, Richard M. and {Bolton}, Adam S. and {Brice{\~n}o}, C{\'e}sar and {Buckley-Geer}, Elizabeth J. and {Butler}, Karen and {Calamida}, Annalisa and {Carlberg}, Raymond G. and {Carter}, Paul and {Casas}, Ricard and {Castander}, Francisco J. and {Choi}, Yumi and {Comparat}, Johan and {Cukanovaite}, Elena and {Delubac}, Timoth{\'e}e and {DeVries}, Kaitlin and {Dey}, Sharmila and {Dhungana}, Govinda and {Dickinson}, Mark and {Ding}, Zhejie and {Donaldson}, John B. and {Duan}, Yutong and {Duckworth}, Christopher J. and {Eftekharzadeh}, Sarah and {Eisenstein}, Daniel J. and {Etourneau}, Thomas and {Fagrelius}, Parker A. and {Farihi}, Jay and {Fitzpatrick}, Mike and {Font-Ribera}, Andreu and {Fulmer}, Leah and {G{\"a}nsicke}, Boris T. and {Gaztanaga}, Enrique and {George}, Koshy and {Gerdes}, David W. and {Gontcho}, Satya Gontcho A. and {Gorgoni}, Claudio and {Green}, Gregory and {Guy}, Julien and {Harmer}, Diane and {Hernandez}, M. and {Honscheid}, Klaus and {Huang}, Lijuan Wendy and {James}, David J. and {Jannuzi}, Buell T. and {Jiang}, Linhua and {Joyce}, Richard and {Karcher}, Armin and {Karkar}, Sonia and {Kehoe}, Robert and {Kneib}, Jean-Paul and {Kueter-Young}, Andrea and {Lan}, Ting-Wen and {Lauer}, Tod R. and {Le Guillou}, Laurent and {Le Van Suu}, Auguste and {Lee}, Jae Hyeon and {Lesser}, Michael and {Perreault Levasseur}, Laurence and {Li}, Ting S. and {Mann}, Justin L. and {Marshall}, Robert and {Mart{\'\i}nez-V{\'a}zquez}, C.~E. and {Martini}, Paul and {du Mas des Bourboux}, H{\'e}lion and {McManus}, Sean and {Meier}, Tobias Gabriel and {M{\'e}nard}, Brice and {Metcalfe}, Nigel and {Mu{\~n}oz-Guti{\'e}rrez}, Andrea and {Najita}, Joan and {Napier}, Kevin and {Narayan}, Gautham and {Newman}, Jeffrey A. and {Nie}, Jundan and {Nord}, Brian and {Norman}, Dara J. and {Olsen}, Knut A.~G. and {Paat}, Anthony and {Palanque-Delabrouille}, Nathalie and {Peng}, Xiyan and {Poppett}, Claire L. and {Poremba}, Megan R. and {Prakash}, Abhishek and {Rabinowitz}, David and {Raichoor}, Anand and {Rezaie}, Mehdi and {Robertson}, A.~N. and {Roe}, Natalie A. and {Ross}, Ashley J. and {Ross}, Nicholas P. and {Rudnick}, Gregory and {Safonova}, Sasha and {Saha}, Abhijit and {S{\'a}nchez}, F. Javier and {Savary}, Elodie and {Schweiker}, Heidi and {Scott}, Adam and {Seo}, Hee-Jong and {Shan}, Huanyuan and {Silva}, David R. and {Slepian}, Zachary and {Soto}, Christian and {Sprayberry}, David and {Staten}, Ryan and {Stillman}, Coley M. and {Stupak}, Robert J. and {Summers}, David L. and {Sien Tie}, Suk and {Tirado}, H. and {Vargas-Maga{\~n}a}, Mariana and {Vivas}, A. Katherina and {Wechsler}, Risa H. and {Williams}, Doug and {Yang}, Jinyi and {Yang}, Qian and {Yapici}, Tolga and {Zaritsky}, Dennis and {Zenteno}, A. and {Zhang}, Kai and {Zhang}, Tianmeng and {Zhou}, Rongpu and {Zhou}, Zhimin},
        title = "{Overview of the DESI Legacy Imaging Surveys}",
      journal = {\aj},
         year = 2019,
        month = may,
       volume = {157},
       number = {5},
          eid = {168},
        pages = {168},
          doi = {10.3847/1538-3881/ab089d},
archivePrefix = {arXiv},
       eprint = {1804.08657},
 primaryClass = {astro-ph.IM},
       adsurl = {https://ui.adsabs.harvard.edu/abs/2019AJ....157..168D}
}

@ARTICLE{gallardo25,
       author = {{Gallardo}, Patricio A. and {Gong}, Yulin and {Hadzhiyska}, Boryana and {Hsu}, Yun-Hsin},
        title = "{Iskay2: Signal Extraction of the Kinematic Sunyaev─Zel'dovich Effect Through the Pairwise Estimator: Pipeline and Validation}",
      journal = {Research Notes of the American Astronomical Society},
         year = 2025,
        month = oct,
       volume = {9},
       number = {10},
          eid = {284},
        pages = {284},
          doi = {10.3847/2515-5172/ae1587},
archivePrefix = {arXiv},
       eprint = {2510.20715},
 primaryClass = {astro-ph.CO},
       adsurl = {https://ui.adsabs.harvard.edu/abs/2025RNAAS...9..284G}
}

@ARTICLE{gong24,
       author = {{Gong}, Yulin and {Bean}, Rachel and {Gallardo}, Patricio A. and {Vavagiakis}, Eve M. and {Battaglia}, Nicholas and {Niemack}, Michael},
        title = "{Pairwise kinematic Sunyaev-Zel'dovich signal extraction efficacy and optical depth estimation}",
      journal = {\prd},
         year = 2024,
        month = jan,
       volume = {109},
       number = {2},
          eid = {023513},
        pages = {023513},
          doi = {10.1103/PhysRevD.109.023513},
archivePrefix = {arXiv},
       eprint = {2307.11894},
 primaryClass = {astro-ph.CO},
       adsurl = {https://ui.adsabs.harvard.edu/abs/2024PhRvD.109b3513G}
}

@ARTICLE{gong25,
       author = {{Gong}, Yulin and {Bean}, Rachel},
        title = "{Cluster optical depth and pairwise velocity estimation using machine learning}",
      journal = {\prd},
         year = 2025,
        month = jul,
       volume = {112},
       number = {2},
          eid = {023527},
        pages = {023527},
          doi = {10.1103/d6tz-tdb8},
archivePrefix = {arXiv},
       eprint = {2505.12720},
 primaryClass = {astro-ph.CO},
       adsurl = {https://ui.adsabs.harvard.edu/abs/2025PhRvD.112b3527G}
}

@ARTICLE{gong26,
       author = {{Gong}, Yulin and {Gallardo}, Patricio A. and {Bean}, Rachel and {Moore}, Jenna and {Vavagiakis}, Eve M. and {Battaglia}, Nicholas and {Hadzhiyska}, Boryana and {Hsu}, Yun-Hsin and {Aguilar}, Jessica Nicole and {Ahlen}, Steven and {Bianchi}, Davide and {Brooks}, David and {Claybaugh}, Todd and {Canning}, Rebecca and {Devlin}, Mark and {Doel}, Peter and {de la Macorra}, Axel and {Ferraro}, Simone and {Font-Ribera}, Andreu and {Forero-Romero}, Jaime E. and {Gazta{\~n}aga}, Enrique and {Gutierrez}, Gaston and {Gontcho}, Satya Gontcho A. and {Guy}, Julien and {Honscheid}, Klaus and {Howlett}, Cullan and {Liu}, R. Henry and {Ishak}, Mustapha and {Joyce}, Dick and {Kremin}, Anthony and {Lamman}, Claire and {Levi}, Michael and {Landriau}, Martin and {Manera}, Marc and {Meisner}, Aaron and {Miquel}, Ramon and {Niemack}, Michael D. and {Nadathur}, Seshadri and {Percival}, Will and {Prada}, Francisco and {Rossi}, Graziano and {Guachalla}, Bernardita Ried and {Sanchez}, Eusebio and {Seo}, Hee-Jong and {Sprayberry}, David and {Schlegel}, David and {Sif{\'o}n}, Crist{\'o}bal and {Schubnell}, Michael and {Silber}, Joseph Harry and {Tarl{\'e}}, Gregory and {Weaver}, Benjamin Alan and {Zhou}, Rongpu and {Zou}, Hu},
        title = "{Detection of the pairwise kinematic Sunyaev-Zel'dovich effect and pairwise velocity with DESI DR1 galaxies and ACT DR6 and Planck CMB data}",
      journal = {\prd},
         year = 2026,
        month = mar,
       volume = {113},
       number = {6},
          eid = {063538},
        pages = {063538},
          doi = {10.1103/kkrh-1svk},
archivePrefix = {arXiv},
       eprint = {2511.23417},
 primaryClass = {astro-ph.CO},
       adsurl = {https://ui.adsabs.harvard.edu/abs/2026PhRvD.113f3538G}
}

@ARTICLE{grayson23,
       author = {{Grayson}, Skylar and {Scannapieco}, Evan and {Dav{\'e}}, Romeel},
        title = "{Distinguishing Active Galactic Nuclei Feedback Models with the Thermal Sunyaev-Zel'dovich Effect}",
      journal = {\apj},
         year = 2023,
        month = nov,
       volume = {957},
       number = {1},
          eid = {17},
        pages = {17},
          doi = {10.3847/1538-4357/acfd26},
archivePrefix = {arXiv},
       eprint = {2310.01502},
 primaryClass = {astro-ph.GA},
       adsurl = {https://ui.adsabs.harvard.edu/abs/2023ApJ...957...17G}
}

@ARTICLE{guy23,
       author = {{Guy}, J. and {Bailey}, S. and {Kremin}, A. and {Alam}, Shadab and {Alexander}, D.~M. and {Allende Prieto}, C. and {BenZvi}, S. and {Bolton}, A.~S. and {Brooks}, D. and {Chaussidon}, E. and {Cooper}, A.~P. and {Dawson}, K. and {de la Macorra}, A. and {Dey}, A. and {Dey}, Biprateep and {Dhungana}, G. and {Eisenstein}, D.~J. and {Font-Ribera}, A. and {Forero-Romero}, J.~E. and {Gazta{\~n}aga}, E. and {Gontcho A Gontcho}, S. and {Green}, D. and {Honscheid}, K. and {Ishak}, M. and {Kehoe}, R. and {Kirkby}, D. and {Kisner}, T. and {Koposov}, Sergey E. and {Lan}, Ting-Wen and {Landriau}, M. and {Le Guillou}, L. and {Levi}, Michael E. and {Magneville}, C. and {Manser}, Christopher J. and {Martini}, P. and {Meisner}, Aaron M. and {Miquel}, R. and {Moustakas}, J. and {Myers}, Adam D. and {Newman}, Jeffrey A. and {Nie}, Jundan and {Palanque-Delabrouille}, N. and {Percival}, W.~J. and {Poppett}, C. and {Prada}, F. and {Raichoor}, A. and {Ravoux}, C. and {Ross}, A.~J. and {Schlafly}, E.~F. and {Schlegel}, D. and {Schubnell}, M. and {Sharples}, Ray M. and {Tarl{\'e}}, Gregory and {Weaver}, B.~A. and {Y{\'e}che}, Christophe and {Zhou}, Rongpu and {Zhou}, Zhimin and {Zou}, H.},
        title = "{The Spectroscopic Data Processing Pipeline for the Dark Energy Spectroscopic Instrument}",
      journal = {\aj},
         year = 2023,
        month = apr,
       volume = {165},
       number = {4},
          eid = {144},
        pages = {144},
          doi = {10.3847/1538-3881/acb212},
archivePrefix = {arXiv},
       eprint = {2209.14482},
 primaryClass = {astro-ph.IM},
       adsurl = {https://ui.adsabs.harvard.edu/abs/2023AJ....165..144G}
}

@ARTICLE{hadzhiyska25.2,
       author = {{Hadzhiyska}, B. and {Ferraro}, S. and {Ried Guachalla}, B. and {Schaan}, E. and {Aguilar}, J. and {Ahlen}, S. and {Battaglia}, N. and {Bond}, J.~R. and {Brooks}, D. and {Calabrese}, E. and {Choi}, S.~K. and {Claybaugh}, T. and {Coulton}, W.~R. and {Dawson}, K. and {Devlin}, M. and {Dey}, B. and {Doel}, P. and {Duivenvoorden}, A.~J. and {Dunkley}, J. and {Farren}, G.~S. and {Font-Ribera}, A. and {Forero-Romero}, J.~E. and {Gallardo}, P.~A. and {Gazta{\~n}aga}, E. and {Gontcho Gontcho}, S. and {Gralla}, M. and {Le Guillou}, L. and {Gutierrez}, G. and {Guy}, J. and {Hill}, J.~C. and {Hlo{\v{z}}ek}, R. and {Honscheid}, K. and {Juneau}, S. and {Kehoe}, R. and {Kisner}, T. and {Kremin}, A. and {Landriau}, M. and {Liu}, R.~H. and {Louis}, T. and {MacCrann}, N. and {de Macorra}, A. and {Madhavacheril}, M. and {Manera}, M. and {Meisner}, A. and {Miquel}, R. and {Moodley}, K. and {Moustakas}, J. and {Mroczkowski}, T. and {Naess}, S. and {Newman}, J. and {Niemack}, M.~D. and {Niz}, G. and {Page}, L. and {Palanque-Delabrouille}, N. and {Partridge}, B. and {Percival}, W.~J. and {Prada}, F. and {Qu}, F.~J. and {Rossi}, G. and {Sanchez}, E. and {Schlegel}, D. and {Schubnell}, M. and {Sherwin}, B. and {Sehgal}, N. and {Seo}, H. and {Sif{\'o}n}, C. and {Spergel}, D. and {Sprayberry}, D. and {Staggs}, S. and {Tarl{\'e}}, G. and {Vargas}, C. and {Vavagiakis}, E.~M. and {Weaver}, B.~A. and {Wollack}, E.~J. and {Zhou}, R. and {Zou}, H.},
        title = "{Evidence for large baryonic feedback at low and intermediate redshifts from kinematic Sunyaev-Zel'dovich observations with ACT and DESI photometric galaxies}",
      journal = {\prd},
         year = 2025,
        month = oct,
       volume = {112},
       number = {8},
          eid = {083509},
        pages = {083509},
          doi = {10.1103/kclp-x5j1},
archivePrefix = {arXiv},
       eprint = {2407.07152},
 primaryClass = {astro-ph.CO},
       adsurl = {https://ui.adsabs.harvard.edu/abs/2025PhRvD.112h3509H}
}

@ARTICLE{hadzhiyska25,
       author = {{Hadzhiyska}, B. and {Gong}, Y. and {Hsu}, Y. and {Gallardo}, P.~A. and {Aguilar}, J. and {Ahlen}, S. and {Alonso}, D. and {Bean}, R. and {Bianchi}, D. and {Brooks}, D. and {Castander}, F.~J. and {Claybaugh}, T. and {Cole}, S. and {Cuceu}, A. and {de la Macorra}, A. and {Dey}, Arjun and {Ferraro}, S. and {Font-Ribera}, A. and {Forero-Romero}, J.~E. and {Gontcho}, S. Gontcho A. and {Gutierrez}, G. and {Guy}, J. and {Herrera-Alcantar}, H.~K. and {Howlett}, C. and {Huterer}, D. and {Ishak}, M. and {Joyce}, R. and {Kisner}, T. and {Kremin}, A. and {Landriau}, M. and {Le Guillou}, L. and {Levi}, M.~E. and {Manera}, M. and {Meisner}, A. and {Miquel}, R. and {Moodley}, K. and {Mroczkowski}, T. and {Nadathur}, S. and {Palanque-Delabrouille}, N. and {Percival}, W.~J. and {Prada}, F. and {Qu}, F.~J. and {P{\'e}rez-R{\`a}fols}, I. and {Guachalla}, B. Ried and {Rossi}, G. and {Sanchez}, E. and {Schaan}, E. and {Schlegel}, D. and {Schubnell}, M. and {Seo}, H. and {Sif{\'o}n}, C. and {Silber}, J. and {Sprayberry}, D. and {Tarl{\'e}}, G. and {Vavagiakis}, E.~M. and {Weaver}, B.~A. and {Zhou}, R. and {Zou}, H.},
        title = "{Probing cosmic velocities with the pairwise kinematic Sunyaev-Zel'dovich signal in DESI Bright Galaxy Sample DR1 and ACT DR6}",
      journal = {\prd},
         year = 2026,
        month = mar,
       volume = {113},
       number = {6},
          eid = {063565},
        pages = {063565},
          doi = {10.1103/9h6g-gxdn},
archivePrefix = {arXiv},
       eprint = {2510.14135},
 primaryClass = {astro-ph.CO},
       adsurl = {https://ui.adsabs.harvard.edu/abs/2026PhRvD.113f3565H}
}

@ARTICLE{hand12,
       author = {{Hand}, Nick and {Addison}, Graeme E. and {Aubourg}, Eric and {Battaglia}, Nick and {Battistelli}, Elia S. and {Bizyaev}, Dmitry and {Bond}, J. Richard and {Brewington}, Howard and {Brinkmann}, Jon and {Brown}, Benjamin R. and {Das}, Sudeep and {Dawson}, Kyle S. and {Devlin}, Mark J. and {Dunkley}, Joanna and {Dunner}, Rolando and {Eisenstein}, Daniel J. and {Fowler}, Joseph W. and {Gralla}, Megan B. and {Hajian}, Amir and {Halpern}, Mark and {Hilton}, Matt and {Hincks}, Adam D. and {Hlozek}, Ren{\'e}e and {Hughes}, John P. and {Infante}, Leopoldo and {Irwin}, Kent D. and {Kosowsky}, Arthur and {Lin}, Yen-Ting and {Malanushenko}, Elena and {Malanushenko}, Viktor and {Marriage}, Tobias A. and {Marsden}, Danica and {Menanteau}, Felipe and {Moodley}, Kavilan and {Niemack}, Michael D. and {Nolta}, Michael R. and {Oravetz}, Daniel and {Page}, Lyman A. and {Palanque-Delabrouille}, Nathalie and {Pan}, Kaike and {Reese}, Erik D. and {Schlegel}, David J. and {Schneider}, Donald P. and {Sehgal}, Neelima and {Shelden}, Alaina and {Sievers}, Jon and {Sif{\'o}n}, Crist{\'o}bal and {Simmons}, Audrey and {Snedden}, Stephanie and {Spergel}, David N. and {Staggs}, Suzanne T. and {Swetz}, Daniel S. and {Switzer}, Eric R. and {Trac}, Hy and {Weaver}, Benjamin A. and {Wollack}, Edward J. and {Yeche}, Christophe and {Zunckel}, Caroline},
        title = "{Evidence of Galaxy Cluster Motions with the Kinematic Sunyaev-Zel'dovich Effect}",
      journal = {\prl},
         year = 2012,
        month = jul,
       volume = {109},
       number = {4},
          eid = {041101},
        pages = {041101},
          doi = {10.1103/PhysRevLett.109.041101},
archivePrefix = {arXiv},
       eprint = {1203.4219},
 primaryClass = {astro-ph.CO},
       adsurl = {https://ui.adsabs.harvard.edu/abs/2012PhRvL.109d1101H}
}

@ARTICLE{henderson16,
       author = {{Henderson}, S.~W. and {Allison}, R. and {Austermann}, J. and {Baildon}, T. and {Battaglia}, N. and {Beall}, J.~A. and {Becker}, D. and {De Bernardis}, F. and {Bond}, J.~R. and {Calabrese}, E. and {Choi}, S.~K. and {Coughlin}, K.~P. and {Crowley}, K.~T. and {Datta}, R. and {Devlin}, M.~J. and {Duff}, S.~M. and {Dunkley}, J. and {D{\"u}nner}, R. and {van Engelen}, A. and {Gallardo}, P.~A. and {Grace}, E. and {Hasselfield}, M. and {Hills}, F. and {Hilton}, G.~C. and {Hincks}, A.~D. and {Hloẑek}, R. and {Ho}, S.~P. and {Hubmayr}, J. and {Huffenberger}, K. and {Hughes}, J.~P. and {Irwin}, K.~D. and {Koopman}, B.~J. and {Kosowsky}, A.~B. and {Li}, D. and {McMahon}, J. and {Munson}, C. and {Nati}, F. and {Newburgh}, L. and {Niemack}, M.~D. and {Niraula}, P. and {Page}, L.~A. and {Pappas}, C.~G. and {Salatino}, M. and {Schillaci}, A. and {Schmitt}, B.~L. and {Sehgal}, N. and {Sherwin}, B.~D. and {Sievers}, J.~L. and {Simon}, S.~M. and {Spergel}, D.~N. and {Staggs}, S.~T. and {Stevens}, J.~R. and {Thornton}, R. and {Van Lanen}, J. and {Vavagiakis}, E.~M. and {Ward}, J.~T. and {Wollack}, E.~J.},
        title = "{Advanced ACTPol Cryogenic Detector Arrays and Readout}",
      journal = {Journal of Low Temperature Physics},
         year = 2016,
        month = aug,
       volume = {184},
       number = {3-4},
        pages = {772-779},
          doi = {10.1007/s10909-016-1575-z},
archivePrefix = {arXiv},
       eprint = {1510.02809},
 primaryClass = {astro-ph.IM},
       adsurl = {https://ui.adsabs.harvard.edu/abs/2016JLTP..184..772H}
}

@ARTICLE{hill18,
       author = {{Hill}, J. Colin and {Baxter}, Eric J. and {Lidz}, Adam and {Greco}, Johnny P. and {Jain}, Bhuvnesh},
        title = "{Two-halo term in stacked thermal Sunyaev-Zel'dovich measurements: Implications for self-similarity}",
      journal = {\prd},
         year = 2018,
        month = apr,
       volume = {97},
       number = {8},
          eid = {083501},
        pages = {083501},
          doi = {10.1103/PhysRevD.97.083501},
archivePrefix = {arXiv},
       eprint = {1706.03753},
 primaryClass = {astro-ph.CO},
       adsurl = {https://ui.adsabs.harvard.edu/abs/2018PhRvD..97h3501H}
}

@unpublished{hsu26,
        author = {{Hsu}, Y-H. and others},
        year = {2026},
        note = {in preparation}
}

@ARTICLE{jubelgas08,
       author = {{Jubelgas}, M. and {Springel}, V. and {En{\ss}lin}, T. and {Pfrommer}, C.},
        title = "{Cosmic ray feedback in hydrodynamical simulations of galaxy formation}",
      journal = {\aap},
         year = 2008,
        month = apr,
       volume = {481},
       number = {1},
        pages = {33-63},
          doi = {10.1051/0004-6361:20065295},
archivePrefix = {arXiv},
       eprint = {astro-ph/0603485},
 primaryClass = {astro-ph},
       adsurl = {https://ui.adsabs.harvard.edu/abs/2008A&A...481...33J}
}

@ARTICLE{kluge24,
       author = {{Kluge}, M. and {Comparat}, J. and {Liu}, A. and {Balzer}, F. and {Bulbul}, E. and {Ider Chitham}, J. and {Ghirardini}, V. and {Garrel}, C. and {Bahar}, Y.~E. and {Artis}, E. and {Bender}, R. and {Clerc}, N. and {Dwelly}, T. and {Fabricius}, M.~H. and {Grandis}, S. and {Hern{\'a}ndez-Lang}, D. and {Hill}, G.~J. and {Joshi}, J. and {Lamer}, G. and {Merloni}, A. and {Nandra}, K. and {Pacaud}, F. and {Predehl}, P. and {Ramos-Ceja}, M.~E. and {Reiprich}, T.~H. and {Salvato}, M. and {Sanders}, J.~S. and {Schrabback}, T. and {Seppi}, R. and {Zelmer}, S. and {Zenteno}, A. and {Zhang}, X.},
        title = "{The SRG/eROSITA All-Sky Survey. Optical identification and properties of galaxy clusters and groups in the western galactic hemisphere}",
      journal = {\aap},
         year = 2024,
        month = aug,
       volume = {688},
          eid = {A210},
        pages = {A210},
          doi = {10.1051/0004-6361/202349031},
archivePrefix = {arXiv},
       eprint = {2402.08453},
 primaryClass = {astro-ph.CO},
       adsurl = {https://ui.adsabs.harvard.edu/abs/2024A&A...688A.210K}
}

@ARTICLE{kornoelje25,
       author = {{Kornoelje}, K. and {Bleem}, L.~E. and {Rykoff}, E.~S. and {Abbott}, T.~M.~C. and {Ade}, P.~A.~R. and {Aguena}, M. and {Alves}, O. and {Anderson}, A.~J. and {Andrade-Oliveira}, F. and {Ansarinejad}, B. and {Archipley}, M. and {Ashby}, M.~L.~N. and {Austermann}, J.~E. and {Bacon}, D. and {Balkenhol}, L. and {Bayliss}, Matthew B. and {Beall}, J.~A. and {Benabed}, K. and {Bender}, A.~N. and {Benson}, B.~A. and {Bianchini}, F. and {Bocquet}, S. and {Bouchet}, F.~R. and {Brooks}, D. and {Burke}, D.~L. and {Calzadilla}, M. and {Campitiello}, M.~G. and {Camphuis}, E. and {Carlstrom}, J.~E. and {Carnero Rosell}, A. and {Carretero}, J. and {Chang}, C.~L. and {Chaubal}, P. and {Chiang}, H.~C. and {Chichura}, P.~M. and {Chokshi}, A. and {Chou}, T.-L. and {Citron}, R. and {Coerver}, A. and {Moran}, C. Corbett and {Costanzi}, M. and {Crawford}, T.~M. and {Crites}, A.~T. and {da Costa}, L.~N. and {Daley}, C. and {de Haan}, T. and {De Vicente}, J. and {Desai}, S. and {Dibert}, K.~R. and {Dobbs}, M.~A. and {Doel}, P. and {Doohan}, M. and {Doussot}, A. and {Dutcher}, D. and {Everett}, W. and {Everett}, S. and {Feng}, C. and {Ferguson}, K.~R. and {Mena-Fern{\'a}ndez}, J. and {Ferrero}, I. and {Fichman}, K. and {Flaugher}, B. and {Floyd}, B. and {Foster}, A. and {Friedel}, D. and {Frieman}, J. and {Galli}, S. and {Gallicchio}, J. and {Gambrel}, A.~E. and {Garc{\'\i}a-Bellido}, J. and {Gardner}, R.~W. and {Gassis}, R. and {Gatti}, M. and {Ge}, F. and {George}, E.~M. and {Giannini}, G. and {Goeckner-Wald}, N. and {Grandis}, S. and {Gruen}, D. and {Gruendl}, R.~A. and {Gualtieri}, R. and {Guidi}, F. and {Guns}, S. and {Gupta}, N. and {Gutierrez}, G. and {Halverson}, N.~W. and {Hinton}, S.~R. and {Hivon}, E. and {Holder}, G.~P. and {Hollowood}, D.~L. and {Holzapfel}, W.~L. and {Honscheid}, K. and {Hood}, J.~C. and {Hrubes}, J.~D. and {Hryciuk}, A. and {Huang}, N. and {Hubmayr}, J. and {Irwin}, K.~D. and {James}, D.~J. and {K{\'e}ruzor{\'e}}, F. and {Khalife}, A.~R. and {Klein}, M. and {Knox}, L. and {Korman}, M. and {Kuehn}, K. and {Kuo}, C.-L. and {Lahav}, O. and {Lee}, A.~T. and {Lee}, S. and {Levy}, K. and {Li}, D. and {Lima}, M. and {Lowitz}, A.~E. and {Lowitz}, A. and {Lu}, C. and {Mahler}, Guillaume and {Maniyar}, A. and {Marshal}, J.~L. and {Marshall}, J.~L. and {Martsen}, E.~S. and {McDonald}, M. and {McMahon}, J.~J. and {Menanteau}, F. and {Millea}, M. and {Miquel}, R. and {Mohr}, J.~J. and {Montgomery}, J. and {Myles}, J. and {Nakato}, Y. and {Natoli}, T. and {Nibarger}, J.~P. and {Noble}, G.~I. and {Novosad}, V. and {Ogando}, R.~L.~C. and {Omori}, Y. and {Ouellette}, A. and {Padin}, S. and {Pan}, Z. and {Patil}, S. and {Pereira}, M.~E.~S. and {Phadke}, K.~A. and {Pieres}, A. and {Plazas Malag\&ooacute} and {n}, A.~A. and {Pollak}, A.~W. and {Prabhu}, K. and {Pryke}, C. and {Quan}, W. and {Rahimi}, M. and {Rahlin}, A. and {Reichardt}, C.~L. and {Rodr{\'\i}guez-Monroy}, M. and {Romer}, A.~K. and {Rouble}, M. and {Ruhl}, J.~E. and {Saliwanchik}, B.~R. and {Salvati}, L. and {Samuroff}, S. and {Sanchez}, E. and {Sarkar}, Arnab and {Saro}, A. and {Schaffer}, K.~K. and {Schiappucci}, E. and {Schrabback}, T. and {Sevilla-Noarbe}, I. and {Sievers}, C. and {Smecher}, G. and {Smith}, M. and {Sobrin}, J.~A. and {Somboonpanyakul}, T. and {Stalder}, B. and {Stark}, A.~A. and {Strazzullo}, V. and {Suchyta}, E. and {Swanson}, M.~E.~C. and {Tandoi}, C. and {Tarle}, G. and {Thorne}, B. and {To}, C. and {Trendafilova}, C. and {Tucker}, C. and {Umilta}, C. and {Veach}, T. and {Vieira}, J.~D. and {Vincenzi}, M. and {Vitrier}, A. and {Wan}, Y. and {Wang}, G. and {Weaverdyck}, N. and {Weller}, J. and {Whitehorn}, N. and {Wiseman}, P. and {Wu}, W.~L.~K. and {Yefremenko}, V. and {Young}, M.~R. and {Zebrowski}, J.~A. and {Zhang}, Y.},
        title = "{The SPT-deep Cluster Catalog: Sunyaev─Zel'dovich Selected Clusters from Combined SPT-3G and SPTpol Measurements over 100 Square Degrees}",
      journal = {\apj},
         year = 2026,
        month = jun,
       volume = {1004},
       number = {1},
          eid = {48},
        pages = {48},
          doi = {10.3847/1538-4357/ae5f7b},
archivePrefix = {arXiv},
       eprint = {2503.17271},
 primaryClass = {astro-ph.CO},
       adsurl = {https://ui.adsabs.harvard.edu/abs/2026ApJ..1004...48K}
}

@ARTICLE{liu25,
       author = {{Liu}, R. Henry and {Ferraro}, Simone and {Schaan}, Emmanuel and {Zhou}, Rongpu and {Aguilar}, Jessica Nicole and {Ahlen}, Steven and {Battaglia}, Nicholas and {Bianchi}, Davide and {Brooks}, David and {Claybaugh}, Todd and {Cole}, Shaun and {Coulton}, William R. and {de la Macorra}, Axel and {Dey}, Arjun and {Fanning}, Kevin and {Forero-Romero}, Jaime E. and {Gazta{\~n}aga}, Enrique and {Gong}, Yulin and {Gontcho}, Satya Gontcho A. and {Gruen}, Daniel and {Gutierrez}, Gaston and {Hadzhiyska}, Boryana and {Honscheid}, Klaus and {Howlett}, Cullan and {Kehoe}, Robert and {Kisner}, Theodore and {Kremin}, Anthony and {Kusiak}, Aleksandra and {Lambert}, Andrew and {Landriau}, Martin and {Le Guillou}, Laurent and {Levi}, Michael and {Lokken}, Martine and {Manera}, Marc and {Martini}, Paul and {Meisner}, Aaron and {Miquel}, Ramon and {Moodley}, Kavilan and {Newman}, Jeffrey A. and {Niz}, Gustavo and {Palanque-Delabrouille}, Nathalie and {Percival}, Will and {Prada}, Francisco and {P{\'e}rez-R{\`a}fols}, Ignasi and {Ried Guachalla}, Bernardita and {Rossi}, Graziano and {Sanchez}, Eusebio and {Schlegel}, David and {Schubnell}, Michael and {Seo}, Hee-Jong and {Sif{\'o}n}, Crist{\'o}bal and {Sprayberry}, David and {Tarl{\'e}}, Gregory and {Vavagiakis}, Eve M. and {Weaver}, Benjamin Alan and {Wollack}, Edward J. and {Zou}, Hu},
        title = "{Measurements of the thermal Sunyaev-Zel'dovich effect with ACT and DESI luminous red galaxies}",
      journal = {\prd},
         year = 2025,
        month = oct,
       volume = {112},
       number = {8},
          eid = {083561},
        pages = {083561},
          doi = {10.1103/jqn8-19gx},
archivePrefix = {arXiv},
       eprint = {2502.08850},
 primaryClass = {astro-ph.CO},
       adsurl = {https://ui.adsabs.harvard.edu/abs/2025PhRvD.112h3561L}
}

@ARTICLE{mcclintock19,
       author = {{McClintock}, T. and {Varga}, T.~N. and {Gruen}, D. and {Rozo}, E. and {Rykoff}, E.~S. and {Shin}, T. and {Melchior}, P. and {DeRose}, J. and {Seitz}, S. and {Dietrich}, J.~P. and {Sheldon}, E. and {Zhang}, Y. and {von der Linden}, A. and {Jeltema}, T. and {Mantz}, A.~B. and {Romer}, A.~K. and {Allen}, S. and {Becker}, M.~R. and {Bermeo}, A. and {Bhargava}, S. and {Costanzi}, M. and {Everett}, S. and {Farahi}, A. and {Hamaus}, N. and {Hartley}, W.~G. and {Hollowood}, D.~L. and {Hoyle}, B. and {Israel}, H. and {Li}, P. and {MacCrann}, N. and {Morris}, G. and {Palmese}, A. and {Plazas}, A.~A. and {Pollina}, G. and {Rau}, M.~M. and {Simet}, M. and {Soares-Santos}, M. and {Troxel}, M.~A. and {Vergara Cervantes}, C. and {Wechsler}, R.~H. and {Zuntz}, J. and {Abbott}, T.~M.~C. and {Abdalla}, F.~B. and {Allam}, S. and {Annis}, J. and {Avila}, S. and {Bridle}, S.~L. and {Brooks}, D. and {Burke}, D.~L. and {Carnero Rosell}, A. and {Carrasco Kind}, M. and {Carretero}, J. and {Castander}, F.~J. and {Crocce}, M. and {Cunha}, C.~E. and {D'Andrea}, C.~B. and {da Costa}, L.~N. and {Davis}, C. and {De Vicente}, J. and {Diehl}, H.~T. and {Doel}, P. and {Drlica-Wagner}, A. and {Evrard}, A.~E. and {Flaugher}, B. and {Fosalba}, P. and {Frieman}, J. and {Garc{\'\i}a-Bellido}, J. and {Gaztanaga}, E. and {Gerdes}, D.~W. and {Giannantonio}, T. and {Gruendl}, R.~A. and {Gutierrez}, G. and {Honscheid}, K. and {James}, D.~J. and {Kirk}, D. and {Krause}, E. and {Kuehn}, K. and {Lahav}, O. and {Li}, T.~S. and {Lima}, M. and {March}, M. and {Marshall}, J.~L. and {Menanteau}, F. and {Miquel}, R. and {Mohr}, J.~J. and {Nord}, B. and {Ogando}, R.~L.~C. and {Roodman}, A. and {Sanchez}, E. and {Scarpine}, V. and {Schindler}, R. and {Sevilla-Noarbe}, I. and {Smith}, M. and {Smith}, R.~C. and {Sobreira}, F. and {Suchyta}, E. and {Swanson}, M.~E.~C. and {Tarle}, G. and {Tucker}, D.~L. and {Vikram}, V. and {Walker}, A.~R. and {Weller}, J. and {DES Collaboration}},
        title = "{Dark Energy Survey Year 1 results: weak lensing mass calibration of redMaPPer galaxy clusters}",
      journal = {\mnras},
         year = 2019,
        month = jan,
       volume = {482},
       number = {1},
        pages = {1352-1378},
          doi = {10.1093/mnras/sty2711},
archivePrefix = {arXiv},
       eprint = {1805.00039},
 primaryClass = {astro-ph.CO},
       adsurl = {https://ui.adsabs.harvard.edu/abs/2019MNRAS.482.1352M}
}

@ARTICLE{meinke21,
       author = {{Meinke}, Jeremy and {B{\"o}ckmann}, Kathrin and {Cohen}, Seth and {Mauskopf}, Philip and {Scannapieco}, Evan and {Sarmento}, Richard and {Lunde}, Emily and {Cottle}, J'Neil},
        title = "{The Thermal Sunyaev-Zel'dovich Effect from Massive, Quiescent 0.5 {\ensuremath{\leq}} z {\ensuremath{\leq}} 1.5 Galaxies}",
      journal = {\apj},
         year = 2021,
        month = jun,
       volume = {913},
       number = {2},
          eid = {88},
        pages = {88},
          doi = {10.3847/1538-4357/abf2b4},
archivePrefix = {arXiv},
       eprint = {2103.01245},
 primaryClass = {astro-ph.CO},
       adsurl = {https://ui.adsabs.harvard.edu/abs/2021ApJ...913...88M}
}

@ARTICLE{meinke23,
       author = {{Meinke}, Jeremy and {Cohen}, Seth and {Moore}, Jenna and {B{\"o}ckmann}, Kathrin and {Mauskopf}, Philip and {Scannapieco}, Evan},
        title = "{Evidence of Extended Dust and Feedback around z {\ensuremath{\approx}} 1 Quiescent Galaxies via Millimeter Observations}",
      journal = {\apj},
         year = 2023,
        month = sep,
       volume = {954},
       number = {2},
          eid = {119},
        pages = {119},
          doi = {10.3847/1538-4357/acdcf4},
archivePrefix = {arXiv},
       eprint = {2306.04760},
 primaryClass = {astro-ph.GA},
       adsurl = {https://ui.adsabs.harvard.edu/abs/2023ApJ...954..119M}
}

@ARTICLE{miller23,
       author = {{Miller}, Timothy N. and {Doel}, Peter and {Gutierrez}, Gaston and {Besuner}, Robert and {Brooks}, David and {Gallo}, Giuseppe and {Heetderks}, Henry and {Jelinsky}, Patrick and {Kent}, Stephen M. and {Lampton}, Michael and {Levi}, Michael E. and {Liang}, Ming and {Meisner}, Aaron and {Sholl}, Michael J. and {Silber}, Joseph Harry and {Sprayberry}, David and {Aguilar}, Jessica Nicole and {de la Macorra}, Axel and {Eisenstein}, Daniel and {Fanning}, Kevin and {Font-Ribera}, Andreu and {Gazta{\~n}aga}, Enrique and {Gontcho A Gontcho}, Satya and {Honscheid}, Klaus and {Jimenez}, Jorge and {Joyce}, Dick and {Kehoe}, Robert and {Kisner}, Theodore and {Kremin}, Anthony and {Landriau}, Martin and {Le Guillou}, Laurent and {Magneville}, Christophe and {Martini}, Paul and {Miquel}, Ramon and {Moustakas}, John and {Nie}, Jundan and {Percival}, Will and {Poppett}, Claire and {Prada}, Francisco and {Rossi}, Graziano and {Schlegel}, David and {Schubnell}, Michael and {Seo}, Hee-Jong and {Sharples}, Ray and {Tarl{\'e}}, Gregory and {Vargas-Maga{\~n}a}, Mariana and {Zhou}, Zhimin and {the DESI Collaboration}},
        title = "{The Optical Corrector for the Dark Energy Spectroscopic Instrument}",
      journal = {\aj},
         year = 2024,
        month = aug,
       volume = {168},
       number = {2},
          eid = {95},
        pages = {95},
          doi = {10.3847/1538-3881/ad45fe},
archivePrefix = {arXiv},
       eprint = {2306.06310},
 primaryClass = {astro-ph.IM},
       adsurl = {https://ui.adsabs.harvard.edu/abs/2024AJ....168...95M}
}

@ARTICLE{moser21,
       author = {{Moser}, Emily and {Amodeo}, Stefania and {Battaglia}, Nicholas and {Alvarez}, Marcelo A. and {Ferraro}, Simone and {Schaan}, Emmanuel},
        title = "{The Impacts of Modeling Choices on the Inference of Circumgalactic Medium Properties from Sunyaev-Zeldovich Observations}",
      journal = {\apj},
         year = 2021,
        month = sep,
       volume = {919},
       number = {1},
          eid = {2},
        pages = {2},
          doi = {10.3847/1538-4357/ac0cea},
archivePrefix = {arXiv},
       eprint = {2103.02469},
 primaryClass = {astro-ph.GA},
       adsurl = {https://ui.adsabs.harvard.edu/abs/2021ApJ...919....2M}
}

@ARTICLE{moser22,
       author = {{Moser}, Emily and {Battaglia}, Nicholas and {Nagai}, Daisuke and {Lau}, Erwin and {Machado Poletti Valle}, Luis Fernando and {Villaescusa-Navarro}, Francisco and {Amodeo}, Stefania and {Angl{\'e}s-Alc{\'a}zar}, Daniel and {Bryan}, Greg L. and {Dave}, Romeel and {Hernquist}, Lars and {Vogelsberger}, Mark},
        title = "{The Circumgalactic Medium from the CAMELS Simulations: Forecasting Constraints on Feedback Processes from Future Sunyaev-Zeldovich Observations}",
      journal = {\apj},
         year = 2022,
        month = jul,
       volume = {933},
       number = {2},
          eid = {133},
        pages = {133},
          doi = {10.3847/1538-4357/ac70c6},
archivePrefix = {arXiv},
       eprint = {2201.02708},
 primaryClass = {astro-ph.CO},
       adsurl = {https://ui.adsabs.harvard.edu/abs/2022ApJ...933..133M}
}

@ARTICLE{moser23,
       author = {{Moser}, Emily and {Battaglia}, Nicholas and {Amodeo}, Stefania},
        title = "{Searching for Systematics in Forward Modeling Sunyaev-Zeldovich Profiles}",
      journal = {arXiv e-prints},
         year = 2023,
        month = jul,
          eid = {arXiv:2307.10919},
        pages = {arXiv:2307.10919},
          doi = {10.48550/arXiv.2307.10919},
archivePrefix = {arXiv},
       eprint = {2307.10919},
 primaryClass = {astro-ph.CO},
       adsurl = {https://ui.adsabs.harvard.edu/abs/2023arXiv230710919M}
}

@ARTICLE{mroczkowski19,
       author = {{Mroczkowski}, Tony and {Nagai}, Daisuke and {Basu}, Kaustuv and {Chluba}, Jens and {Sayers}, Jack and {Adam}, R{\'e}mi and {Churazov}, Eugene and {Crites}, Abigail and {Di Mascolo}, Luca and {Eckert}, Dominique and {Macias-Perez}, Juan and {Mayet}, Fr{\'e}d{\'e}ric and {Perotto}, Laurence and {Pointecouteau}, Etienne and {Romero}, Charles and {Ruppin}, Florian and {Scannapieco}, Evan and {ZuHone}, John},
        title = "{Astrophysics with the Spatially and Spectrally Resolved Sunyaev-Zeldovich Effects. A Millimetre/Submillimetre Probe of the Warm and Hot Universe}",
      journal = {\ssr},
         year = 2019,
        month = feb,
       volume = {215},
       number = {1},
          eid = {17},
        pages = {17},
          doi = {10.1007/s11214-019-0581-2},
archivePrefix = {arXiv},
       eprint = {1811.02310},
 primaryClass = {astro-ph.CO},
       adsurl = {https://ui.adsabs.harvard.edu/abs/2019SSRv..215...17M}
}

@ARTICLE{mueller15.2,
       author = {{Mueller}, Eva-Maria and {de Bernardis}, Francesco and {Bean}, Rachel and {Niemack}, Michael D.},
        title = "{Constraints on massive neutrinos from the pairwise kinematic Sunyaev-Zel'dovich effect}",
      journal = {\prd},
         year = 2015,
        month = sep,
       volume = {92},
       number = {6},
          eid = {063501},
        pages = {063501},
          doi = {10.1103/PhysRevD.92.063501},
archivePrefix = {arXiv},
       eprint = {1412.0592},
 primaryClass = {astro-ph.CO},
       adsurl = {https://ui.adsabs.harvard.edu/abs/2015PhRvD..92f3501M}
}

@ARTICLE{mueller15.1,
       author = {{Mueller}, Eva-Maria and {de Bernardis}, Francesco and {Bean}, Rachel and {Niemack}, Michael D.},
        title = "{Constraints on Gravity and Dark Energy from the Pairwise Kinematic Sunyaev-Zel'dovich Effect}",
      journal = {\apj},
         year = 2015,
        month = jul,
       volume = {808},
       number = {1},
          eid = {47},
        pages = {47},
          doi = {10.1088/0004-637X/808/1/47},
archivePrefix = {arXiv},
       eprint = {1408.6248},
 primaryClass = {astro-ph.CO},
       adsurl = {https://ui.adsabs.harvard.edu/abs/2015ApJ...808...47M}
}

@article{naess21,
    author = {{Naess}, Sigurd and {Madhavacheril}, Mathew and {Hasselfield}, Matthew},
    title = {Pixell: Rectangular pixel map manipulation and harmonic analysis library},
    journal = {Astrophysics Source Code Library, record ascl:2102.003},
    year = 2021,
    month = feb,
    eid = {ascl:2102.003},
    archivePrefix = {ascl},
    adsurl = {https://ui.adsabs.harvard.edu/abs/2021ascl.soft02003N}
    
}

@ARTICLE{naess25,
       author = {{Naess}, Sigurd and {Guan}, Yilun and {Duivenvoorden}, Adriaan J. and {Hasselfield}, Matthew and {Wang}, Yuhan and {Abril-Cabezas}, Irene and {Addison}, Graeme E. and {Ade}, Peter A.~R. and {Aiola}, Simone and {Alford}, Tommy and {Alonso}, David and {Amiri}, Mandana and {An}, Rui and {Atkins}, Zachary and {Austermann}, Jason E. and {Barbavara}, Eleonora and {Battaglia}, Nicholas and {Battistelli}, Elia Stefano and {Beall}, James A. and {Bean}, Rachel and {Beheshti}, Ali and {Beringue}, Benjamin and {Bhandarkar}, Tanay and {Biermann}, Emily and {Bolliet}, Boris and {Bond}, J. Richard and {Calabrese}, Erminia and {Capalbo}, Valentina and {Carrero}, Felipe and {Chen}, Stephen and {Chesmore}, Grace and {Cho}, Hsiao-mei and {Choi}, Steve K. and {Clark}, Susan E. and {Rosado}, Rodrigo Cordova and {Cothard}, Nicholas F. and {Coughlin}, Kevin and {Coulton}, William and {Crichton}, Devin and {Crowley}, Kevin T. and {Devlin}, Mark J. and {Dicker}, Simon and {Duell}, Cody J. and {Duff}, Shannon M. and {Dunkley}, Jo and {Dunner}, Rolando and {Embil Villagra}, Carmen and {Fankhanel}, Max and {Farren}, Gerrit S. and {Ferraro}, Simone and {Foster}, Allen and {Freundt}, Rodrigo and {Fuzia}, Brittany and {Gallardo}, Patricio A. and {Garrido}, Xavier and {Giardiello}, Serena and {Gill}, Ajay and {Givans}, Jahmour and {Gluscevic}, Vera and {Golec}, Joseph E. and {Gong}, Yulin and {Halpern}, Mark and {Harrison}, Ian and {Healy}, Erin and {Henderson}, Shawn and {Hensley}, Brandon and {Herv{\'\i}as-Caimapo}, Carlos and {Hill}, J. Colin and {Hilton}, Gene C. and {Hilton}, Matt and {Hincks}, Adam D. and {Hlo{\v{z}}ek}, Ren{\'e}e and {Ho}, Shuay-Pwu Patty and {Hood}, John and {Hornecker}, Erika and {Huber}, Zachary B. and {Hubmayr}, Johannes and {Huffenberger}, Kevin M. and {Hughes}, John P. and {Ikape}, Margaret and {Irwin}, Kent and {Isopi}, Giovanni and {Jense}, Hidde T. and {Joshi}, Neha and {Keller}, Ben and {Kim}, Joshua and {Knowles}, Kenda and {Koopman}, Brian J. and {Kosowsky}, Arthur and {Kramer}, Darby and {Kusiak}, Aleksandra and {La Posta}, Adrien and {Lagu{\"e}}, Alex and {Lakey}, Victoria and {Lee}, Eunseong and {Li}, Yaqiong and {Li}, Zack and {Limon}, Michele and {Lokken}, Martine and {Louis}, Thibaut and {Lungu}, Marius and {MacCrann}, Niall and {MacInnis}, Amanda and {Madhavacheril}, Mathew S. and {Maldonado}, Diego and {Maldonado}, Felipe and {Mallaby-Kay}, Maya and {Marques}, Gabriela A. and {van Marrewijk}, Joshiwa and {McCarthy}, Fiona and {McMahon}, Jeff and {Mehta}, Yogesh and {Menanteau}, Felipe and {Moodley}, Kavilan and {Morris}, Thomas W. and {Mroczkowski}, Tony and {Namikawa}, Toshiya and {Nati}, Federico and {Nerval}, Simran K. and {Newburgh}, Laura and {Nicola}, Andrina and {Niemack}, Michael D. and {Nolta}, Michael R. and {Orlowski-Scherer}, John and {Page}, Lyman A. and {Pandey}, Shivam and {Partridge}, Bruce and {Perez Sarmiento}, Karen and {Prince}, Heather and {Puddu}, Roberto and {Qu}, Frank J. and {Ragavan}, Damien C. and {Ried Guachalla}, Bernardita and {Rogers}, Keir K. and {Rojas}, Felipe and {Sakuma}, Tai and {Schaan}, Emmanuel and {Schmitt}, Benjamin L. and {Sehgal}, Neelima and {Shaikh}, Shabbir and {Sherwin}, Blake D. and {Sierra}, Carlos and {Sievers}, Jon and {Sif{\'o}n}, Crist{\'o}bal and {Simon}, Sara and {Sonka}, Rita and {London}, Alexander Spencer and {Spergel}, David N. and {Staggs}, Suzanne T. and {Storer}, Emilie and {Surrao}, Kristen and {Switzer}, Eric R. and {Tampier}, Niklas and {Thornton}, Robert and {Trac}, Hy and {Tucker}, Carole and {Ullom}, Joel and {Vale}, Leila R. and {Van Engelen}, Alexander and {Van Lanen}, Jeff and {Vargas}, Cristian and {Vavagiakis}, Eve M. and {Wagoner}, Kasey and {Wenzl}, Lukas and {Wollack}, Edward J. and {Zheng}, Kaiwen and {The Atacama Cosmology Telescope collaboration}},
        title = "{The Atacama Cosmology Telescope: DR6 maps}",
      journal = {\jcap},
         year = 2025,
        month = nov,
       volume = {2025},
       number = {11},
          eid = {061},
        pages = {061},
          doi = {10.1088/1475-7516/2025/11/061},
archivePrefix = {arXiv},
       eprint = {2503.14451},
 primaryClass = {astro-ph.CO},
       adsurl = {https://ui.adsabs.harvard.edu/abs/2025JCAP...11..061N}
}

@ARTICLE{schaan21,
       author = {{Schaan}, Emmanuel and {Ferraro}, Simone and {Amodeo}, Stefania and {Battaglia}, Nicholas and {Aiola}, Simone and {Austermann}, Jason E. and {Beall}, James A. and {Bean}, Rachel and {Becker}, Daniel T. and {Bond}, Richard J. and {Calabrese}, Erminia and {Calafut}, Victoria and {Choi}, Steve K. and {Denison}, Edward V. and {Devlin}, Mark J. and {Duff}, Shannon M. and {Duivenvoorden}, Adriaan J. and {Dunkley}, Jo and {D{\"u}nner}, Rolando and {Gallardo}, Patricio A. and {Guan}, Yilun and {Han}, Dongwon and {Hill}, J. Colin and {Hilton}, Gene C. and {Hilton}, Matt and {Hlo{\v{z}}ek}, Ren{\'e}e and {Hubmayr}, Johannes and {Huffenberger}, Kevin M. and {Hughes}, John P. and {Koopman}, Brian J. and {MacInnis}, Amanda and {McMahon}, Jeff and {Madhavacheril}, Mathew S. and {Moodley}, Kavilan and {Mroczkowski}, Tony and {Naess}, Sigurd and {Nati}, Federico and {Newburgh}, Laura B. and {Niemack}, Michael D. and {Page}, Lyman A. and {Partridge}, Bruce and {Salatino}, Maria and {Sehgal}, Neelima and {Schillaci}, Alessandro and {Sif{\'o}n}, Crist{\'o}bal and {Smith}, Kendrick M. and {Spergel}, David N. and {Staggs}, Suzanne and {Storer}, Emilie R. and {Trac}, Hy and {Ullom}, Joel N. and {Van Lanen}, Jeff and {Vale}, Leila R. and {van Engelen}, Alexander and {Maga{\~n}a}, Mariana Vargas and {Vavagiakis}, Eve M. and {Wollack}, Edward J. and {Xu}, Zhilei and {Atacama Cosmology Telescope Collaboration}},
        title = "{Atacama Cosmology Telescope: Combined kinematic and thermal Sunyaev-Zel'dovich measurements from BOSS CMASS and LOWZ halos}",
      journal = {\prd},
         year = 2021,
        month = mar,
       volume = {103},
       number = {6},
          eid = {063513},
        pages = {063513},
          doi = {10.1103/PhysRevD.103.063513},
archivePrefix = {arXiv},
       eprint = {2009.05557},
 primaryClass = {astro-ph.CO},
       adsurl = {https://ui.adsabs.harvard.edu/abs/2021PhRvD.103f3513S}
}

@ARTICLE{poppett24,
       author = {{Poppett}, Claire and {Tyas}, Luke and {Aguilar}, J. and {Bebek}, Christopher and {Bramall}, D. and {Claybaugh}, T. and {Edelstein}, J. and {Fagrelius}, P. and {Heetderks}, H. and {Jelinsky}, P. and {Jelinsky}, S. and {Lafever}, Robin and {Lambert}, A. and {Lampton}, M. and {Levi}, Michael E. and {Martini}, P. and {Rockosi}, C. and {Schmoll}, J. and {Sharples}, Ray M. and {Sirk}, Martin and {Wishnow}, Edward and {Yu}, Jiaxi and {Ahlen}, S. and {Bault}, A. and {BenZvi}, S. and {Brooks}, D. and {Cole}, S. and {de la Macorra}, A. and {Dey}, Arjun and {Doel}, P. and {Fanning}, K. and {Font-Ribera}, A. and {Forero-Romero}, J.~E. and {Gazta{\~n}aga}, E. and {Gontcho A Gontcho}, S. and {Gonzalez-Morales}, A.~X. and {Hahn}, C. and {Honscheid}, K. and {Jimenez}, J. and {Juneau}, S. and {Kirkby}, D. and {Kremin}, A. and {Landriau}, M. and {Le Guillou}, L. and {Manera}, M. and {Meisner}, A. and {Miquel}, R. and {Moustakas}, J. and {Mueller}, E. and {Mu{\~n}oz-Guti{\'e}rrez}, A. and {Myers}, A.~D. and {Nie}, J. and {Niz}, G. and {Palanque-Delabrouille}, N. and {Percival}, W.~J. and {Prada}, F. and {Rabinowitz}, D. and {Rezaie}, M. and {Rossi}, G. and {Sanchez}, E. and {Schlafly}, Edward F. and {Schlegel}, D. and {Schubnell}, M. and {Seo}, H. and {Sprayberry}, D. and {Tarl{\'e}}, G. and {Vargas-Maga{\~n}a}, M. and {Weaver}, B.~A. and {Zhou}, R.},
        title = "{Overview of the Fiber System for the Dark Energy Spectroscopic Instrument}",
      journal = {\aj},
         year = 2024,
        month = dec,
       volume = {168},
       number = {6},
          eid = {245},
        pages = {245},
          doi = {10.3847/1538-3881/ad76a4},
       adsurl = {https://ui.adsabs.harvard.edu/abs/2024AJ....168..245P}
}

@ARTICLE{schlafly23,
       author = {{Schlafly}, Edward F. and {Kirkby}, David and {Schlegel}, David J. and {Myers}, Adam D. and {Raichoor}, Anand and {Dawson}, Kyle and {Aguilar}, Jessica and {Allende Prieto}, Carlos and {Bailey}, Stephen and {BenZvi}, Segev and {Bermejo-Climent}, Jose and {Brooks}, David and {de la Macorra}, Axel and {Dey}, Arjun and {Doel}, Peter and {Fanning}, Kevin and {Font-Ribera}, Andreu and {Forero-Romero}, Jaime E. and {Garc{\'\i}a-Bellido}, Juan and {Gontcho A Gontcho}, Satya and {Guy}, Julien and {Hahn}, ChangHoon and {Honscheid}, Klaus and {Ishak}, Mustapha and {Juneau}, St{\'e}phanie and {Kehoe}, Robert and {Kisner}, Theodore and {Kremin}, Anthony and {Landriau}, Martin and {Lang}, Dustin A. and {Lasker}, James and {Levi}, Michael E. and {Magneville}, Christophe and {Manser}, Christopher J. and {Martini}, Paul and {Meisner}, Aaron M. and {Miquel}, Ramon and {Moustakas}, John and {Newman}, Jeffrey A. and {Nie}, Jundan and {Palanque-Delabrouille}, Nathalie. and {Percival}, Will J. and {Poppett}, Claire and {Rockosi}, Constance and {Ross}, Ashley J. and {Rossi}, Graziano and {Tarl{\'e}}, Gregory and {Weaver}, Benjamin A. and {Y{\`e}che}, Christophe and {Zhou}, Rongpu and {DESI Collaboration}},
        title = "{Survey Operations for the Dark Energy Spectroscopic Instrument}",
      journal = {\aj},
         year = 2023,
        month = dec,
       volume = {166},
       number = {6},
          eid = {259},
        pages = {259},
          doi = {10.3847/1538-3881/ad0832},
archivePrefix = {arXiv},
       eprint = {2306.06309},
 primaryClass = {astro-ph.CO},
       adsurl = {https://ui.adsabs.harvard.edu/abs/2023AJ....166..259S}
}

@ARTICLE{siudek24,
       author = {{Siudek}, M. and {Pucha}, R. and {Mezcua}, M. and {Juneau}, S. and {Aguilar}, J. and {Ahlen}, S. and {Brooks}, D. and {Circosta}, C. and {Claybaugh}, T. and {Cole}, S. and {Dawson}, K. and {de la Macorra}, A. and {Dey}, A. and {Dey}, B. and {Doel}, P. and {Font-Ribera}, A. and {Forero-Romero}, J.~E. and {Gazta{\~n}aga}, E. and {Gontcho A Gontcho}, S. and {Gutierrez}, G. and {Honscheid}, K. and {Howlett}, C. and {Ishak}, M. and {Kehoe}, R. and {Kirkby}, D. and {Kisner}, T. and {Kremin}, A. and {Lambert}, A. and {Landriau}, M. and {Le Guillou}, L. and {Manera}, M. and {Martini}, P. and {Meisner}, A. and {Miquel}, R. and {Moustakas}, J. and {Newman}, J.~A. and {Niz}, G. and {Pan}, Z. and {Percival}, W.~J. and {Poppett}, C. and {Prada}, F. and {Rossi}, G. and {Saintonge}, A. and {Sanchez}, E. and {Schlegel}, D. and {Scholte}, D. and {Schubnell}, M. and {Seo}, H. and {Speranza}, F. and {Sprayberry}, D. and {Tarl{\'e}}, G. and {Weaver}, B.~A. and {Zou}, H.},
        title = "{Value-added catalog of physical properties for more than 1.3 million galaxies from the DESI survey}",
      journal = {\aap},
         year = 2024,
        month = nov,
       volume = {691},
          eid = {A308},
        pages = {A308},
          doi = {10.1051/0004-6361/202451761},
archivePrefix = {arXiv},
       eprint = {2409.19066},
 primaryClass = {astro-ph.GA},
       adsurl = {https://ui.adsabs.harvard.edu/abs/2024A&A...691A.308S}
}

@ARTICLE{siudek25,
       author = {{Siudek}, M. and {Mezcua}, M. and {Circosta}, C. and {Maraston}, C. and {Moustakas}, J. and {Zou}, H. and {Aguilar}, J. and {Ahlen}, S. and {Bianchi}, D. and {Brooks}, D. and {Claybaugh}, T. and {Dawson}, K.~S. and {de la Macorra}, A. and {Dey}, A. and {Doel}, P. and {Forero-Romero}, J.~E. and {Gazta{\~n}aga}, E. and {Gontcho A Gontcho}, S. and {Gutierrez}, G. and {Ishak}, M. and {Juneau}, S. and {Kirkby}, D. and {Kisner}, T. and {Kremin}, A. and {Lambert}, A. and {Landriau}, M. and {Le Guillou}, L. and {Meisner}, A. and {Miquel}, R. and {Prada}, F. and {P{\'e}rez-R{\`a}fols}, I. and {Rossi}, G. and {Sanchez}, E. and {Schlegel}, D. and {Schubnell}, M. and {Seo}, H. and {Sprayberry}, D. and {Tarl{\'e}}, G. and {Weaver}, B.~A.},
        title = "{Beyond traditional diagnostics: Identifying active galactic nuclei using spectral energy distribution fitting in DESI data}",
      journal = {\aap},
         year = 2025,
        month = aug,
       volume = {700},
          eid = {A209},
        pages = {A209},
          doi = {10.1051/0004-6361/202555463},
archivePrefix = {arXiv},
       eprint = {2506.09143},
 primaryClass = {astro-ph.GA},
       adsurl = {https://ui.adsabs.harvard.edu/abs/2025A&A...700A.209S}
}

@ARTICLE{soergel18,
       author = {{Soergel}, Bjoern and {Saro}, Alexandro and {Giannantonio}, Tommaso and {Efstathiou}, George and {Dolag}, Klaus},
        title = "{Cosmology with the pairwise kinematic SZ effect: calibration and validation using hydrodynamical simulations}",
      journal = {\mnras},
         year = 2018,
        month = aug,
       volume = {478},
       number = {4},
        pages = {5320-5335},
          doi = {10.1093/mnras/sty1324},
archivePrefix = {arXiv},
       eprint = {1712.05714},
 primaryClass = {astro-ph.CO},
       adsurl = {https://ui.adsabs.harvard.edu/abs/2018MNRAS.478.5320S}
}

@ARTICLE{so19,
       author = {{Ade}, Peter and {Aguirre}, James and {Ahmed}, Zeeshan and {Aiola}, Simone and {Ali}, Aamir and {Alonso}, David and {Alvarez}, Marcelo A. and {Arnold}, Kam and {Ashton}, Peter and {Austermann}, Jason and {Awan}, Humna and {Baccigalupi}, Carlo and {Baildon}, Taylor and {Barron}, Darcy and {Battaglia}, Nick and {Battye}, Richard and {Baxter}, Eric and {Bazarko}, Andrew and {Beall}, James A. and {Bean}, Rachel and {Beck}, Dominic and {Beckman}, Shawn and {Beringue}, Benjamin and {Bianchini}, Federico and {Boada}, Steven and {Boettger}, David and {Bond}, J. Richard and {Borrill}, Julian and {Brown}, Michael L. and {Bruno}, Sarah Marie and {Bryan}, Sean and {Calabrese}, Erminia and {Calafut}, Victoria and {Calisse}, Paolo and {Carron}, Julien and {Challinor}, Anthony and {Chesmore}, Grace and {Chinone}, Yuji and {Chluba}, Jens and {Cho}, Hsiao-Mei Sherry and {Choi}, Steve and {Coppi}, Gabriele and {Cothard}, Nicholas F. and {Coughlin}, Kevin and {Crichton}, Devin and {Crowley}, Kevin D. and {Crowley}, Kevin T. and {Cukierman}, Ari and {D'Ewart}, John M. and {D{\"u}nner}, Rolando and {de Haan}, Tijmen and {Devlin}, Mark and {Dicker}, Simon and {Didier}, Joy and {Dobbs}, Matt and {Dober}, Bradley and {Duell}, Cody J. and {Duff}, Shannon and {Duivenvoorden}, Adri and {Dunkley}, Jo and {Dusatko}, John and {Errard}, Josquin and {Fabbian}, Giulio and {Feeney}, Stephen and {Ferraro}, Simone and {Flux{\`a}}, Pedro and {Freese}, Katherine and {Frisch}, Josef C. and {Frolov}, Andrei and {Fuller}, George and {Fuzia}, Brittany and {Galitzki}, Nicholas and {Gallardo}, Patricio A. and {Tomas Galvez Ghersi}, Jose and {Gao}, Jiansong and {Gawiser}, Eric and {Gerbino}, Martina and {Gluscevic}, Vera and {Goeckner-Wald}, Neil and {Golec}, Joseph and {Gordon}, Sam and {Gralla}, Megan and {Green}, Daniel and {Grigorian}, Arpi and {Groh}, John and {Groppi}, Chris and {Guan}, Yilun and {Gudmundsson}, Jon E. and {Han}, Dongwon and {Hargrave}, Peter and {Hasegawa}, Masaya and {Hasselfield}, Matthew and {Hattori}, Makoto and {Haynes}, Victor and {Hazumi}, Masashi and {He}, Yizhou and {Healy}, Erin and {Henderson}, Shawn W. and {Hervias-Caimapo}, Carlos and {Hill}, Charles A. and {Hill}, J. Colin and {Hilton}, Gene and {Hilton}, Matt and {Hincks}, Adam D. and {Hinshaw}, Gary and {Hlo{\v{z}}ek}, Ren{\'e}e and {Ho}, Shirley and {Ho}, Shuay-Pwu Patty and {Howe}, Logan and {Huang}, Zhiqi and {Hubmayr}, Johannes and {Huffenberger}, Kevin and {Hughes}, John P. and {Ijjas}, Anna and {Ikape}, Margaret and {Irwin}, Kent and {Jaffe}, Andrew H. and {Jain}, Bhuvnesh and {Jeong}, Oliver and {Kaneko}, Daisuke and {Karpel}, Ethan D. and {Katayama}, Nobuhiko and {Keating}, Brian and {Kernasovskiy}, Sarah S. and {Keskitalo}, Reijo and {Kisner}, Theodore and {Kiuchi}, Kenji and {Klein}, Jeff and {Knowles}, Kenda and {Koopman}, Brian and {Kosowsky}, Arthur and {Krachmalnicoff}, Nicoletta and {Kuenstner}, Stephen E. and {Kuo}, Chao-Lin and {Kusaka}, Akito and {Lashner}, Jacob and {Lee}, Adrian and {Lee}, Eunseong and {Leon}, David and {Leung}, Jason S.-Y. and {Lewis}, Antony and {Li}, Yaqiong and {Li}, Zack and {Limon}, Michele and {Linder}, Eric and {Lopez-Caraballo}, Carlos and {Louis}, Thibaut and {Lowry}, Lindsay and {Lungu}, Marius and {Madhavacheril}, Mathew and {Mak}, Daisy and {Maldonado}, Felipe and {Mani}, Hamdi and {Mates}, Ben and {Matsuda}, Frederick and {Maurin}, Lo{\"\i}c and {Mauskopf}, Phil and {May}, Andrew and {McCallum}, Nialh and {McKenney}, Chris and {McMahon}, Jeff and {Meerburg}, P. Daniel and {Meyers}, Joel and {Miller}, Amber and {Mirmelstein}, Mark and {Moodley}, Kavilan and {Munchmeyer}, Moritz and {Munson}, Charles and {Naess}, Sigurd and {Nati}, Federico and {Navaroli}, Martin and {Newburgh}, Laura and {Nguyen}, Ho Nam and {Niemack}, Michael and {Nishino}, Haruki and {Orlowski-Scherer}, John and {Page}, Lyman and {Partridge}, Bruce and {Peloton}, Julien and {Perrotta}, Francesca and {Piccirillo}, Lucio and {Pisano}, Giampaolo and {Poletti}, Davide and {Puddu}, Roberto and {Puglisi}, Giuseppe and {Raum}, Chris and {Reichardt}, Christian L. and {Remazeilles}, Mathieu and {Rephaeli}, Yoel and {Riechers}, Dominik and {Rojas}, Felipe and {Roy}, Anirban and {Sadeh}, Sharon and {Sakurai}, Yuki and {Salatino}, Maria and {Sathyanarayana Rao}, Mayuri and {Schaan}, Emmanuel and {Schmittfull}, Marcel and {Sehgal}, Neelima and {Seibert}, Joseph},
       collaboration = "Simons Observatory",
        title = "{The Simons Observatory: science goals and forecasts}",
      journal = {\jcap},
         year = 2019,
        month = feb,
       volume = {2019},
       number = {2},
          eid = {056},
        pages = {056},
          doi = {10.1088/1475-7516/2019/02/056},
archivePrefix = {arXiv},
       eprint = {1808.07445},
 primaryClass = {astro-ph.CO},
       adsurl = {https://ui.adsabs.harvard.edu/abs/2019JCAP...02..056A}
}

@ARTICLE{so25,
       author = {{Abitbol}, M. and {Abril-Cabezas}, I. and {Adachi}, S. and {Ade}, P. and {Adler}, A.~E. and {Agrawal}, P. and {Aguirre}, J. and {Ahmed}, Z. and {Aiola}, S. and {Alford}, T. and {Ali}, A. and {Alonso}, D. and {Alvarez}, M.~A. and {An}, R. and {Arnold}, K. and {Ashton}, P. and {Atkins}, Z. and {Austermann}, J. and {Azzoni}, S. and {Baccigalupi}, C. and {Baleato Lizancos}, A. and {Barron}, D. and {Barry}, P. and {Bartlett}, J. and {Battaglia}, N. and {Battye}, R. and {Baxter}, E. and {Bazarko}, A. and {Beall}, J.~A. and {Bean}, R. and {Beck}, D. and {Beckman}, S. and {Begin}, J. and {Beheshti}, A. and {Beringue}, B. and {Bhandarkar}, T. and {Bhimani}, S. and {Bianchini}, F. and {Biermann}, E. and {Biquard}, S. and {Bixler}, B. and {Boada}, S. and {Boettger}, D. and {Bolliet}, B. and {Bond}, J.~R. and {Borrill}, J. and {Borrow}, J. and {Braithwaite}, C. and {Brien}, T.~L.~R. and {Brown}, M.~L. and {Bruno}, S.~M. and {Bryan}, S. and {Bustos}, R. and {Cai}, H. and {Calabrese}, E. and {Calafut}, V. and {Carl}, F.~M. and {Carones}, A. and {Carron}, J. and {Challinor}, A. and {Chanial}, P. and {Chen}, N. and {Cheung}, K. and {Chiang}, B. and {Chinone}, Y. and {Chluba}, J. and {Cho}, H.~S. and {Choi}, S.~K. and {Chu}, M. and {Clancy}, J. and {Clark}, S.~E. and {Clarke}, P. and {Cleary}, J. and {Clements}, D.~L. and {Connors}, J. and {Contaldi}, C. and {Coppi}, G. and {Corbett}, L. and {Cothard}, N.~F. and {Coulton}, W. and {Crowley}, K.~D. and {Crowley}, K.~T. and {Cukierman}, A. and {D'Ewart}, J.~M. and {Dachlythra}, K. and {Datta}, R. and {Day-Weiss}, S. and {de Haan}, T. and {Devlin}, M. and {Di Mascolo}, L. and {Dicker}, S. and {Dober}, B. and {Doux}, C. and {Dow}, P. and {Doyle}, S. and {Duell}, C.~J. and {Duff}, S.~M. and {Duivenvoorden}, A.~J. and {Dunkley}, J. and {Dutcher}, D. and {D{\"u}nner}, R. and {Edenton}, M. and {El Bouhargani}, H. and {Errard}, J. and {Fabbian}, G. and {Fanfani}, V. and {Farren}, G.~S. and {Fergusson}, J. and {Ferraro}, S. and {Flauger}, R. and {Foster}, A. and {Freese}, K. and {Frisch}, J.~C. and {Frolov}, A. and {Fuller}, G. and {Galitzki}, N. and {Gallardo}, P.~A. and {Galvez Ghersi}, J.~T. and {Ganga}, K. and {Gao}, J. and {Garrido}, X. and {Gawiser}, E. and {Gerbino}, M. and {Gerras}, R. and {Giardiello}, S. and {Gill}, A. and {Gilles}, V. and {Giri}, U. and {Gleave}, E. and {Gluscevic}, V. and {Goeckner-Wald}, N. and {Golec}, J.~E. and {Gordon}, S. and {Gralla}, M. and {Gratton}, S. and {Green}, D. and {Groh}, J.~C. and {Groppi}, C. and {Guan}, Y. and {Gupta}, N. and {Gudmundsson}, J.~E. and {Hagstotz}, S. and {Hargrave}, P. and {Haridas}, S. and {Harrington}, K. and {Harrison}, I. and {Hasegawa}, M. and {Hasselfield}, M. and {Haynes}, V. and {Hazumi}, M. and {He}, A. and {Healy}, E. and {Henderson}, S.~W. and {Hensley}, B.~S. and {Hertig}, E. and {Herv{\'\i}as-Caimapo}, C. and {Higuchi}, M. and {Hill}, C.~A. and {Hill}, J.~C. and {Hilton}, G. and {Hilton}, M. and {Hincks}, A.~D. and {Hinshaw}, G. and {Hlo{\v{z}}ek}, R. and {Ho}, A.~Y.~Q. and {Ho}, S. and {Ho}, S.~P. and {Hoang}, T.~D. and {Hoh}, J. and {Hornecker}, E. and {Hornsby}, A.~L. and {Hotinli}, S.~C. and {Huang}, Z. and {Huber}, Z.~B. and {Hubmayr}, J. and {Huffenberger}, K. and {Hughes}, J.~P. and {Idicherian Lonappan}, A. and {Ikape}, M. and {Irwin}, K. and {Iuliano}, J. and {Jaffe}, A.~H. and {Jain}, B. and {Jense}, H.~T. and {Jeong}, O. and {Johnson}, A. and {Johnson}, B.~R. and {Johnson}, M. and {Jones}, M. and {Jost}, B. and {Kaneko}, D. and {Karpel}, E.~D. and {Kasai}, Y. and {Katayama}, N. and {Keating}, B. and {Keller}, B. and {Keskitalo}, R. and {Kim}, J. and {Kisner}, T. and {Kiuchi}, K.},
        title = "{The Simons Observatory: science goals and forecasts for the enhanced Large Aperture Telescope}",
        collaboration = "Simons Observatory",
      journal = {\jcap},
         year = 2025,
        month = aug,
       volume = {2025},
       number = {8},
          eid = {034},
        pages = {034},
          doi = {10.1088/1475-7516/2025/08/034},
archivePrefix = {arXiv},
       eprint = {2503.00636},
 primaryClass = {astro-ph.IM},
       adsurl = {https://ui.adsabs.harvard.edu/abs/2025JCAP...08..034A}
}

@article{springel03,
    author = {Springel, Volker and Hernquist, Lars},
    title = {Cosmological smoothed particle hydrodynamics simulations: a hybrid multiphase model for star formation},
    journal = {Monthly Notices of the Royal Astronomical Society},
    volume = {339},
    number = {2},
    pages = {289-311},
    year = {2003},
    month = {02},
    issn = {0035-8711},
    doi = {10.1046/j.1365-8711.2003.06206.x},
    url = {https://doi.org/10.1046/j.1365-8711.2003.06206.x},
    eprint = {https://academic.oup.com/mnras/article-pdf/339/2/289/3230545/339-2-289.pdf},
}

@ARTICLE{sunyaev72,
       author = {{Sunyaev}, R.~A. and {Zeldovich}, Ya. B.},
        title = "{The Observations of Relic Radiation as a Test of the Nature of X-Ray Radiation from the Clusters of Galaxies}",
      journal = {Comments on Astrophysics and Space Physics},
         year = 1972,
        month = nov,
       volume = {4},
        pages = {173},
       adsurl = {https://ui.adsabs.harvard.edu/abs/1972CoASP...4..173S}
}

@ARTICLE{swetz11,
       author = {{Swetz}, D.~S. and {Ade}, P.~A.~R. and {Amiri}, M. and {Appel}, J.~W. and {Battistelli}, E.~S. and {Burger}, B. and {Chervenak}, J. and {Devlin}, M.~J. and {Dicker}, S.~R. and {Doriese}, W.~B. and {D{\"u}nner}, R. and {Essinger-Hileman}, T. and {Fisher}, R.~P. and {Fowler}, J.~W. and {Halpern}, M. and {Hasselfield}, M. and {Hilton}, G.~C. and {Hincks}, A.~D. and {Irwin}, K.~D. and {Jarosik}, N. and {Kaul}, M. and {Klein}, J. and {Lau}, J.~M. and {Limon}, M. and {Marriage}, T.~A. and {Marsden}, D. and {Martocci}, K. and {Mauskopf}, P. and {Moseley}, H. and {Netterfield}, C.~B. and {Niemack}, M.~D. and {Nolta}, M.~R. and {Page}, L.~A. and {Parker}, L. and {Staggs}, S.~T. and {Stryzak}, O. and {Switzer}, E.~R. and {Thornton}, R. and {Tucker}, C. and {Wollack}, E. and {Zhao}, Y.},
        title = "{Overview of the Atacama Cosmology Telescope: Receiver, Instrumentation, and Telescope Systems}",
      journal = {\apjs},
         year = 2011,
        month = jun,
       volume = {194},
       number = {2},
          eid = {41},
        pages = {41},
          doi = {10.1088/0067-0049/194/2/41},
archivePrefix = {arXiv},
       eprint = {1007.0290},
 primaryClass = {astro-ph.IM},
       adsurl = {https://ui.adsabs.harvard.edu/abs/2011ApJS..194...41S}
}

@ARTICLE{tinker08,
       author = {{Tinker}, Jeremy and {Kravtsov}, Andrey V. and {Klypin}, Anatoly and {Abazajian}, Kevork and {Warren}, Michael and {Yepes}, Gustavo and {Gottl{\"o}ber}, Stefan and {Holz}, Daniel E.},
        title = "{Toward a Halo Mass Function for Precision Cosmology: The Limits of Universality}",
      journal = {\apj},
         year = 2008,
        month = dec,
       volume = {688},
       number = {2},
        pages = {709-728},
          doi = {10.1086/591439},
archivePrefix = {arXiv},
       eprint = {0803.2706},
 primaryClass = {astro-ph},
       adsurl = {https://ui.adsabs.harvard.edu/abs/2008ApJ...688..709T}
}

@ARTICLE{tinker10,
       author = {{Tinker}, Jeremy L. and {Robertson}, Brant E. and {Kravtsov}, Andrey V. and {Klypin}, Anatoly and {Warren}, Michael S. and {Yepes}, Gustavo and {Gottl{\"o}ber}, Stefan},
        title = "{The Large-scale Bias of Dark Matter Halos: Numerical Calibration and Model Tests}",
      journal = {\apj},
         year = 2010,
        month = dec,
       volume = {724},
       number = {2},
        pages = {878-886},
          doi = {10.1088/0004-637X/724/2/878},
archivePrefix = {arXiv},
       eprint = {1001.3162},
 primaryClass = {astro-ph.CO},
       adsurl = {https://ui.adsabs.harvard.edu/abs/2010ApJ...724..878T}
}

@ARTICLE{vavagiakis21,
       author = {{Vavagiakis}, E.~M. and {Gallardo}, P.~A. and {Calafut}, V. and {Amodeo}, S. and {Aiola}, S. and {Austermann}, J.~E. and {Battaglia}, N. and {Battistelli}, E.~S. and {Beall}, J.~A. and {Bean}, R. and {Bond}, J.~R. and {Calabrese}, E. and {Choi}, S.~K. and {Cothard}, N.~F. and {Devlin}, M.~J. and {Duell}, C.~J. and {Duff}, S.~M. and {Duivenvoorden}, A.~J. and {Dunkley}, J. and {Dunner}, R. and {Ferraro}, S. and {Guan}, Y. and {Hill}, J.~C. and {Hilton}, G.~C. and {Hilton}, M. and {Hlo{\v{z}}ek}, R. and {Huber}, Z.~B. and {Hubmayr}, J. and {Huffenberger}, K.~M. and {Hughes}, J.~P. and {Koopman}, B.~J. and {Kosowsky}, A. and {Li}, Y. and {Lokken}, M. and {Madhavacheril}, M. and {McMahon}, J. and {Moodley}, K. and {Naess}, S. and {Nati}, F. and {Newburgh}, L.~B. and {Niemack}, M.~D. and {Page}, L.~A. and {Partridge}, B. and {Schaan}, E. and {Schillaci}, A. and {Sif{\'o}n}, C. and {Spergel}, D.~N. and {Staggs}, S.~T. and {Ullom}, J.~N. and {Vale}, L.~R. and {Van Engelen}, A. and {Van Lanen}, J. and {Wollack}, E.~J. and {Xu}, Z.},
        title = "{The Atacama Cosmology Telescope: Probing the baryon content of SDSS DR15 galaxies with the thermal and kinematic Sunyaev-Zel'dovich effects}",
      journal = {\prd},
         year = 2021,
        month = aug,
       volume = {104},
       number = {4},
          eid = {043503},
        pages = {043503},
          doi = {10.1103/PhysRevD.104.043503},
archivePrefix = {arXiv},
       eprint = {2101.08373},
 primaryClass = {astro-ph.CO},
       adsurl = {https://ui.adsabs.harvard.edu/abs/2021PhRvD.104d3503V}
}

@ARTICLE{vikram17,
       author = {{Vikram}, Vinu and {Lidz}, Adam and {Jain}, Bhuvnesh},
        title = "{A Measurement of the Galaxy Group-Thermal Sunyaev-Zel'dovich Effect Cross-Correlation Function}",
      journal = {\mnras},
         year = 2017,
        month = may,
       volume = {467},
       number = {2},
        pages = {2315-2330},
          doi = {10.1093/mnras/stw3311},
archivePrefix = {arXiv},
       eprint = {1608.04160},
 primaryClass = {astro-ph.CO},
       adsurl = {https://ui.adsabs.harvard.edu/abs/2017MNRAS.467.2315V}
}

@ARTICLE{zeldovich69,
       author = {{Zeldovich}, Ya. B. and {Sunyaev}, R.~A.},
        title = "{The Interaction of Matter and Radiation in a Hot-Model Universe}",
      journal = {\apss},
         year = 1969,
        month = jul,
       volume = {4},
       number = {3},
        pages = {301-316},
          doi = {10.1007/BF00661821},
       adsurl = {https://ui.adsabs.harvard.edu/abs/1969Ap&SS...4..301Z}
}

@ARTICLE{zhou20,
       author = {{Zhou}, Rongpu and {Newman}, Jeffrey A. and {Dawson}, Kyle S. and {Eisenstein}, Daniel J. and {Brooks}, David D. and {Dey}, Arjun and {Dey}, Biprateep and {Duan}, Yutong and {Eftekharzadeh}, Sarah and {Gazta{\~n}aga}, Enrique and {Kehoe}, Robert and {Landriau}, Martin and {Levi}, Michael E. and {Licquia}, Timothy C. and {Meisner}, Aaron M. and {Moustakas}, John and {Myers}, Adam D. and {Palanque-Delabrouille}, Nathalie and {Poppett}, Claire and {Prada}, Francisco and {Raichoor}, Anand and {Schlegel}, David J. and {Schubnell}, Michael and {Staten}, Ryan and {Tarl{\'e}}, Gregory and {Y{\`e}che}, Christophe},
        title = "{Preliminary Target Selection for the DESI Luminous Red Galaxy (LRG) Sample}",
      journal = {Research Notes of the American Astronomical Society},
         year = 2020,
        month = oct,
       volume = {4},
       number = {10},
          eid = {181},
        pages = {181},
          doi = {10.3847/2515-5172/abc0f4},
archivePrefix = {arXiv},
       eprint = {2010.11282},
 primaryClass = {astro-ph.CO},
       adsurl = {https://ui.adsabs.harvard.edu/abs/2020RNAAS...4..181Z}
}

@ARTICLE{zhou23.1,
       author = {{Zhou}, Rongpu and {Dey}, Biprateep and {Newman}, Jeffrey A. and {Eisenstein}, Daniel J. and {Dawson}, K. and {Bailey}, S. and {Berti}, A. and {Guy}, J. and {Lan}, Ting-Wen and {Zou}, H. and {Aguilar}, J. and {Ahlen}, S. and {Alam}, Shadab and {Brooks}, D. and {de la Macorra}, A. and {Dey}, A. and {Dhungana}, G. and {Fanning}, K. and {Font-Ribera}, A. and {Gontcho}, S. Gontcho A. and {Honscheid}, K. and {Ishak}, Mustapha and {Kisner}, T. and {Kov{\'a}cs}, A. and {Kremin}, A. and {Landriau}, M. and {Levi}, Michael E. and {Magneville}, C. and {Manera}, Marc and {Martini}, P. and {Meisner}, Aaron M. and {Miquel}, R. and {Moustakas}, J. and {Myers}, Adam D. and {Nie}, Jundan and {Palanque-Delabrouille}, N. and {Percival}, W.~J. and {Poppett}, C. and {Prada}, F. and {Raichoor}, A. and {Ross}, A.~J. and {Schlafly}, E. and {Schlegel}, D. and {Schubnell}, M. and {Tarl{\'e}}, Gregory and {Weaver}, B.~A. and {Wechsler}, R.~H. and {Y{\'e}che}, Christophe and {Zhou}, Zhimin},
        title = "{Target Selection and Validation of DESI Luminous Red Galaxies}",
      journal = {\aj},
         year = 2023,
        month = feb,
       volume = {165},
       number = {2},
          eid = {58},
        pages = {58},
          doi = {10.3847/1538-3881/aca5fb},
archivePrefix = {arXiv},
       eprint = {2208.08515},
 primaryClass = {astro-ph.CO},
       adsurl = {https://ui.adsabs.harvard.edu/abs/2023AJ....165...58Z}
}

@ARTICLE{zhou23.2,
       author = {{Zhou}, Rongpu and {Ferraro}, Simone and {White}, Martin and {DeRose}, Joseph and {Sailer}, Noah and {Aguilar}, Jessica and {Ahlen}, Steven and {Bailey}, Stephen and {Brooks}, David and {Claybaugh}, Todd and {Dawson}, Kyle and {de la Macorra}, Axel and {Dey}, Biprateep and {Doel}, Peter and {Font-Ribera}, Andreu and {Forero-Romero}, Jaime E. and {Gontcho A Gontcho}, Satya and {Guy}, Julien and {Kremin}, Anthony and {Lambert}, Andrew and {Le Guillou}, Laurent and {Levi}, Michael and {Magneville}, Christophe and {Manera}, Marc and {Meisner}, Aaron and {Miquel}, Ramon and {Moustakas}, John and {Myers}, Adam D. and {Newman}, Jeffrey A. and {Nie}, Jundan and {Percival}, Will and {Rezaie}, Mehdi and {Rossi}, Graziano and {Sanchez}, Eusebio and {Schlegel}, David and {Schubnell}, Michael and {Seo}, Hee-Jong and {Tarl{\'e}}, Gregory and {Zhou}, Zhimin},
        title = "{DESI luminous red galaxy samples for cross-correlations}",
      journal = {\jcap},
         year = 2023,
        month = nov,
       volume = {2023},
       number = {11},
          eid = {097},
        pages = {097},
          doi = {10.1088/1475-7516/2023/11/097},
archivePrefix = {arXiv},
       eprint = {2309.06443},
 primaryClass = {astro-ph.CO},
       adsurl = {https://ui.adsabs.harvard.edu/abs/2023JCAP...11..097Z}
}

@article{popik_impacts_2025,
    title = {On the impacts of halo model implementations in {Sunyaev}-{Zeldovich} cross-correlation analyses},
    volume = {2025},
    issn = {1475-7516},
    url = {https://ui.adsabs.harvard.edu/abs/2025JCAP...10..051P},
    doi = {10.1088/1475-7516/2025/10/051},
    urldate = {2026-07-17},
    journal = {Journal of Cosmology and Astroparticle Physics},
    publisher = {IOP},
    author = {Popik, Chad and Battaglia, Nicholas and Kusiak, Aleksandra and Bolliet, Boris and Colin Hill, J.},
    month = oct,
    year = {2025},
    note = {ADS Bibcode: 2025JCAP...10..051P},
    pages = {051},
}

\end{document}